\documentclass[referee]{raa}           
\usepackage{graphicx,times}
\usepackage{natbib}
\usepackage{amssymb,amsmath}
\bibpunct{(}{)}{;}{a}{}{,}

\usepackage[a4paper=true,dvipdfm=true,pagebackref=true]{hyperref}
\hypersetup{pdftitle = The title of my PDF, pdfauthor = My name, pdfsubject= The subject, pdfkeywords = keyword1 keyword2 keyword3} 
\hypersetup{colorlinks = true, linkcolor = green, anchorcolor = red, citecolor = blue, filecolor = red, pagecolor = red, urlcolor = red}

\begin{document}

   \title{Simulation of old open clusters for UVIT on ASTROSAT}


 \volnopage{ {\bf 2012} Vol.\ {\bf X} No. {\bf XX}, 000--000}
   \setcounter{page}{1}

   \author{Sindhu N\inst{1}, Annapurni Subramaniam\inst{2}, C Anu Radha\inst{1}
   }

   \institute{School of Advanced Sciences, VIT University , Vellore 632014, India; \\
        \and
             Indian Institute of Astrophysics, Koramangala, II block, Bangalore 560034, India; {purni@iiap.res.in}\\
	\vs \no
}

\abstract{ Ultra Violet Imaging Telescope (UVIT) is one of the payloads on the first Indian multi wavelength satellite ASTROSAT expected to be launched by Indian Space Research Organisation (ISRO) in the year 2015. We have performed simulations of UV studies of old open clusters for the UVIT. The colour magnitude diagrams (CMDs) and spatial appearances have been created using 10 filters of FUV channel ( 130 - 180 nm) and NUV channel (200 - 300 nm) available for observations on the UVIT, for three old open clusters M67, NGC 188 and NGC 6791. The CMDs are simulated for different filter combinations, and they are used to identify the  loci of various evolutionary sequences, white dwarfs, blue stragglers, red giants, sub giants, turn off stars and the main sequence of the clusters. The present work helps in identifying the potential area of study in the case of three old open clusters, by considering the availability of filters and the detection limits of the instrument. We also recommend filter combinations, which can be used to detect and study the above mentioned evolutionary stages. The simulations and the results presented here are essential for the optimal use of the UVIT for studies of old open clusters.
}

   \authorrunning{Sindhu. N, Annapurni. Subramaniam, C. Anu Radha }            
   \titlerunning{Simulation of old open clusters for UVIT on ASTROSAT}  
   \maketitle

%
\section{Introduction}           
\label{sect:intro}
Ultraviolet (UV) studies have a vital role in the understanding of stellar evolutionary stages such as blue stragglers, white dwarfs WDs, hot sub dwarfs and other exotic population of hot stars. Previous studies in the UV region have shown spectacular results and  have added vital information to stellar evolutionary studies. The UV images of older clusters provide one of the best means to study low mass hot stars, sub dwarfs,  WDs, blue stragglers and horizontal branch stars. In the UV range, most of the cool stars appear very faint, resulting in lesser crowding of stars in the cluster region. Hence one can easily identify peculiar stars that emit in the UV and study their properties. With the need to study the celestial objects in the UV band, many missions have been previously launched and UVIT will add to the list with its unique characteristics of having a wide field of view of 0.5 degree, high spatial resolution of 1.8 arc seconds and a large set of filters will allow scientific community to pursue major scientific research in the UV region, both new and as well as follow ups of recent discoveries. \cite{seigel14}, in their table 1, lists out previous UV missions and their characteristics.
														
UVIT has a twin 38 cm telescope with three channels: far-ultraviolet (FUV:130 - 180 nm), near-ultraviolet (NUV:200 - 300 nm) and Visible (350 - 550 nm). One of the telescopes carries the FUV channel and other telescope carries both NUV and Visible channels. The incoming beam is divided by the dichroic beam splitter, where the visible beam is transmitted and NUV beam is reflected. The visible channel will primarily provide pointing information. All three channels use Micro Channel Plate (MCP) based intensified Star250 CMOS imaging detectors to record frames at the rate of 29frames/sec. UVIT aims to provide flux calibrated images. The images on the detector are recorded in either photon counting mode (high gain) or integrating mode (low gain). 

The UVIT carries a set of filters mounted on the filter wheel located close to the focal plane between the primary mirror and the detector, for selection of the spectral band within each of the three channels (viz., FUV, NUV and VIS). 5 filters and 2 gratings for the FUV channel, 6 filters and a grating for the NUV channel and 5 filters for the visible channel are placed on the filter wheels of diameter 50 mm. The wheel also carries a blind to block any radiations from reaching directly on the detector.  The filter wheels  have an independent drive mechanism to bring the desired filter into the optical path of the telescope.The wavelength range of the 5 FUV filters are BaF2(137 - 175), Sapphire (146 - 175), Silica (166 - 178), CaF2-1 ( 130 - 175) and CaF2-2 ( 130 - 175)nm. The wavelength range of the 5 NUV filters are Silica(200 - 280), NUVB13 (230 - 260), NUVB15 (206 - 233), NUVB4 (250 - 278) and NUVN2 (275 - 284)nm. The pass band wavelength of all the filters as well as the effective area are given in \cite{annapurni12}.

For the present study, we have selected three well known old open clusters M67, NGC 188 and NGC 6791. This study presents estimations of flux  of various stars in these clusters in the UVIT filters and creates the fundamental colour magnitude diagram (CMD) in various filter combinations. The simulated CMDs developed and presented in this study will be used as diagnostic tools when the actual data is obtained from the mission. We have summarised the previous studies of the 3 open clusters below and the basic parameters are tabulated in table 1.


\subsection{M67}
An old open cluster, located at a distance of 908 pc in the constellation of cancer. The age of the cluster is found to be 4Gyr (\citealt{vandenberg04}). This old cluster has evolved phases of single as well as binary star evolution, such as blue stragglers,  WDs, cataclysmic variables, yellow giant with a  WD companion etc. \cite{landsman98} reported the analysis of the images obtained with the Ultraviolet Imaging Telescope (UIT) of the old open cluster M67, NGC 188 and NGC 6791. The UIT detected 20 stars in M67 cluster, which include 11 blue stragglers, 7  WDs and the yellow giant –  WD binary. A hot luminous companion for two blue stragglers (S975 and S1082) has been suggested by the Ultraviolet photometry. The study also shows the domination of the blue stragglers in the integrated spectrum at wavelength shorter than 2600 \AA. \cite{seigel14} presented the Swift Ultra Violet and Optical Telescope (UVOT) observations of this cluster, in the optical and NUV region and their CMDs. \cite{montgomery93} investigated the central one half degree of the old open cluster M67 using U, B, V and I photometric data to a limiting magnitude V=20. We use this photometry for our simulations. They have determined the possibility of 38\% of the stars in the cluster to be in binary systems. 

\subsection{NGC 188}
NGC 188 is one of the older clusters in our galaxy, studied widely due to its richness in stars and also heavy elements, age and its location in the galactic plane. The cluster is located at a distance of 2047 pc. The age of the cluster is determined to be 6.0 Gyr (\citealt{caputo90})and more recently found to be 6.8 Gyr (\citealt{vandenberg04}). The metallicity of the cluster is found to be [Fe/H] = 0.075$\pm$0.045 from the analysis of low resolution spectra of K giants (\citealt{Worthey03}). This cluster is also one of the important clusters chosen by the WIYN Open Cluster Study (\citealt{mathieu00}). The cluster has been studied to understand the origin of blue stragglers and the existence of contact binaries (\citealt{leonard92}). The central concentration of blue stragglers is greater than the giants and 15 blue stragglers are observed in NGC 188 cluster (\citealt{McClure77}). In our study, we have used the photometric data of 1514 stars from (\citealt{Sarajedini99}). Recently, \cite{seigel14} presented the Swift UVOT observations of this cluster, in the optical and NUV region and their CMDs.

\subsection{NGC 6791}
NGC 6791 is one of the oldest open cluster candidates with an age of 8 Gyr (\citealt{chaboyer99}). It has been studied for its rich metal elements, large population of stars and evolved hot stars (\citealt{liebert94}). A photometric study of the cluster field in V and B led to the discovery of 17 variable stars (\citealt{kaluzny93}) with the majority being in the turn off region, of which eight were contact binaries and nine other variables. Two of the contact binaries which exhibit various evolutionary stages were observed to be blue stragglers. They suspect one of the variables to be a binary composed of a red dwarf and hot sub dwarf. \cite{montgomery94} used photoelectric and charge-coupled device (CCD) photometry to study this cluster and we have used this data for our simulations. The cluster has also been studied in the UV region by \cite{buzzoni12}, which shows a phenomenon of UV upturn similar to that observed in the most active UV- upturn ellipticals. This phenomenon is  primarily observed due to the hot stellar component with T$_{\textit{eff}}$ $\geq$10000K, which contributes UV luminosity of the cluster shortward of 2500\AA. The cluster also has a dozen of hot stars along the extreme horizontal-branch.
	
\begin{table}
\bc
\begin{minipage}[]{100mm}
\caption[]{Basic cluster parameters of the three old open clusters.\label{tab1}}\end{minipage}
\setlength{\tabcolsep}{2.5pt}
\small
\begin{tabular}{cccc}
\hline\noalign{\smallskip}
Parameter & M67 & NGC 188 & NGC 6791\\
\hline\noalign{\smallskip}
R.A(2000)&8h 51m 18s&00h 47m 28s&19h 20m 53s \\
Dec (2000)&+11˚ 48' 00''&+85˚ 15' 18''&+37˚ 46' 18''\\
Gal. longitude&215.696&122.843&69.959\\
Gal. latitude&31.896&22.384&10.904\\
Log age&9.409&9.632&9.643\\
E(B$-$V)(mag)&0.059&0.082&0.117\\
Distance(pc)&908&2047&4100\\
Metallicity&Solar&-0.02&0.15\\
\noalign{\smallskip}\hline
\end{tabular}
\ec
\end{table}

\begin{figure}
   \centering
   \includegraphics[width=10cm, angle=0]{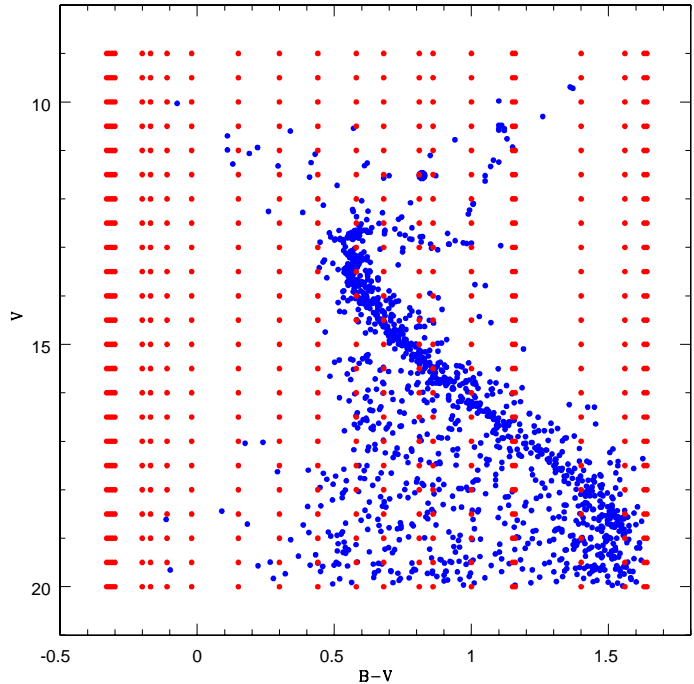}
   \caption{Parameter grid to obtain the count rate and magnitude for the old open cluster M67.} 
   \label{fig1}
   \end{figure}
   
The paper is arranged as follows. The section 2 deals with the methods used for the estimation of FUV and NUV magnitudes and simulations of CMDs, section 3 presents and discusses the FUV CMDs, NUV CMDs and NUV-Visual CMDs, section 4 discusses the colour-colour diagrams, section 5 presents the simulated spatial appearance of the clusters in a few of the FUV and NUV filters, section 6 provides a comparison with Galaxy Evolution Explorer: GALEX  and Swift - UVOT. The discussion and conclusions are presented in section 7.

\section{Simulation of magnitudes and CMDs}
\label{sect: Simulation}
We first estimate the expected flux for stars in all the UVIT filters in each of the open clusters, using the optical data mentioned in the previous section. We have used the V magnitude and (B$-$V) colour of stars to estimate the flux. In order to achieve this, we first estimate the expected count rate using the UVIT exposure time calculator (ETC) developed by Indian Institute of Astrophysics, available in the web site  http://uvit.iiap.res.in/ . 
The tool predicts the count rates for stars of specific spectral type and magnitude, by convolving the effective area curve of the filters with the kurucz model for the spectral type. The web site also gives the effective area curves of the filters used in this study. The tool estimates the count rates for one star at a time for a given magnitude and spectral type. As the star clusters have a large number of stars, with a range of magnitude and spectral type, it is not efficient to use this tool directly. Hence, we created a grid of count rates for a range of magnitudes (eg., V=10, 11, 12, 13, 14 etc..) and spectral types or its equivalent (B$-$V) (eg., (B$-$V) = -0.5, -0.3, -0.1, 0, ...), such that the cluster stars have magnitudes and colours within this range. As the ETC does not have the input as (B$-$V), the colour index, the data are taken from \cite{Martin90} for all the spectral types. That is, we generated count rates for all filters, for these fixed values of V and (B$-$V) and  created a grid as shown in figure \ref{fig1}, where we have also overplotted the CMD of M67. For each point on the grid, the count rates in all the 10 UVIT filters are obtained using the ETC. As stars in the cluster can have any value of V and (B$-$V), (in between the grid points, as shown in figure \ref{fig1}), we created a two dimensional interpolation method to estimate the count rate for any value of V and (B$-$V), within the grid. The interpolation should be able to handle multiple stars at a time, as the clusters have several hundreds to a few thousand stars. Thus the tool we have developed helps to compute the expected counts of a large number of stars with  known V and (B$-$V) values, in one go. The tool needs the grid values, in the form of a table, as input to the program. This tool will be made available to the community in a web based form.
 	
The present work utilizes the V and (B$-$V) data of 1271 stars of the old open cluster M67 from \cite{montgomery93} and similar data of 2 blue stragglers and 4  WD stars of the cluster from \cite{landsman98}. For the cluster NGC 188, similar data of 1514 stars are taken from \cite{Sarajedini99} and for NGC 6791, the data of 1654 stars from \cite{montgomery94}.
The above mentioned tool is used to estimate the count rates of all these stars in the 10 UVIT filters. Next task is to estimate the magnitudes based on the count rates. 
The FUV and NUV magnitudes are calculated using the formulae given below:
\begin{equation}
FUV mag = -2.5 * log (count rate) + 15
\end{equation}
\begin{equation}
NUV mag = -2.5 * log (count rate) + 16 
\end{equation}

The FUV and NUV magnitudes are derived by assuming the above zero-point. This zero-point is assumed based on the expected count rates for known sources. This number will be refined after the launch and is expected to change. Thus, the magnitudes and colours estimated and shown in the CMD are only indicative . As we do not use these for any quantitative estimates, but only as pointers to plan our observations, these estimations are considered for further analysis. We expect to obtain an AB magnitude of 20 mag in the FUV for an integration time of 200s. We expect to detect sources up to 20 mag in all filters, and may be a bit more fainter in the broad band filters.

The magnitudes estimated above are used to plot the cluster CMDs in various filter combinations and spatial appearance of the cluster in various filters. The filter combinations are selected based on their wavelength coverage. The cluster evolutionary sequences look different in the FUV and NUV filter combinations with respect to the optical CMD. The following section presents the UV CMDs, identification and discussion of various evolutionary sequences.

\begin{figure} [here]
\centerline{\includegraphics[width=7.5cm]{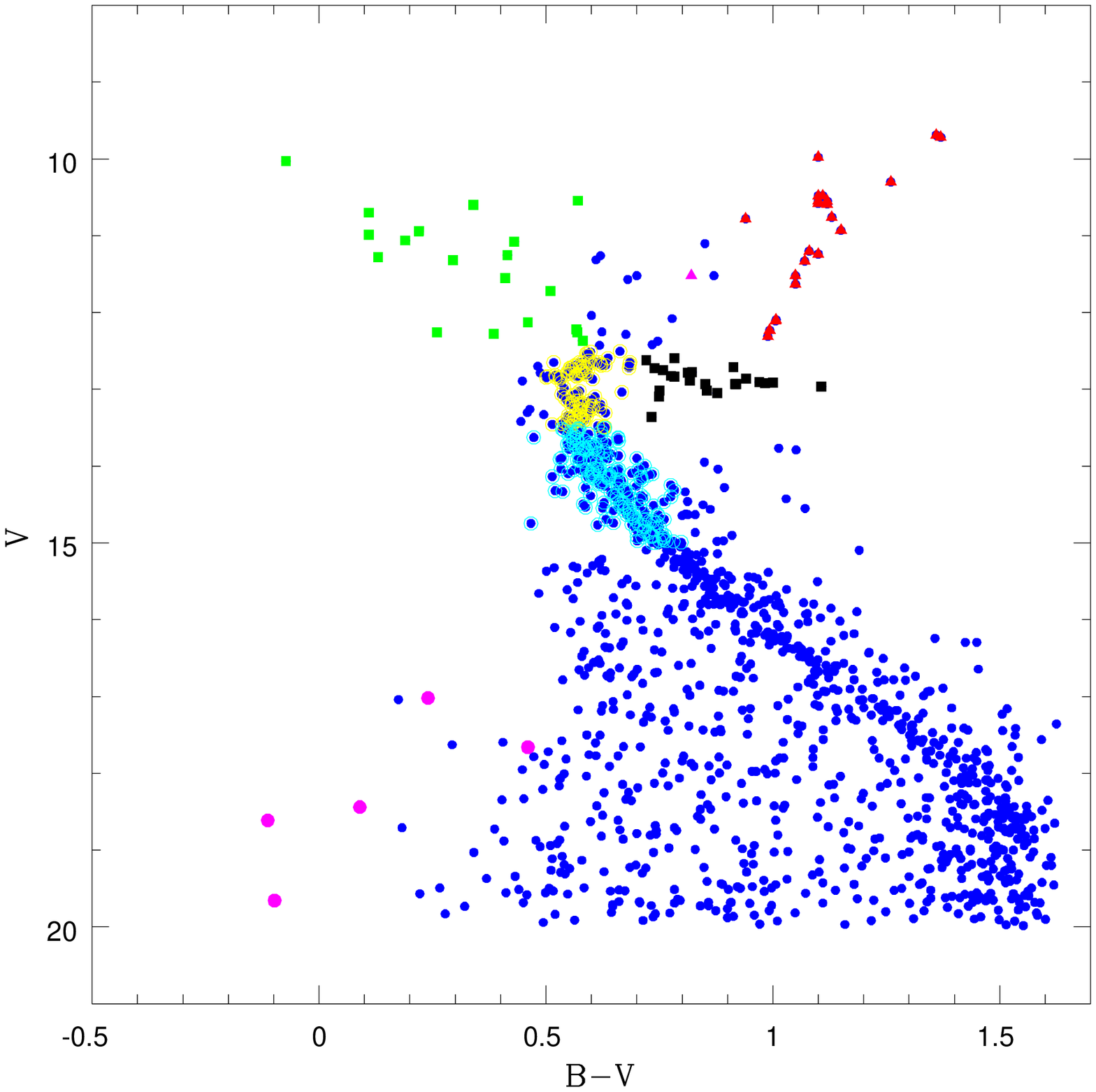} \qquad
     \includegraphics[width=7.5cm]{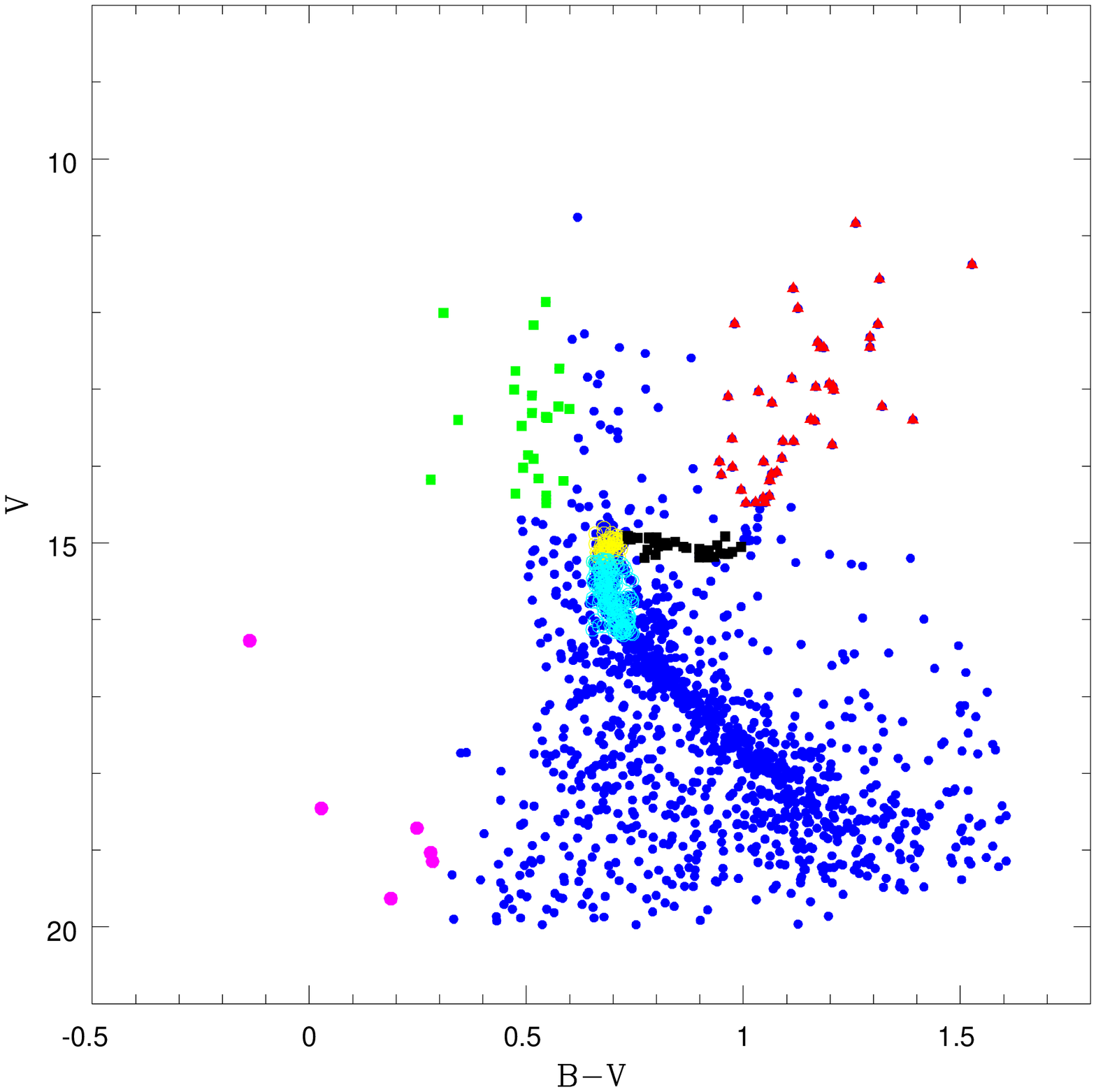}}
\centerline{\includegraphics[width=7.5cm]{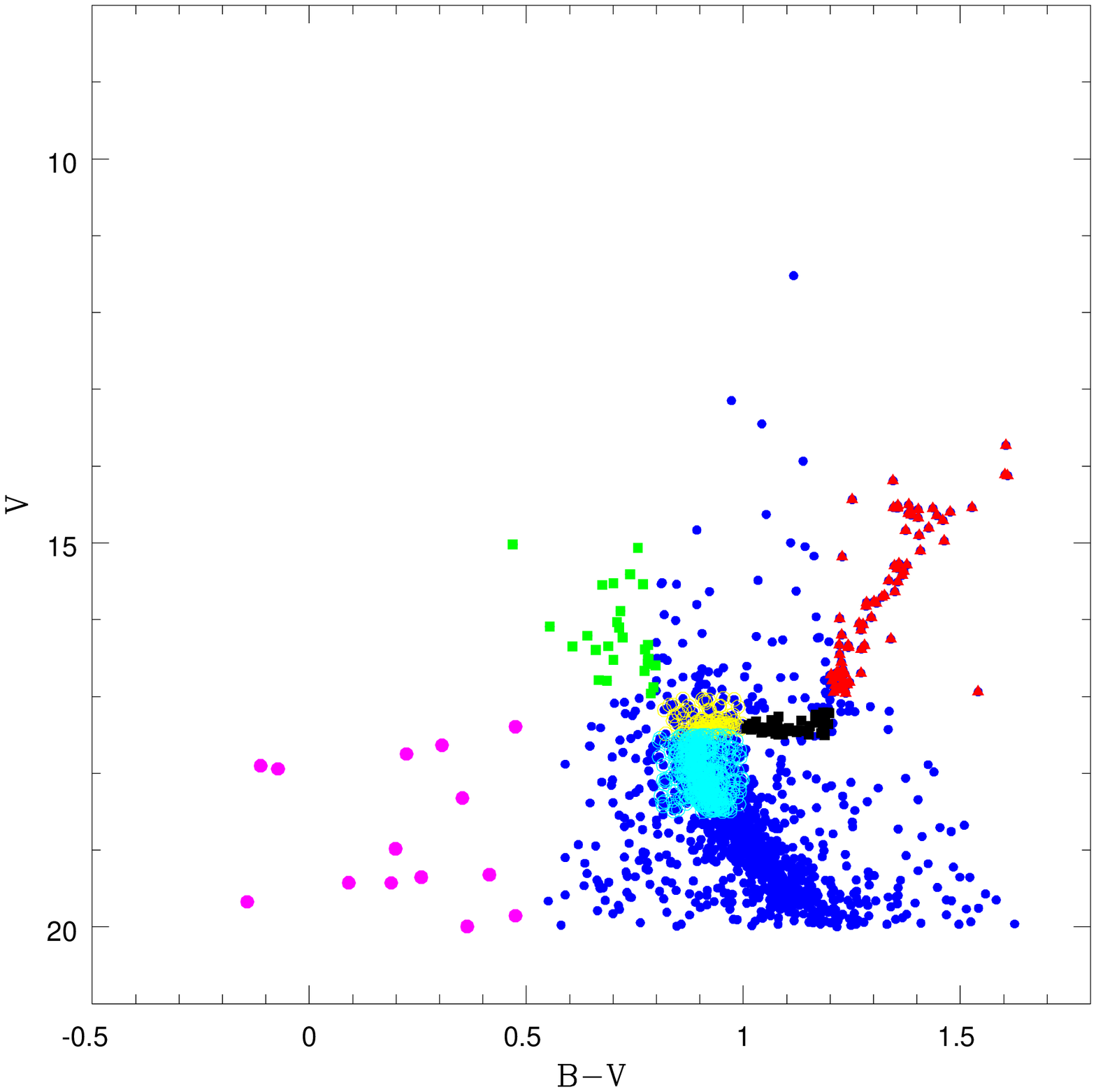}} \qquad
\caption{Optical Colour Magnitude diagram of M67 (top left), NGC 188 (top right) and NGC 6791 (bottom) , stars selected for the study are blue stragglers (green filled squares), white dwarf(magenta filled circle), red giants( red filled triangle), sub giants (black filled square), turn off (yellow open circle) and top main sequence (cyan open circle).\label{fig2}}

\end{figure}

\section{Results: Simulated UV CMDs of M67, NGC 188 and NGC 6791}
\label{sect:results}
One of the main aims of this study is to identify the locations of various evolutionary sequences in the UV CMD with the help of known clusters and their population. In order to achieve this, we need to mark stars in various sequences in the optical CMD. Thus,
a CMD is plotted with V on the vertical axis and B$-$V on the horizontal axis (fig. \ref{fig2}) using the above mentioned data. We have selected stars with magnitude V $<$ 20. In the Figure \ref{fig2}, blue stragglers,  WDs, red  giant stars have been identified, along with the sub giants, turn off and main sequence stars for the old open clusters M67, NGC 188 and NGC 6791. The different coloured symbols are used in the plot to cross identify them in FUV and NUV plots. These symbols have the same meaning in all the FUV and NUV  CMDs. 

The region between V $<$ 12.4 and (B$-$V) $<$ 0.6 is identified as the location of blue stragglers for M67, V $\leq$ 14.5 and (B$-$V) $\leq$ 0.6 for the cluster NGC 188 and V $<$ 17 and (B$-$V) $<$ 0.7 for the cluster NGC 6791 as seen in the figure \ref{fig2}. Blue stragglers are located within a magnitude range of 3 mag. In the figure \ref{fig2} the sub giant branch is identified within the region 0.7 $<$ (B$-$V) $<$ 1.2 and 12.5 $<$ V $<$ 13.5 for M67, 0.73 $<$ (B$-$V) $<$1.4 and 14.8 $<$ V $<$ 15.2 for NGC 188 and 1 $<$ (B$-$V) $<$ 1.2 and 17.2 $<$ V $<$ 17.5 for NGC 6791. The location of the red giant branch is in the  region demarcated by (B$-$V) $>$ 0.9 and V $<$ 12.5 for the M67 cluster, (B$-$V) $\geq$ 0.9 and V $\leq$ 14.5 for NGC 188 cluster and (B$-$V) $\geq$ 1.2 and V$<$ 17 for NGC 6791 cluster. The  WDs are in the hotter and fainter part of the CMD. The yellow giant  WD binary is located at a V = 11.52 and (B$-$V) = 0.82 in the M67 cluster.

\subsection{FUV colour magnitude diagrams}
FUV channel has 5 filters BaF2, CaF2-1, CaF2-2, Sapphire and Silica. The FUV filter wavelength range lies between  130 to 180 nm. As the two CaF2 filters are similar, we used only CaF2-1 for our analysis. We plotted CMDs for various combinations of these filters. For  this study,  we present the CMD up to 23 magnitude. We estimated the expected error to be ~0.2 mag at FUV~20 mag, based on the error in the corresponding V and (B$-$V) values. We do not show the errors as it makes it difficult to interpret the CMD. In all the FUV CMDs, we could detect blue stragglers,  WDs and tip of the turn off stars for the cluster M67. The tip of the turn off stars are not observed in NGC 188 and only the blue stragglers are observed in the cluster NGC 6791. In M67, we identify stars near the turn-off (yellow points) and those below the turn-off (cyan points) located in the very similar area of the CMD. We also notice that these stars are located either very close or mixed up making it difficult to identify a distinct location of these evolutionary phases.   
\subsubsection{CMDs using BaF2 filter}
\begin{figure} [here]
\centerline{\includegraphics[width=7.5cm]{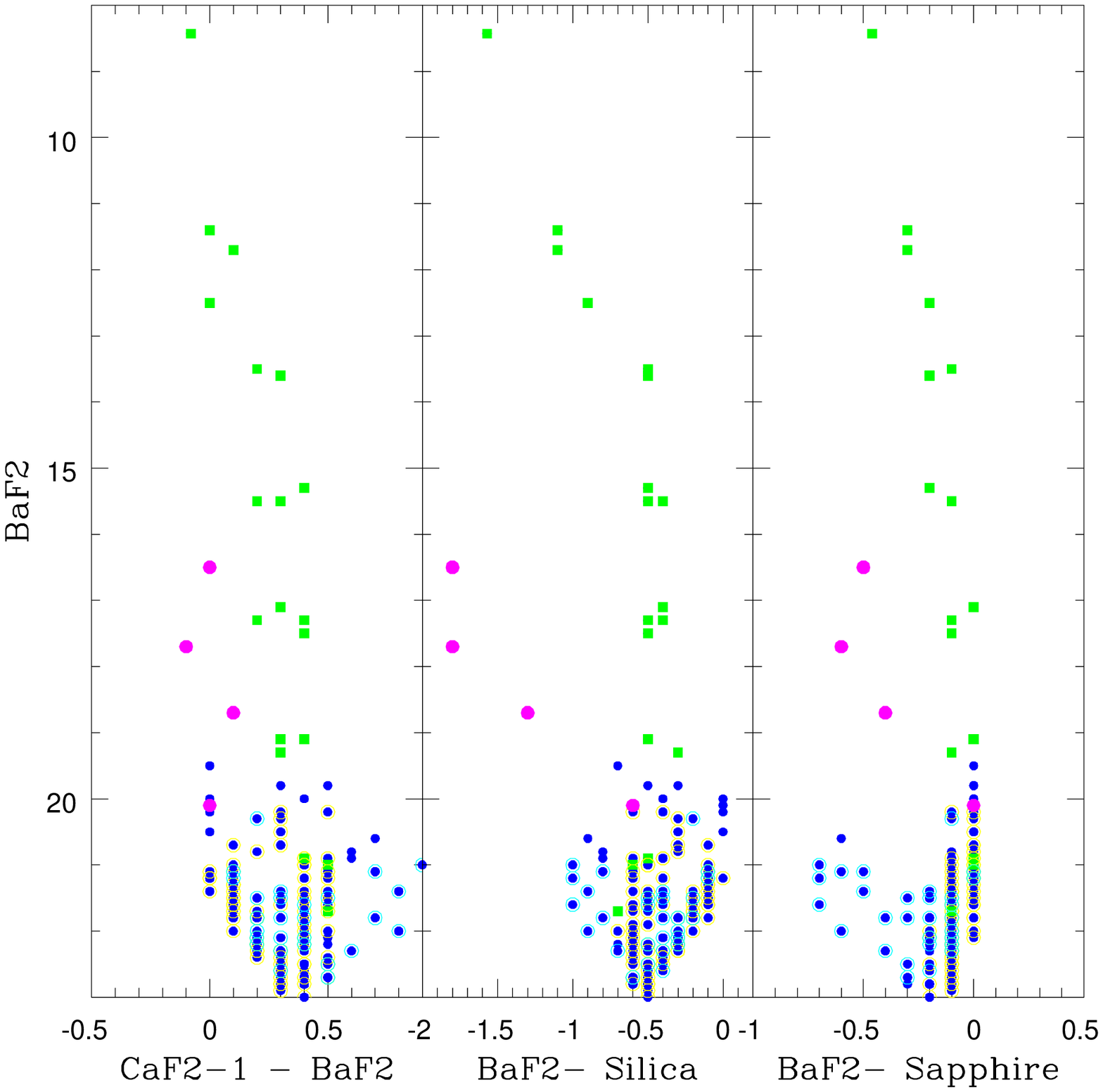} \qquad
     \includegraphics[width=7.5cm]{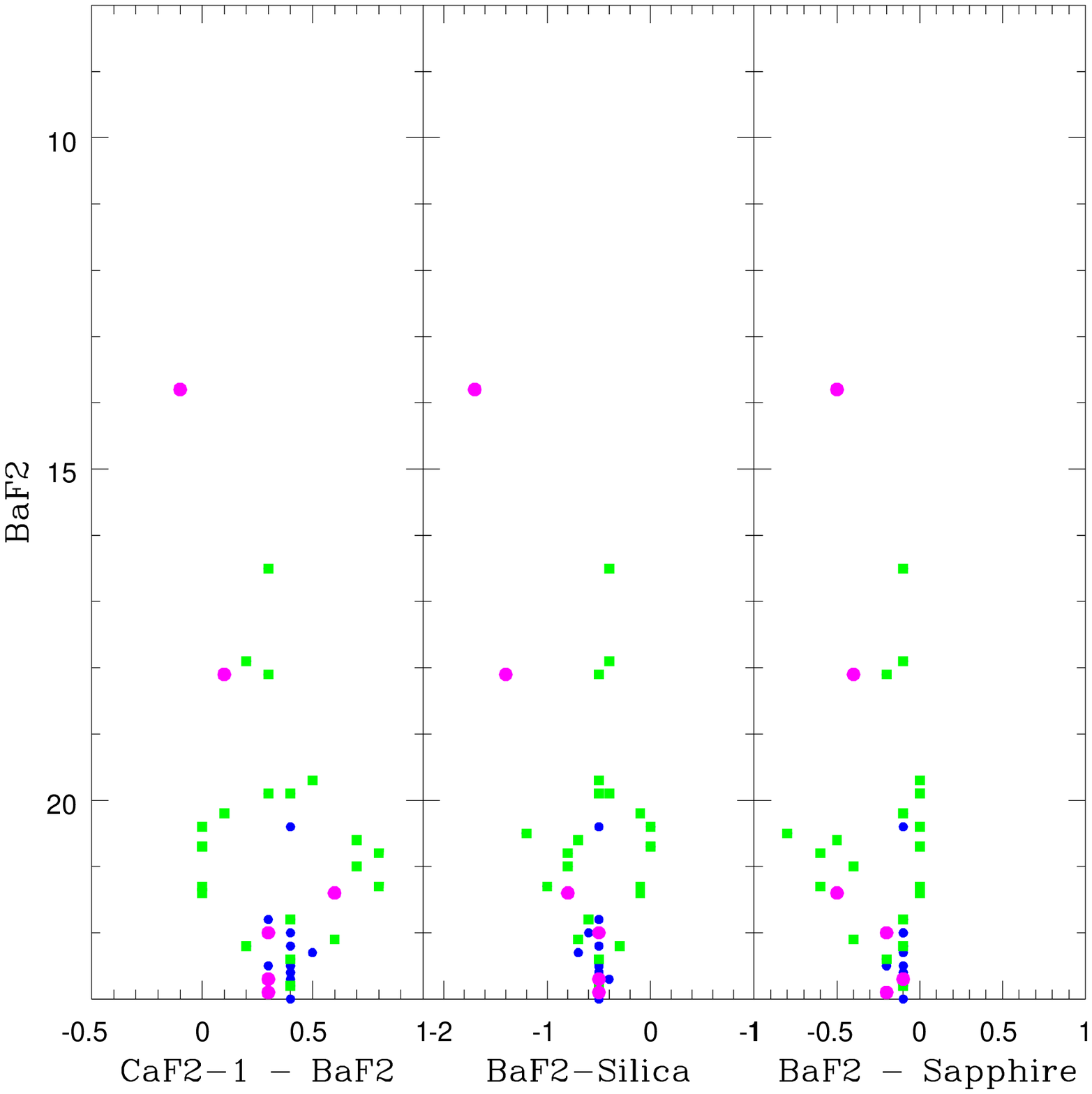}}
\centerline{\includegraphics[width=7.5cm]{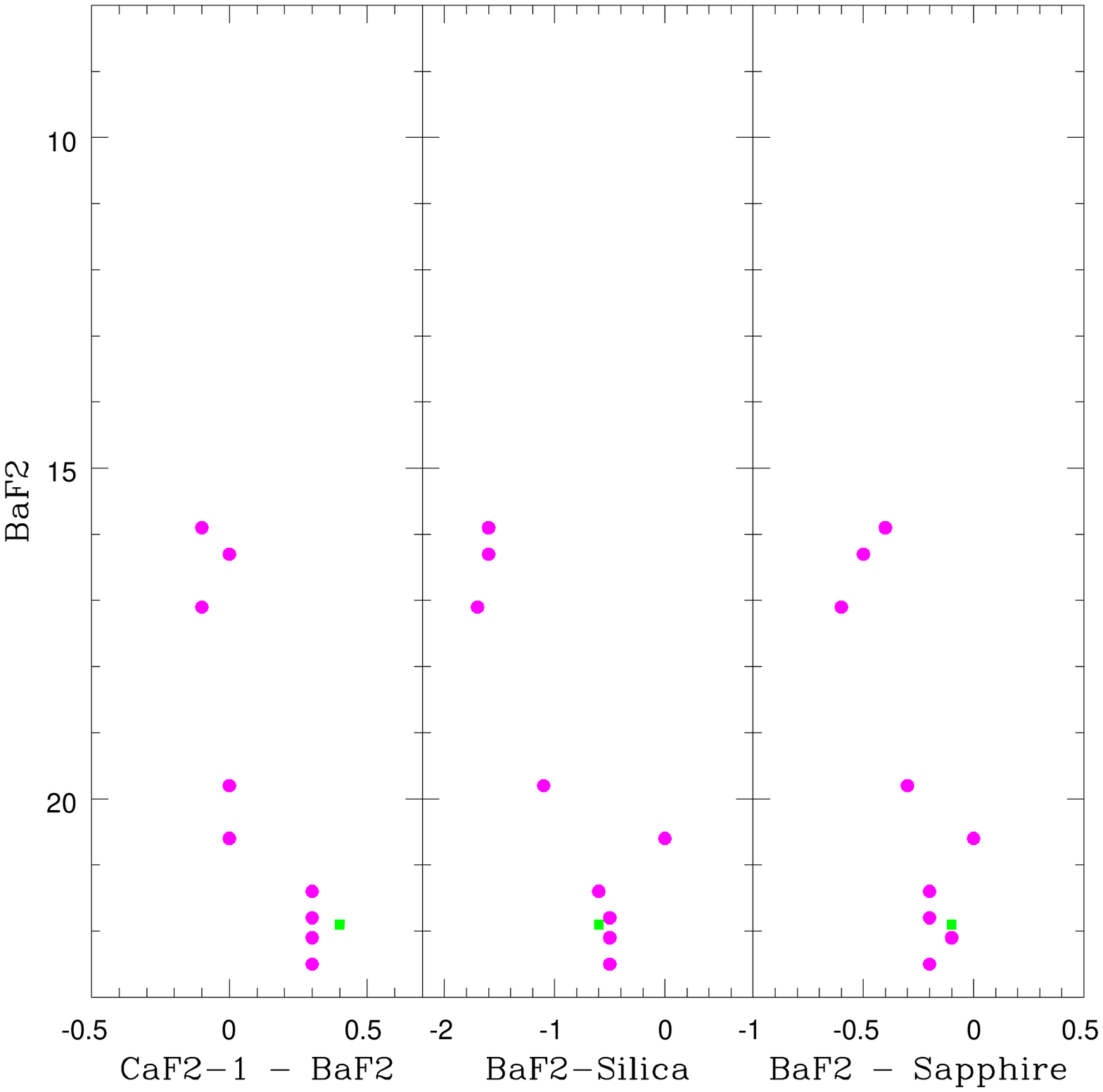}} \qquad
\caption{FUV CMD of M67 (top left), NGC 188 (top right) and NGC 6791 (bottom) (a) CaF2-1$-$BaF2, (b) BaF2$-$Silica and (c) BaF2$-$Sapphire vs. BaF2\label{fig3}}

\end{figure}

We created three CMDs using BaF2 magnitudes and 3 colours estimated using CaF2-1, Sapphire and Silica. The three CMDs for each cluster are shown in figure \ref{fig3} . We observe that the magnitude range occupied by the blue stragglers get stretched from $\approx$ 3 magnitude in the optical CMD, to about 13 magnitudes in M67 and 7 magnitudes in NGC 188. We detect only one blue straggler in the NGC 6791 cluster. It is to be noted that the blue stragglers are treated as normal main sequence(MS) stars, when the count rates are estimated. The  WDs and the blue stragglers are seen to be well separated when the BaF2-Silica colour is used, as illustrated in the middle CMD of M67 and NGC 188. This filter combination also provides a large colour range for the WDs ($\approx$ 1.5 magnitude).
In these CMDs, only the tip of the turn off stars are observed in M67 and whereas only the blue stragglers and  WDs are observed in NGC 188 and NGC 6791.
\subsubsection{CMDs using CaF2 filter}
\begin{figure} [here]
\centerline{\includegraphics[width=7.5cm]{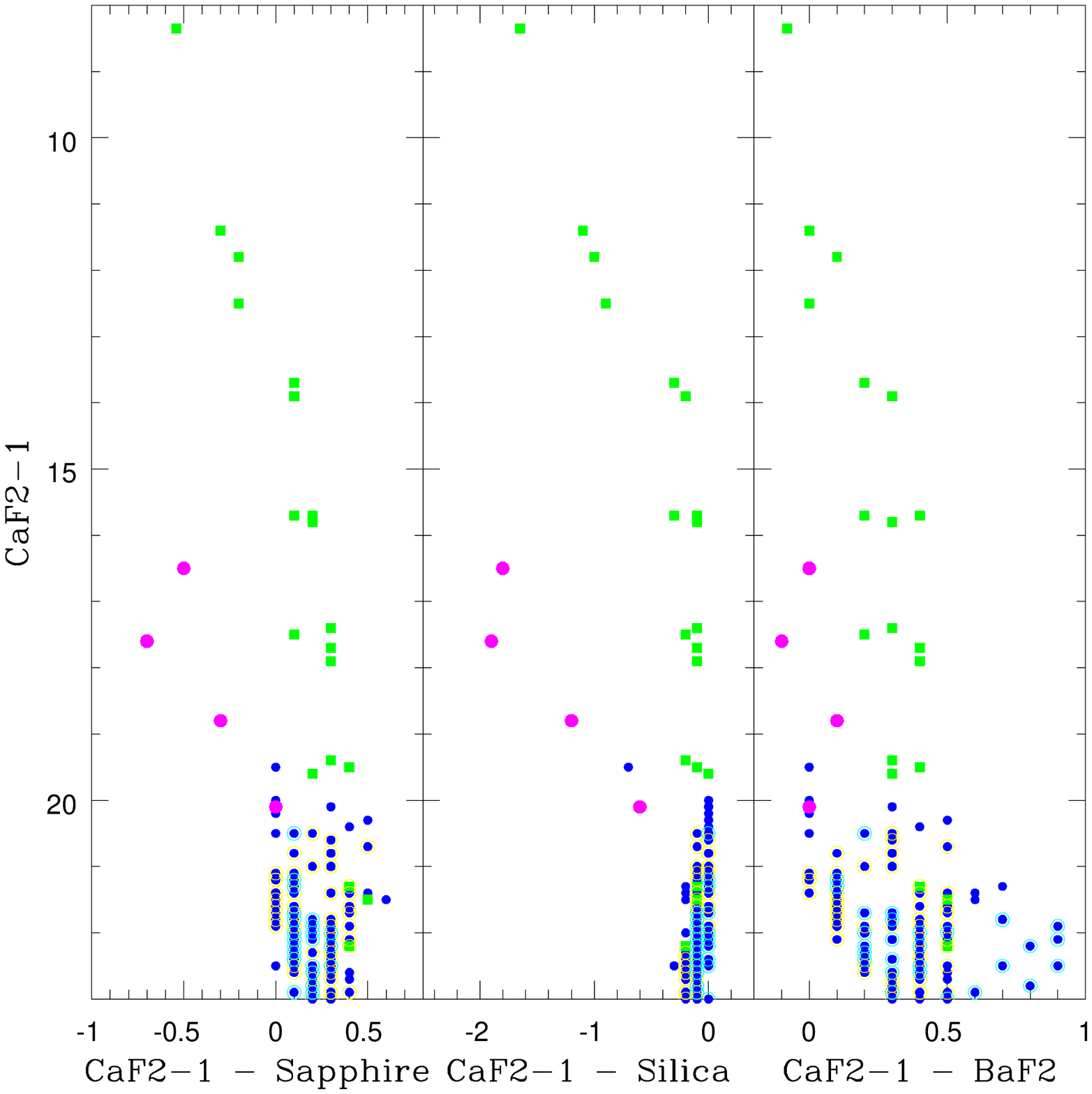} \qquad
     \includegraphics[width=7.5cm]{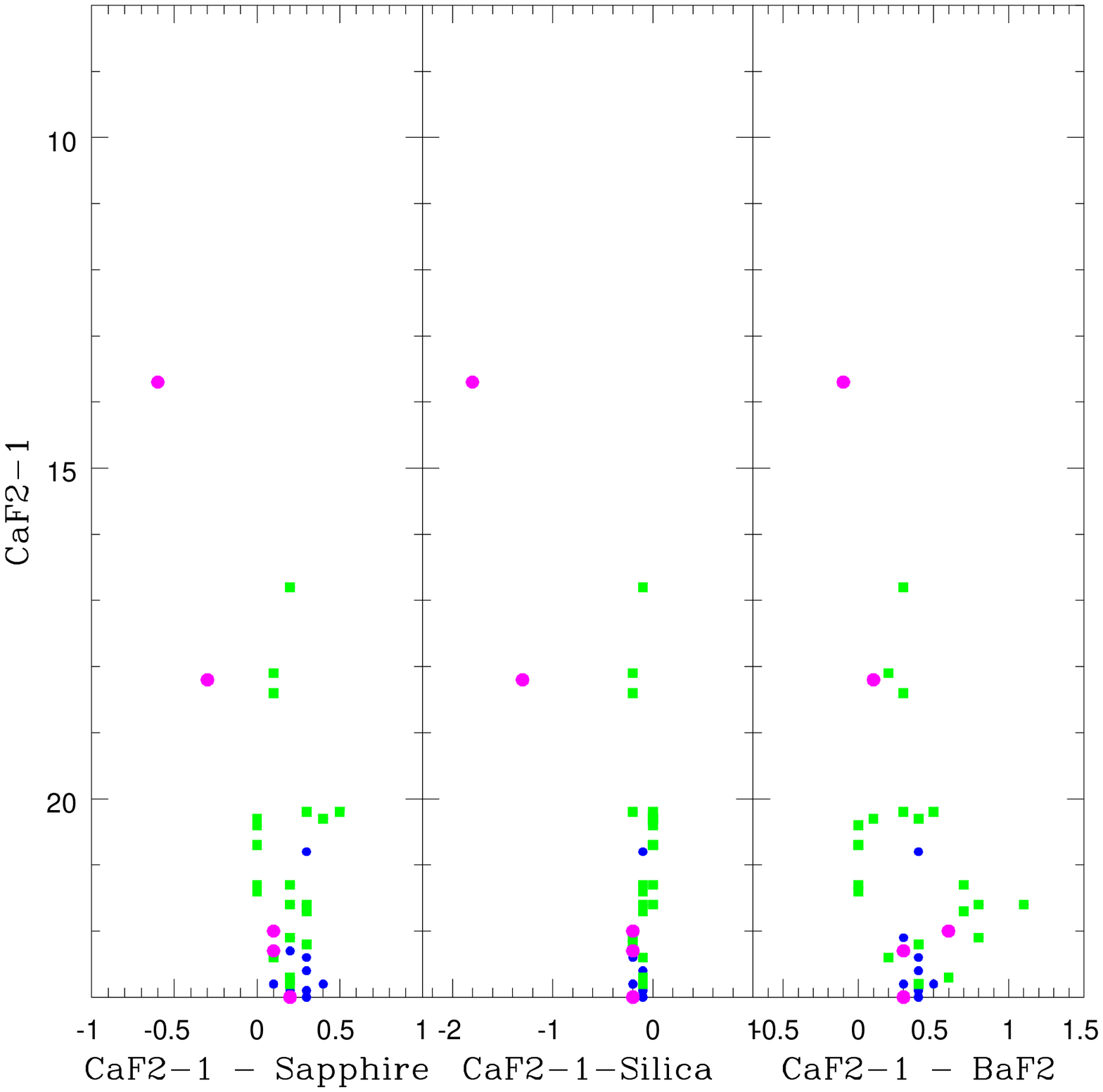}}
\centerline{\includegraphics[width=7.5cm]{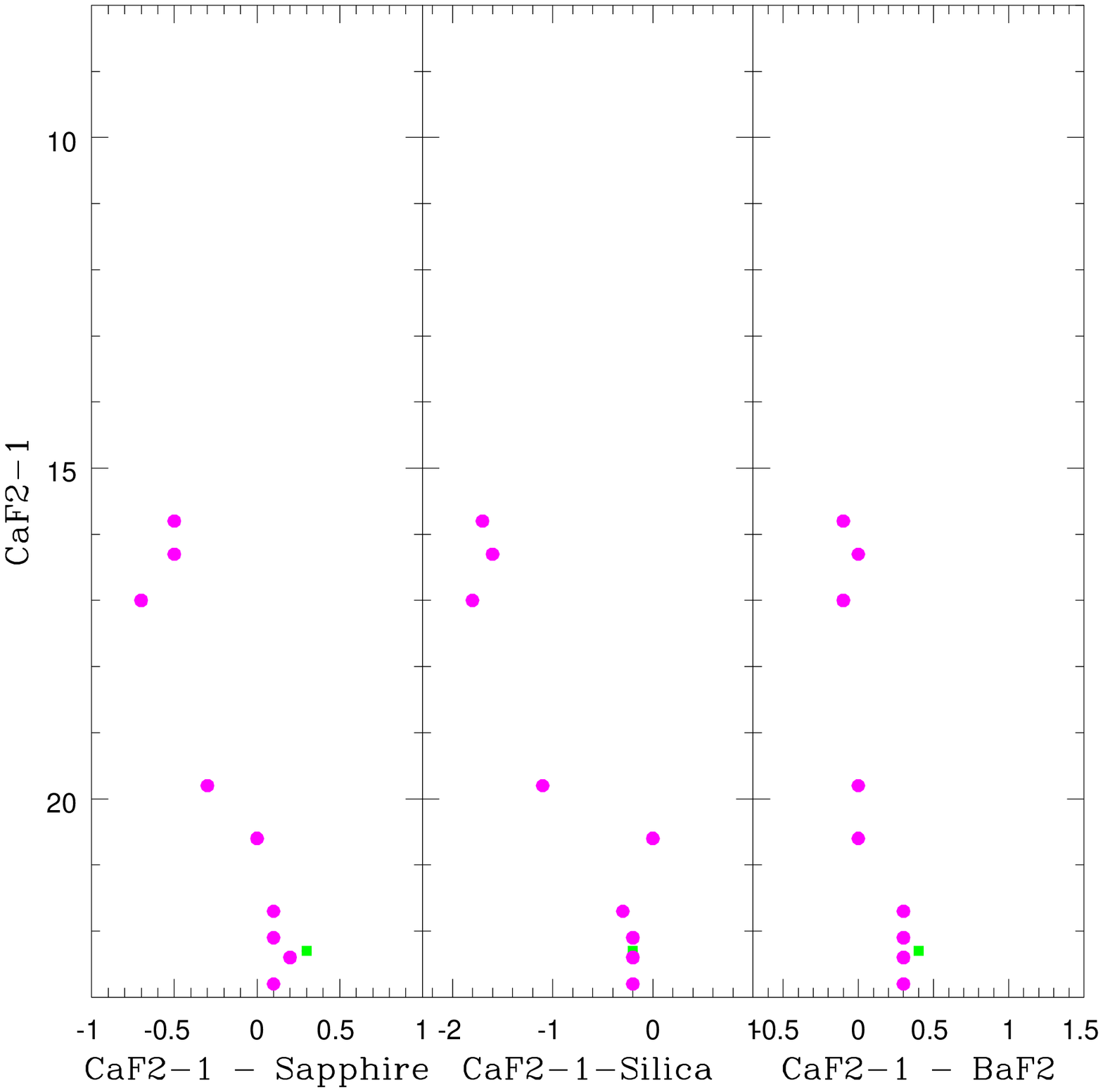}} \qquad
\caption{(i)FUV CMD of M67 (top left), NGC 188 (top right) and NGC 6791 (bottom) (a) CaF2-1$-$Sapphire, (b) CaF2-1$-$Silica and (c) CaF2-1$-$BaF2 vs. CaF2-1,\label{fig4}}

\end{figure}

In these CMDs, the CaF2-1 filter is taken as the magnitude in the vertical axis and plotted with a combination of other three FUV filters as the colour (fig. \ref{fig4}). We observe that the number of blue stragglers and  WDs does not change for different FUV filters combinations though there is a slight increase in the magnitude range for blue stragglers and  WDs in all CaF2-1 filter combinations. The turn-off region of M67 is just fainter than $\sim$ 20 magnitude. It is interesting to note that the turn-off region has more colour range ($\sim$ 1 mag) in the CaF2$-$BaF2 filter combination, with decreasing colour range in CaF2$-$Sapphire and CaF2$-$Silica combinations. Hence we suggest that CaF2$-$BaF2 colour combination is best suited to study the turn-off region of similar clusters. The CaF2$-$Silica combination has the maximum colour range and thus helps to separate WDs from blue stragglers, as suggested in the M67 CMD. Thus, this combination is likely to help in identifying new WDs.
\subsubsection{CMDs using Sapphire and Silica filters}
\begin{figure} [here]
\centerline{\includegraphics[width=7.5cm]{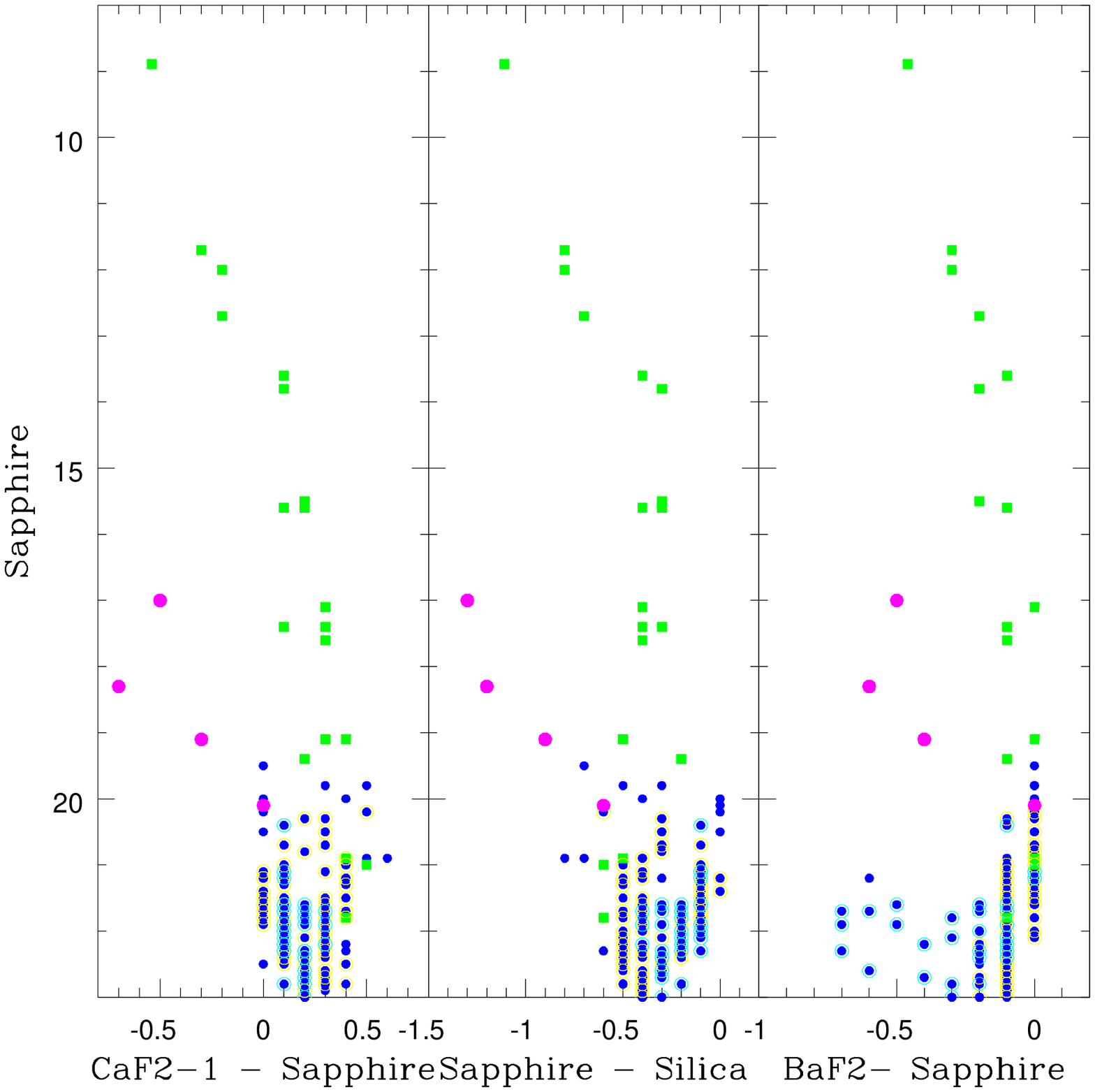} \qquad
     \includegraphics[width=7.5cm]{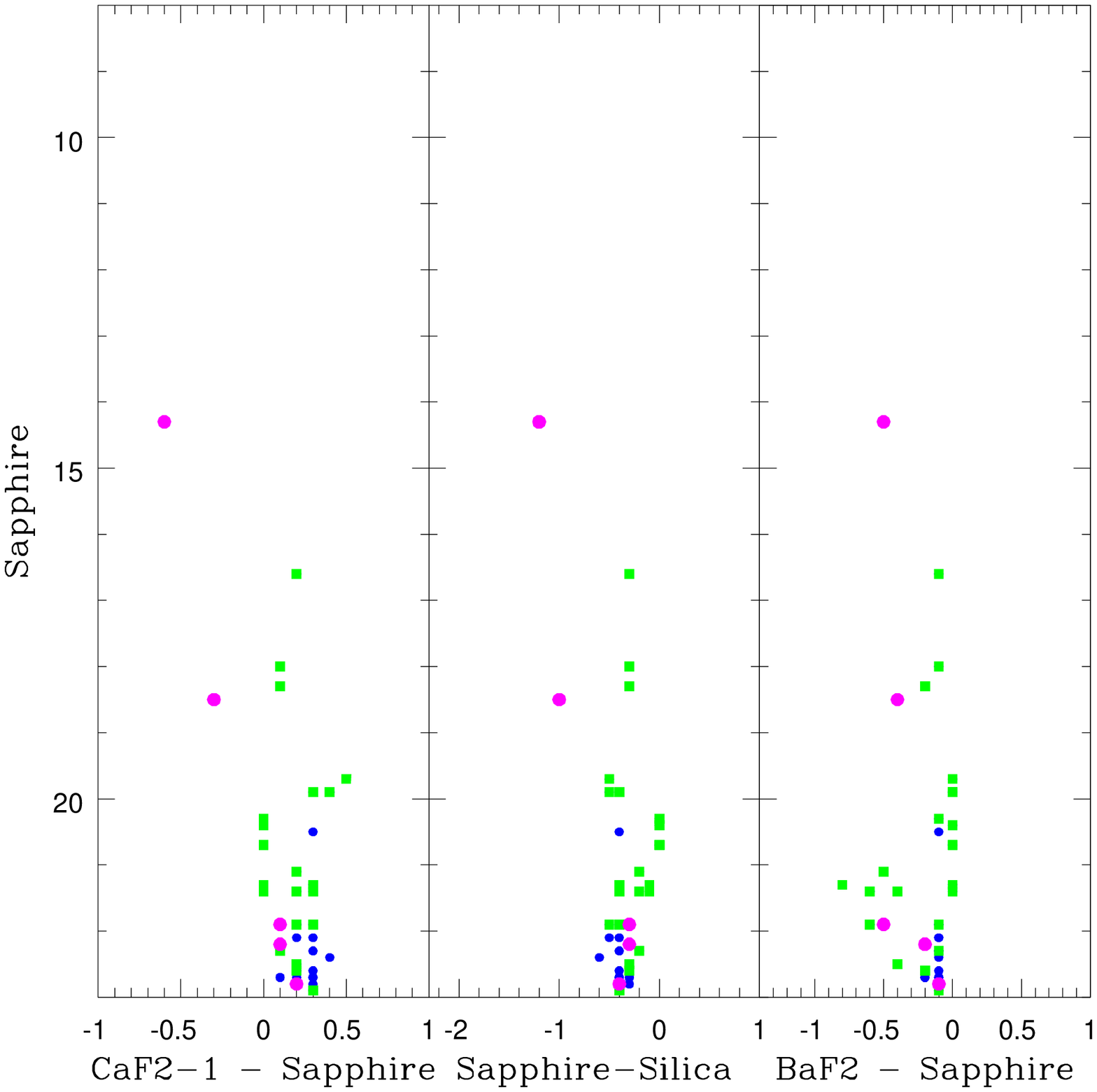}}
\centerline{\includegraphics[width=7.5cm]{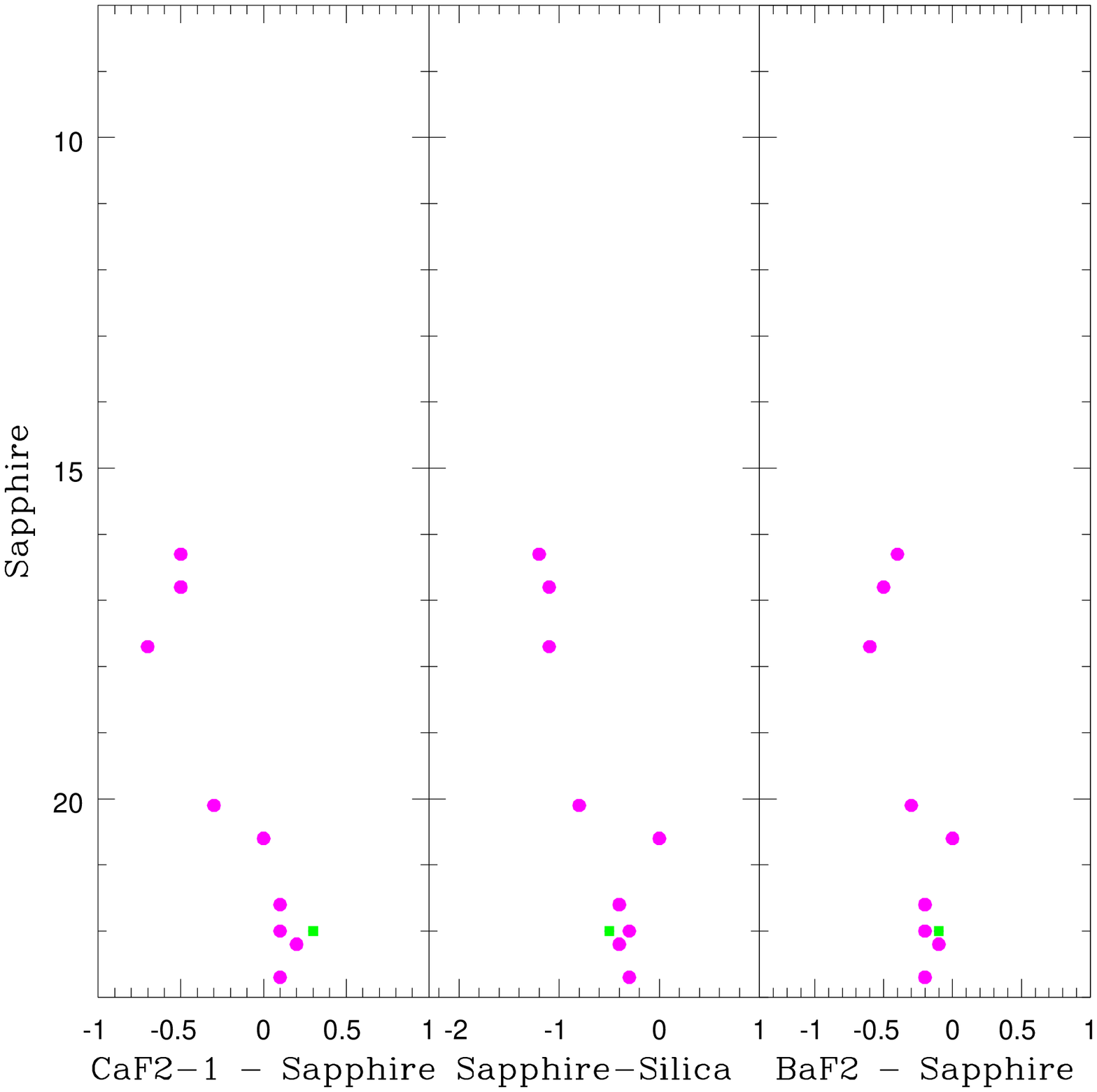}} \qquad
\caption{FUV CMD of M67 (top left), NGC 188 (top right) and NGC 6791 (bottom)((a) CaF2-1$-$Sapphire, (b) Sapphire$-$Silica and (c) BaF2$-$Sapphire vs. Sapphire.\label{fig5}}

\end{figure}

\begin{figure} [here]
\centerline{\includegraphics[width=7.5cm]{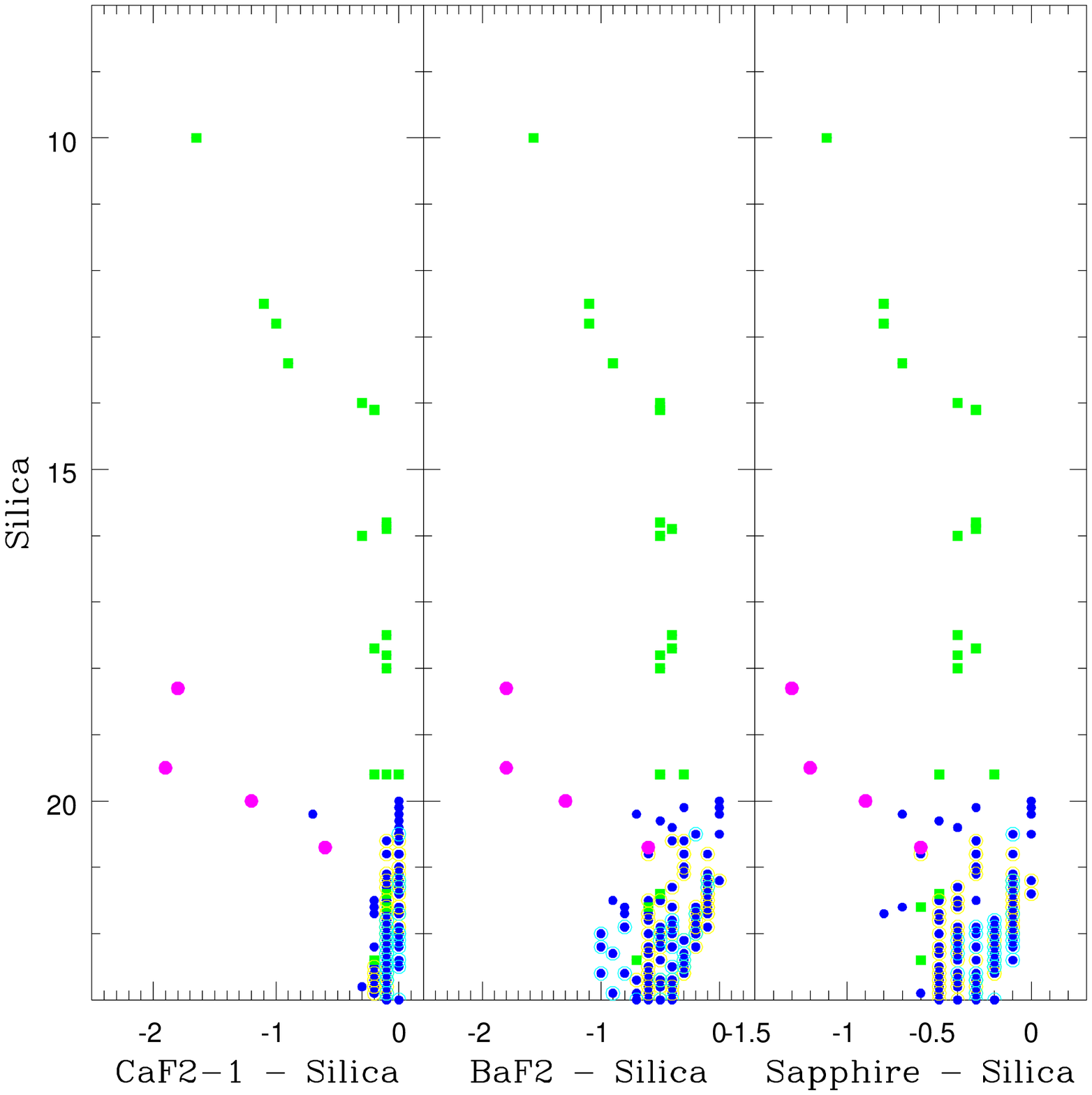} \qquad
     \includegraphics[width=7.5cm]{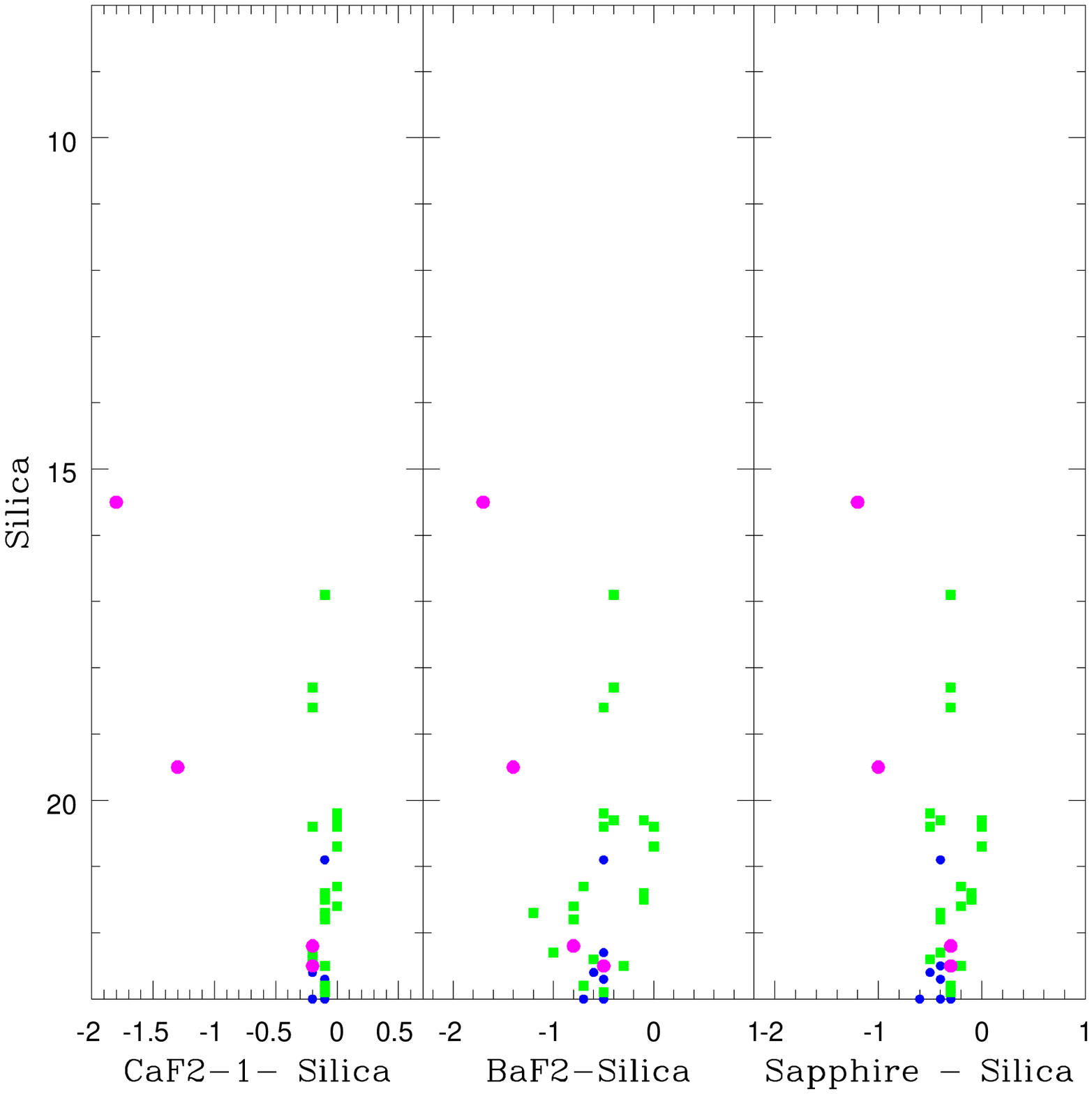}}
\centerline{\includegraphics[width=7.5cm]{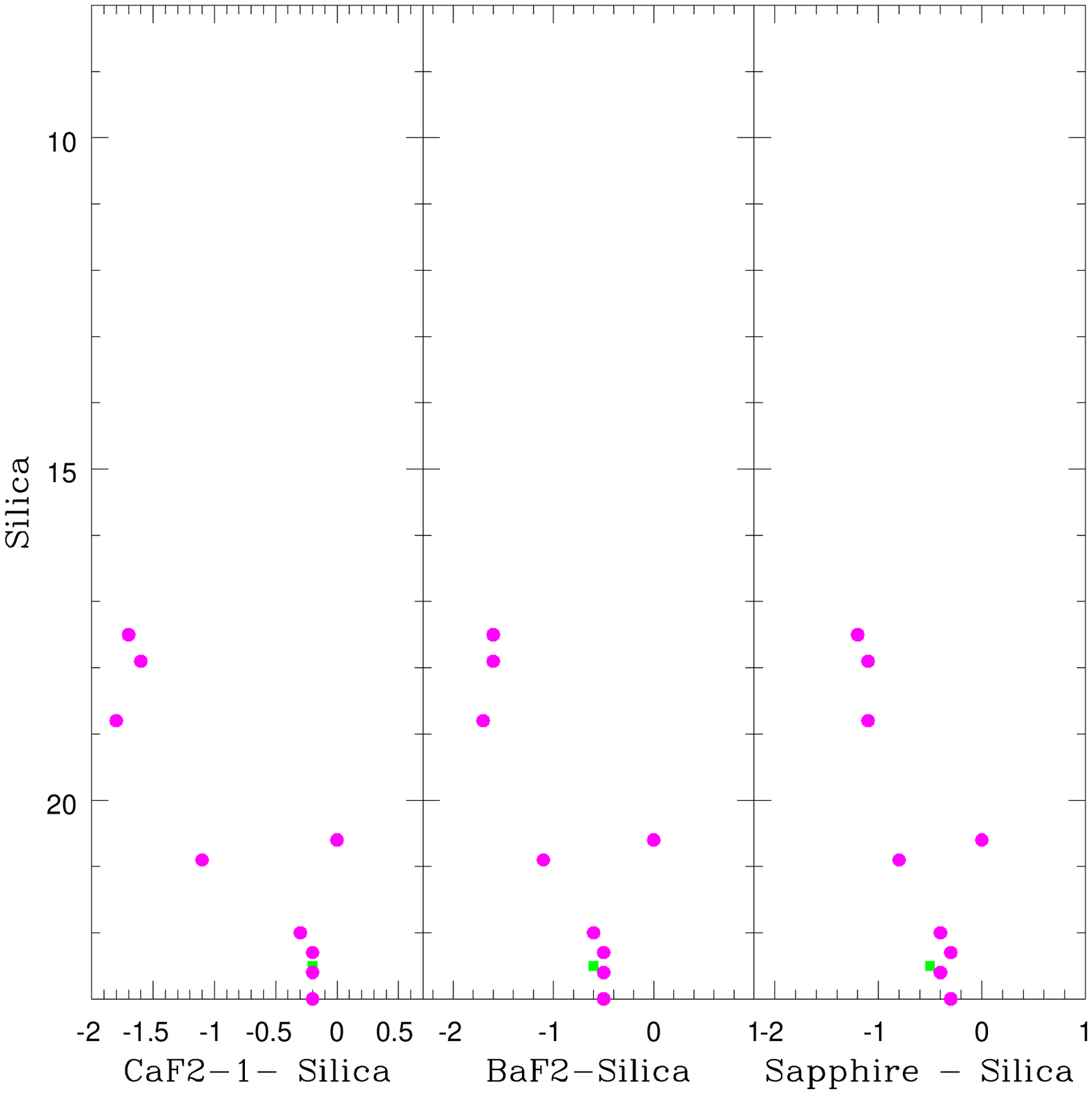}} \qquad
\caption{FUV CMD of M67 (top left), NGC 188 (top right) and NGC 6791 (bottom)(a) CaF2-1$-$Silica, (b) BaF2$-$Silica and (c) Sapphire$-$Silica vs. Silica.\label{fig6}}

\end{figure}

The CMDs with Sapphire and Silica magnitudes are presented in figures \ref{fig5} and \ref{fig6} respectively. These CMDs appear similar to the CMDs created with BaF2 and CaF2 filters. In the case of Sapphire filter CMDs, we observe that the turn-off stars and the tip of the main sequence in all the FUV CMD are bluer, due to the wavelength coverage of Sapphire filter with respect to other filters. The Silica filter CMDs once again suggest that the CaF2$-$Silica colour combination has a large colour range, which helps to separate WDs. We do not notice any other specific advantage in these CMDs.


\subsection{NUV colour magnitude diagrams}
NUV channel has 5 filters NUVB13, NUVN2, NUVB4, NUVB15 and NUV Silica. We derived the magnitudes of stars in the three open clusters and are used to create the CMDs using various combinations of these filters. For this study, we have  shown the CMD upto 23 magnitude, such that we could consider detectable characteristics. We estimated the expected error to be ~0.2 mag at NUV~20 mag, based on the error in the corresponding V and (B$-$V) values. We do not show the errors as it makes it difficult to interpret the CMD. The NUV filter wavelength range is from 200 to 300 nm. We observe blue stragglers, red giants,  WDs, sub giants, turn-off and main sequence stars. For simplicity, the following notations are used for the filters: B13(NUVB13), N2(NUVN2), B4(NUVB4), B15(NUVB15) and Silica(NUV Silica). Among these filters, B15 has a relatively poor efficiency. 

In the optical CMD of M67, 20 blue stragglers, 5 WDs, a yellow giant –  WD) binary, 22 red giant, sub giant branch, turn-off stars were identified. We identify the same number of stars in all the NUV CMDs, though with variation in their brightness and colour. In the figure \ref{fig7}, we have plotted with B13 filter as magnitude on vertical axis and combining it with different NUV filters to get the colour on the horizontal axis. Various evolutionary phases can be easily identified with the help of colour code. The MS just below the turn-off (cyan points) is stretched up to 5 magnitudes, whereas it is stretched to $\sim$1.5 magnitude in the V filter. Thus, this stretching of the MS near the turn-off region can help in estimating the ages more accurately. The turn-off stars can be identified as located parallel and close to cyan points, followed by the sub giant stars (black points). This suggests that the sub giant branch appears almost vertical and parallel to the MS, in the NUV CMDs. The giant branch starts at the faintest end of the sub giant branch and appears as a vertical distribution. Thus, unless previously known or identified, it will not be possible to separate the MS stars from sub giants. 
\subsubsection{CMDs using B13 filter}
\begin{figure} [here]
\centerline{\includegraphics[width=7.5cm]{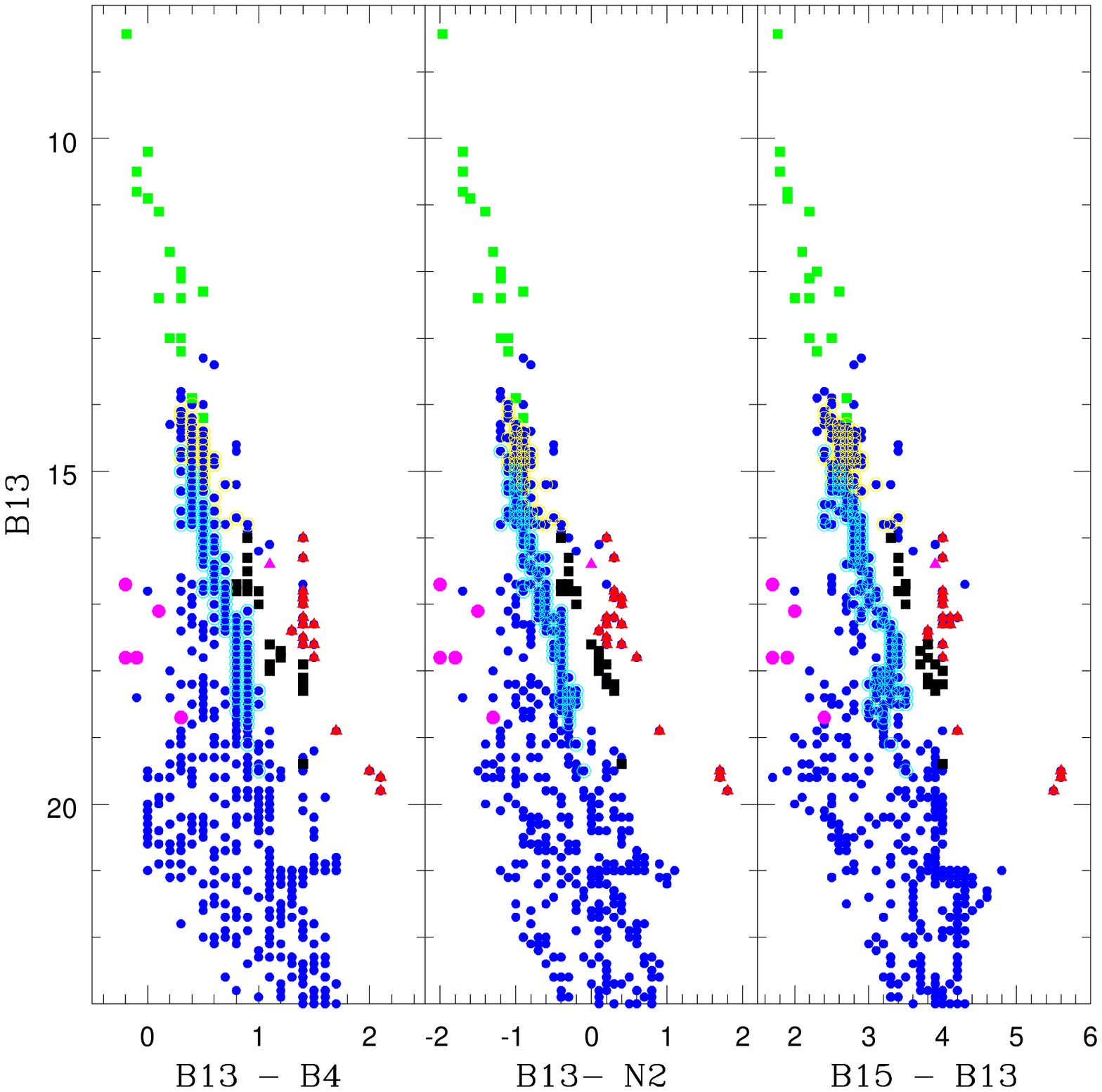} \qquad
     \includegraphics[width=7.5cm]{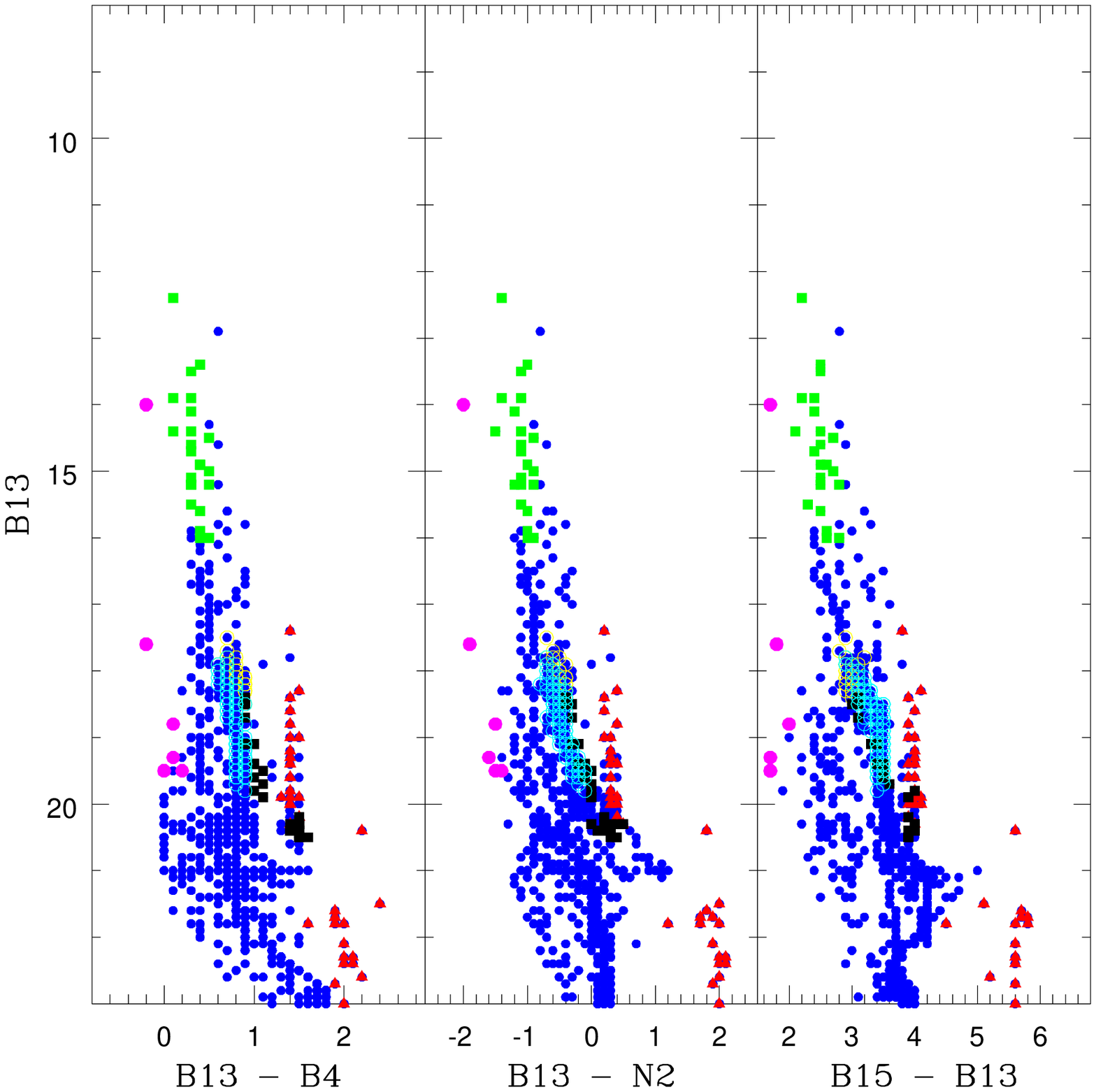}}
\centerline{\includegraphics[width=7.5cm]{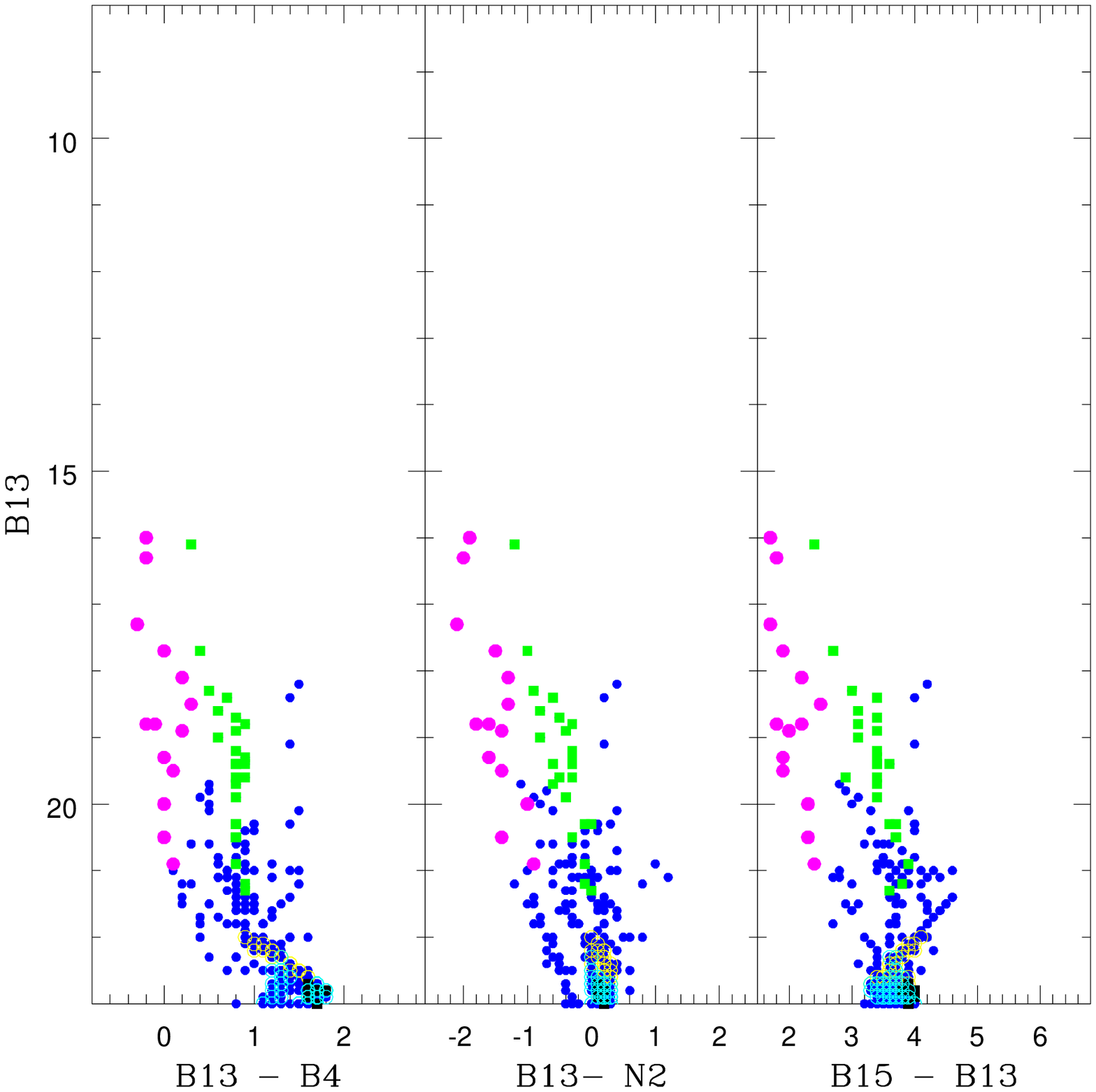}} \qquad
\caption{(i)NUV CMD of M67 (top left), NGC 188 (top right) and NGC 6791 (bottom) (a) B13$-$B4, (b) B13$-$N2 and (c) B15$-$B13 vs. NUVB13.\label{fig7}}

\end{figure}

In figure \ref{fig7}, we have shown the CMDs with respect to the B13 filter. As expected, M67 has the brightest turn-off and hence a good portion of the MS can be detected. In the case of NGC 188, the MS near the turn-off is clearly seen. On the other hand, only the  blue stragglers above the turn-off can be detected in the case of NGC 6791. The colour range which can be observed for the combinations shown suggest that, the filter B13 with N2 has a better colour separation to study blue stragglers. Also from figure \ref{fig7} (b) we observe that these stars appear bluer in this NUV CMD. Though from the figure \ref{fig7}(c), has a good colour separation, B15 filter may not be able to detect stars which appear red. The NUV CMD of B13-N2 vs. B13 is better for the studies of sub giant branch stars, as they have better colour separation. 
\subsubsection{CMDs using N2 filter}
\begin{figure} [here]
\centerline{\includegraphics[width=7.5cm]{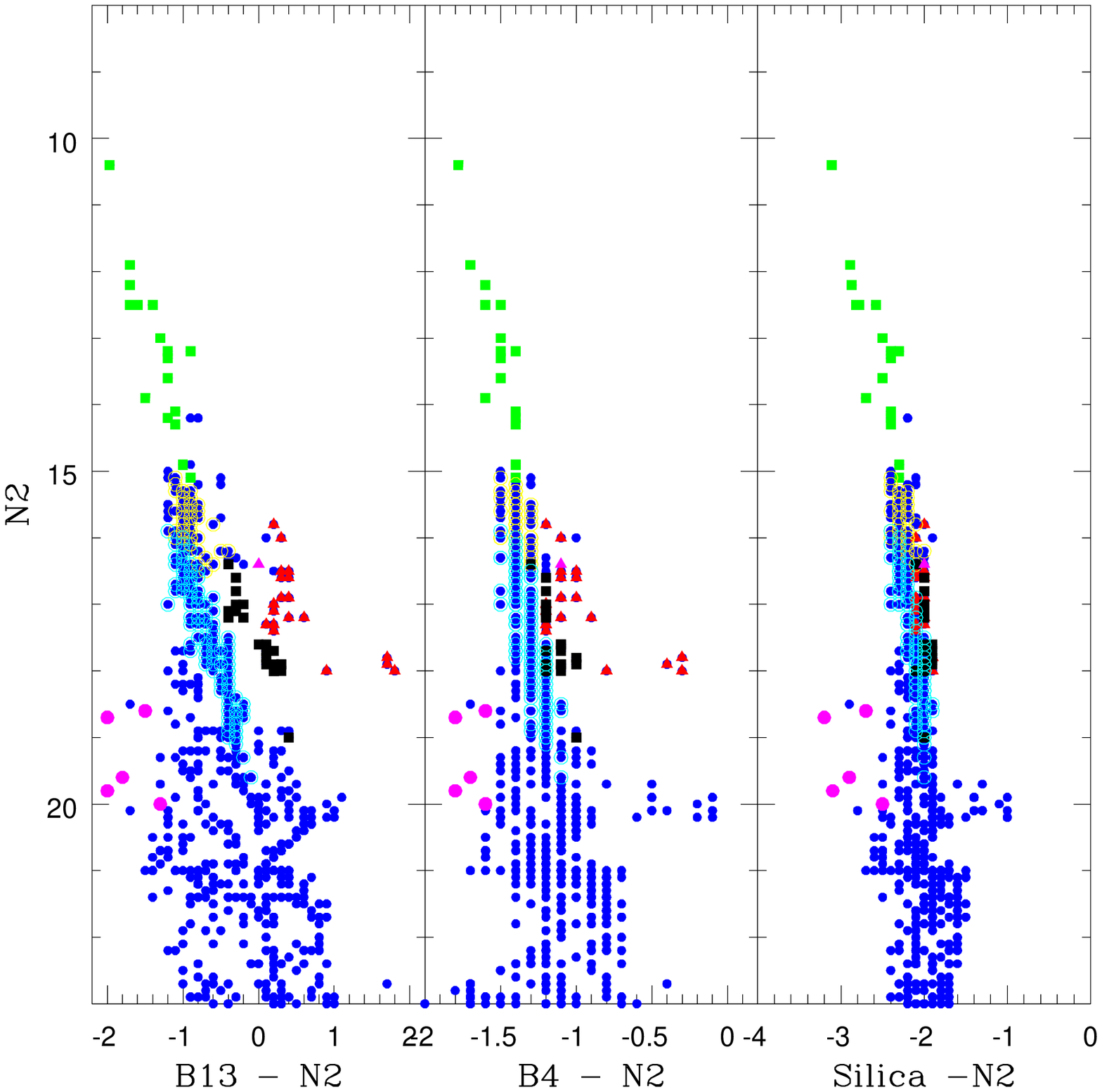} \qquad
     \includegraphics[width=7.5cm]{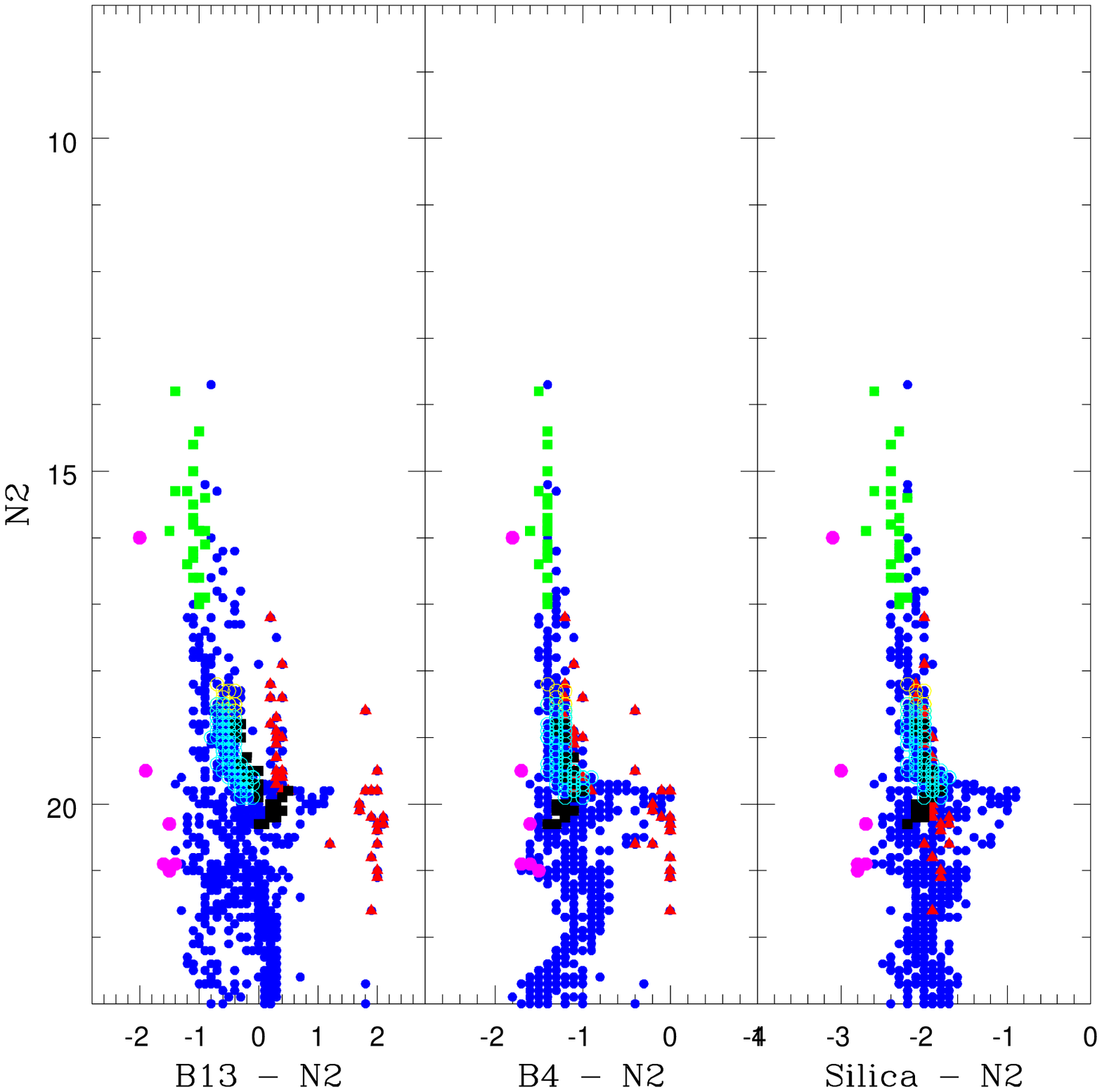}}
\centerline{\includegraphics[width=7.5cm]{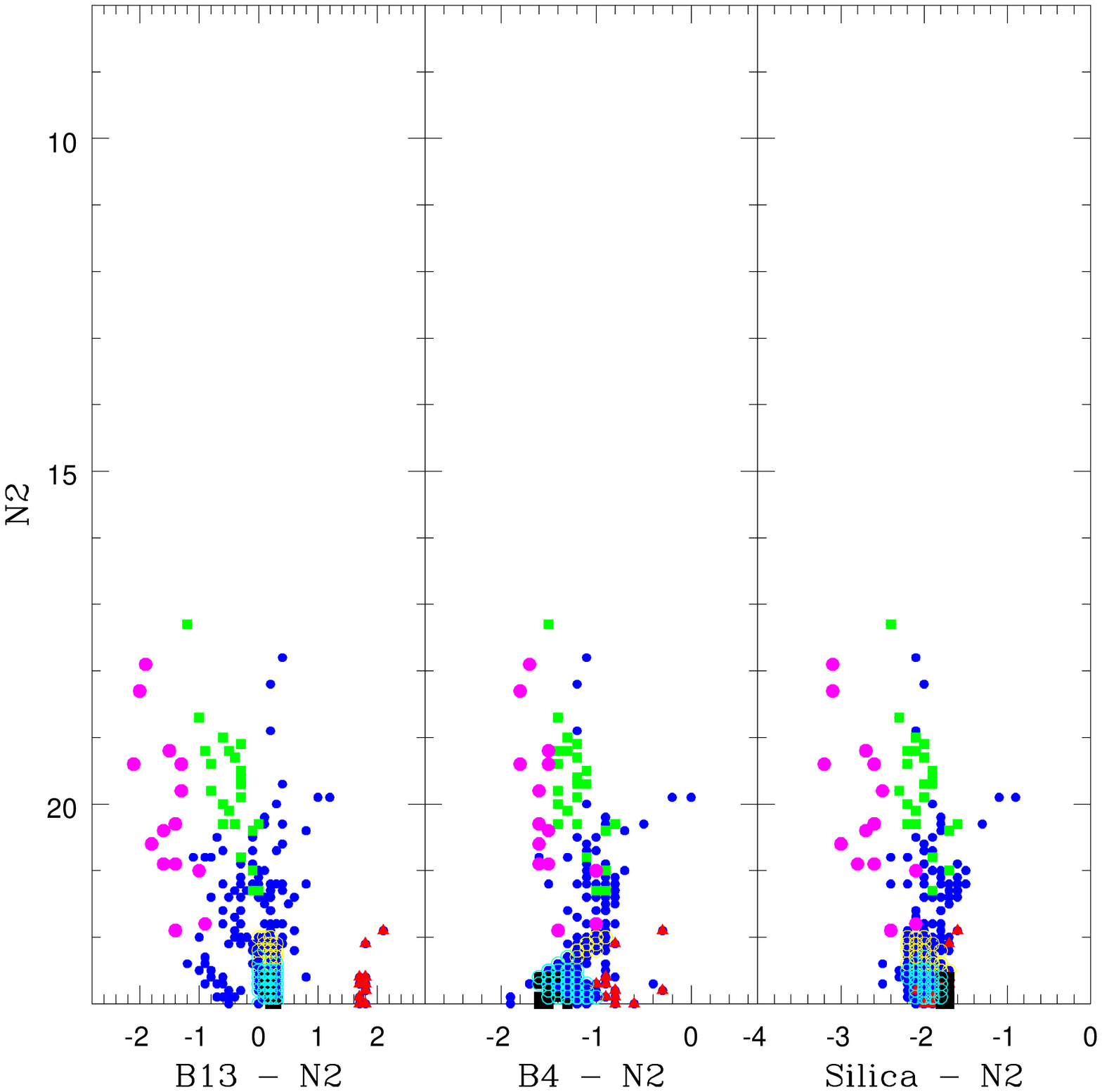}} \qquad
\caption{(i)NUV CMD of M67 (top left), NGC 188 (top right) and NGC 6791 (bottom)(a) B13$-$N2 (b) B4$-$N2 and (c) Silica$-$N2 vs. N2.\label{fig8}}

\end{figure}

The NUV CMD for the cluster M67, with B13$-$N2 vs. N2 (fig. \ref{fig8}) also support the above inferences. In figure \ref{fig8}, we can also notice that the upper MS, turn-off, sub giant branch and the giants get narrowed down to a small range in colour. In the case of Silica$-$N2 colour, the  WDs appear bluer, along with the larger range of colour for entire stellar sequence. 
\subsubsection{CMDs using Silica filter}
\begin{figure} [here]
\centerline{\includegraphics[width=7.5cm]{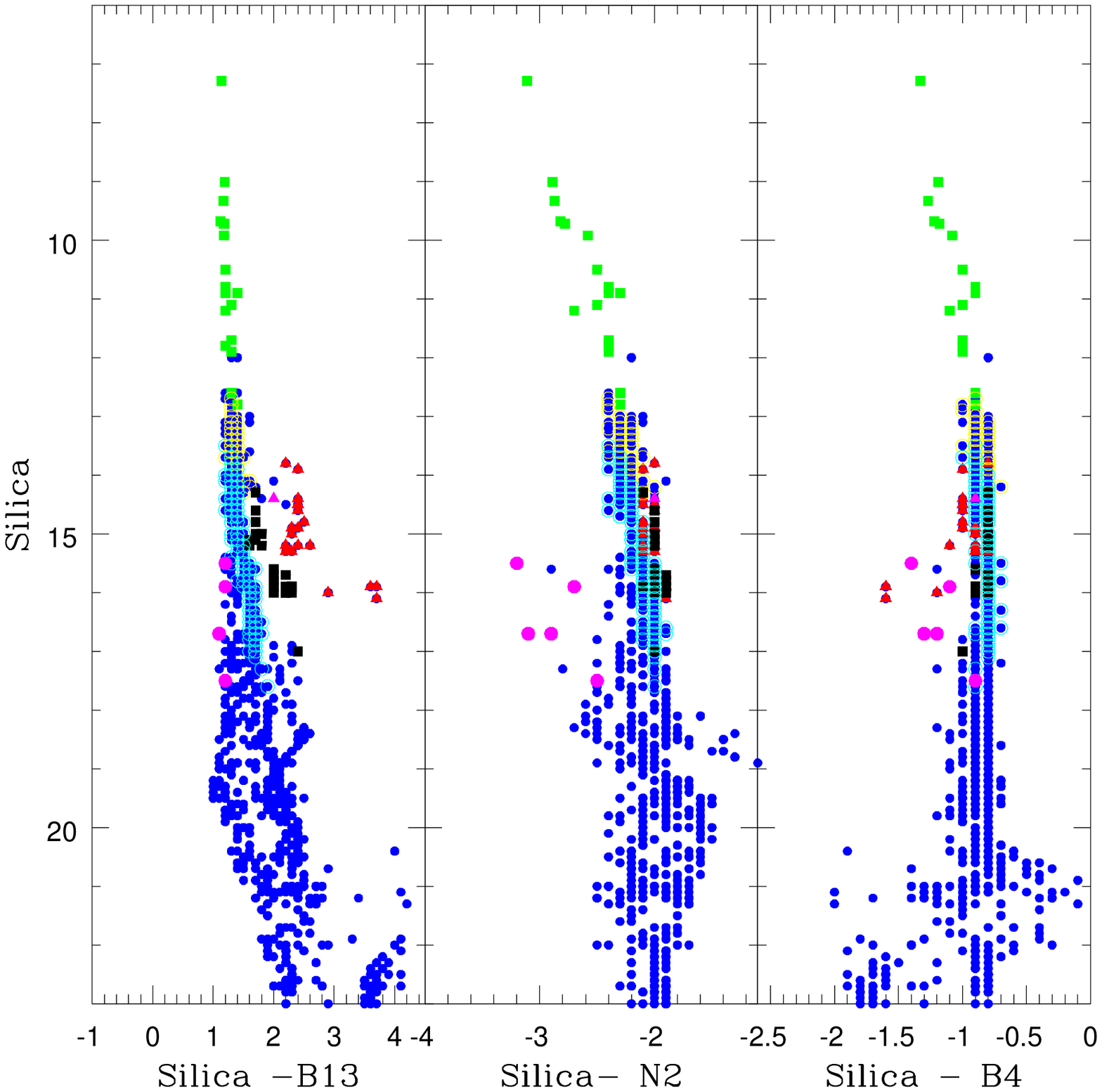} \qquad
     \includegraphics[width=7.5cm]{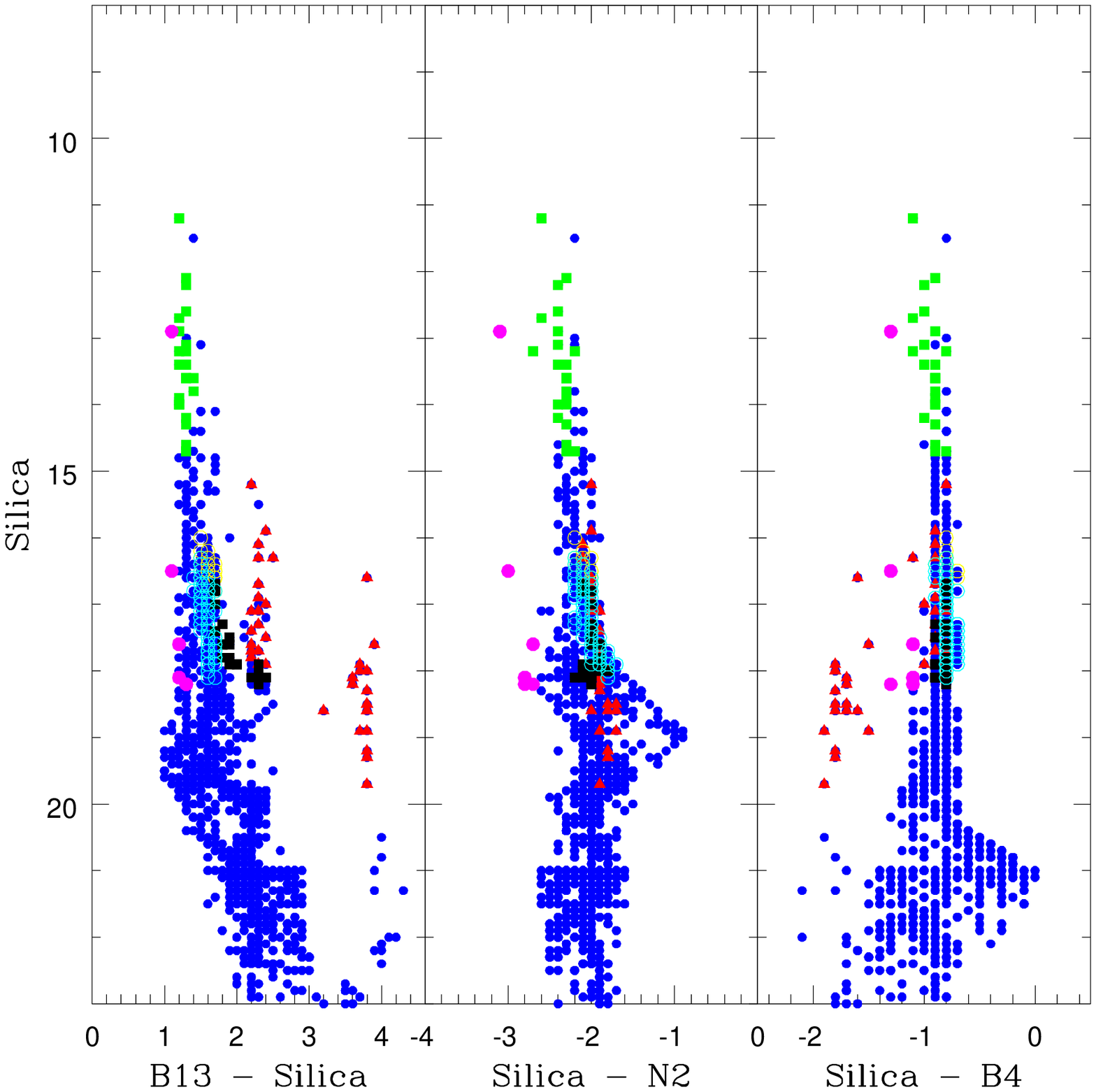}}
\centerline{\includegraphics[width=7.5cm]{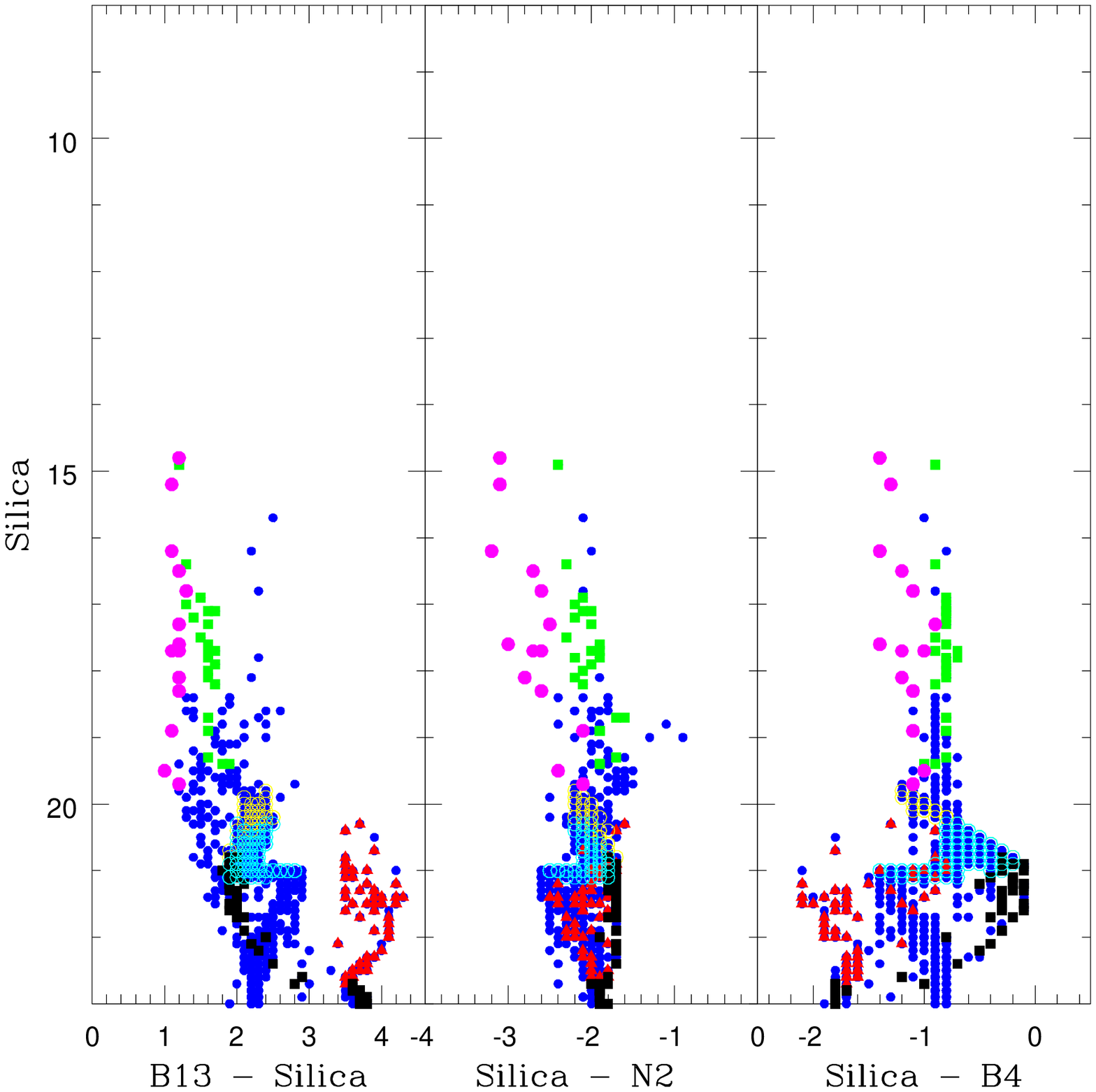}} \qquad
\caption{(i)NUV CMD of M67 (top left), NGC 188 (top right) and NGC 6791 (bottom) (a) B13$-$Silica (b) Silica$-$N2 (c) Silica$-$B4 vs. Silica.\label{fig9}}

\end{figure}

In the Silica CMDs, shown in figure \ref{fig9}, we observe a large colour range for Silica$-$N2 colour. The CMD using B13$-$Silica colour separates the sub giant stars from the upper MS and red giants, but the  WDs do not get separated from the MS. Also, the red giants are found to have very red colour, separating them from sub giants. The CMD with Silica$-$N2 colour separates the  WDs, but the MS, sub giants and red giants do not get separated. The Silica$-$B4 colour makes the red giants bluer than the MS, which is incorrect. This may be due to the combination of input spectra and the coverage. We plan to look into this problem. 
\subsubsection{CMDs using B4 filter}
\begin{figure} [here]
\centerline{\includegraphics[width=7.5cm]{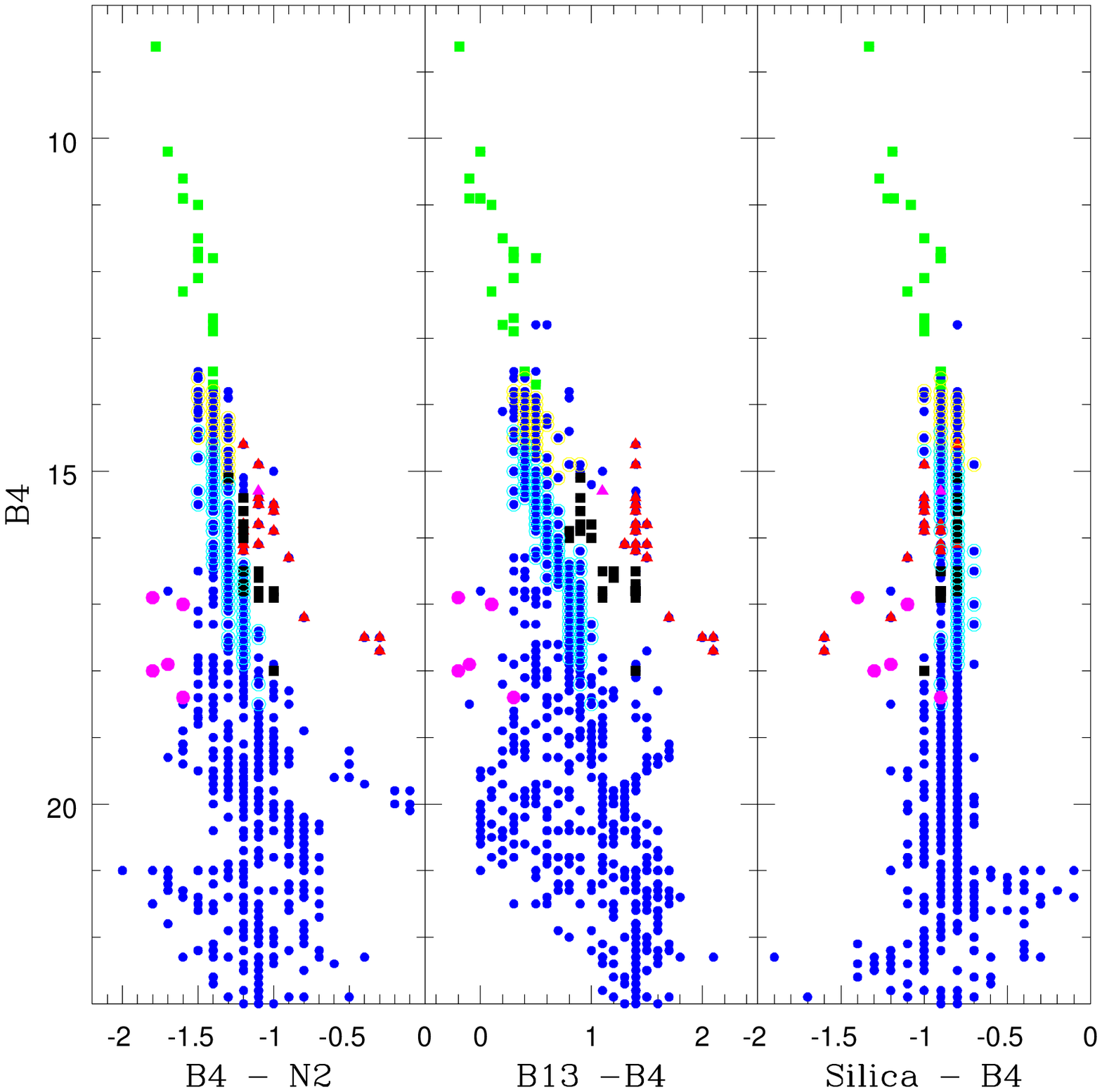} \qquad
     \includegraphics[width=7.5cm]{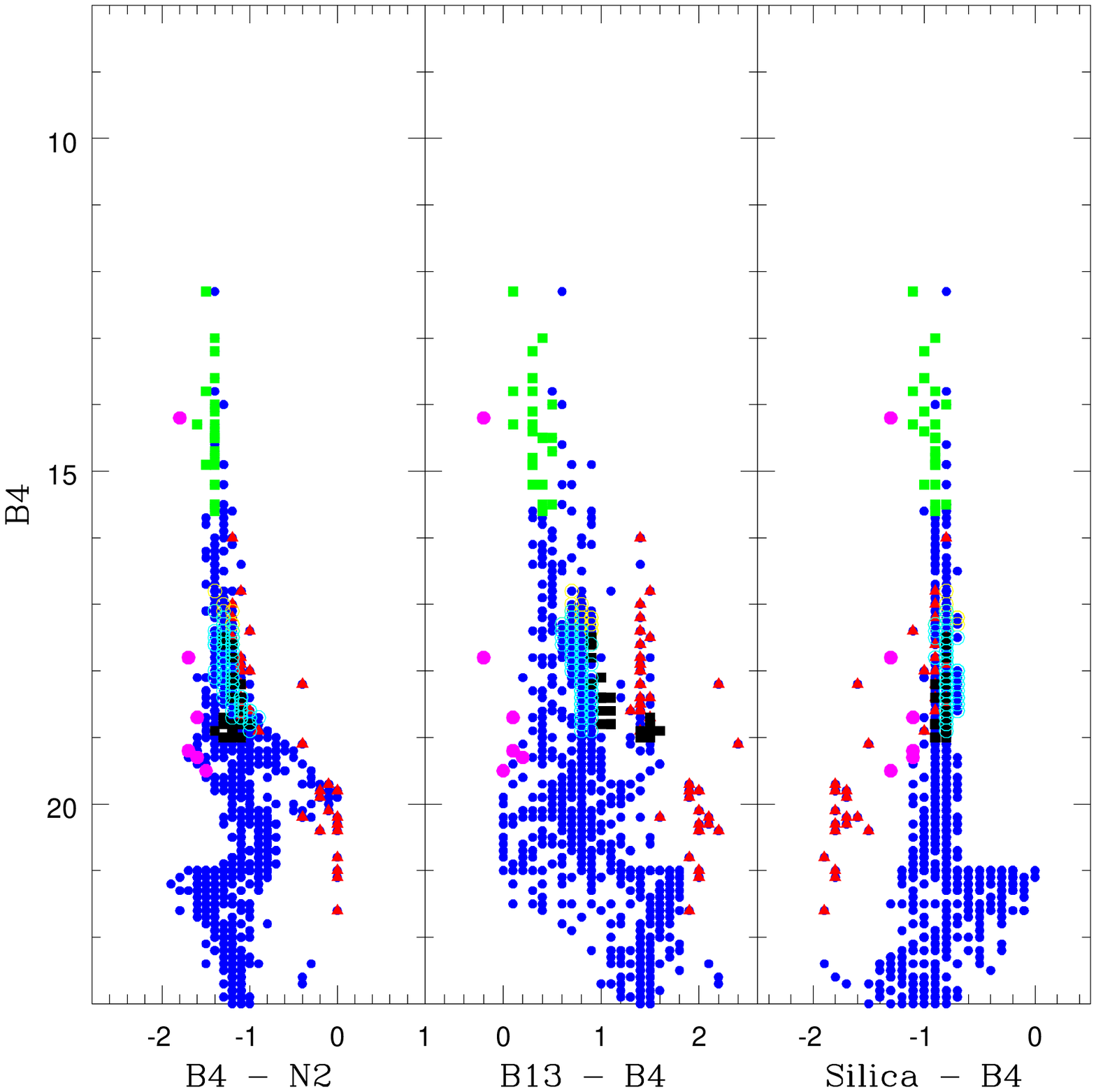}}
\centerline{\includegraphics[width=7.5cm]{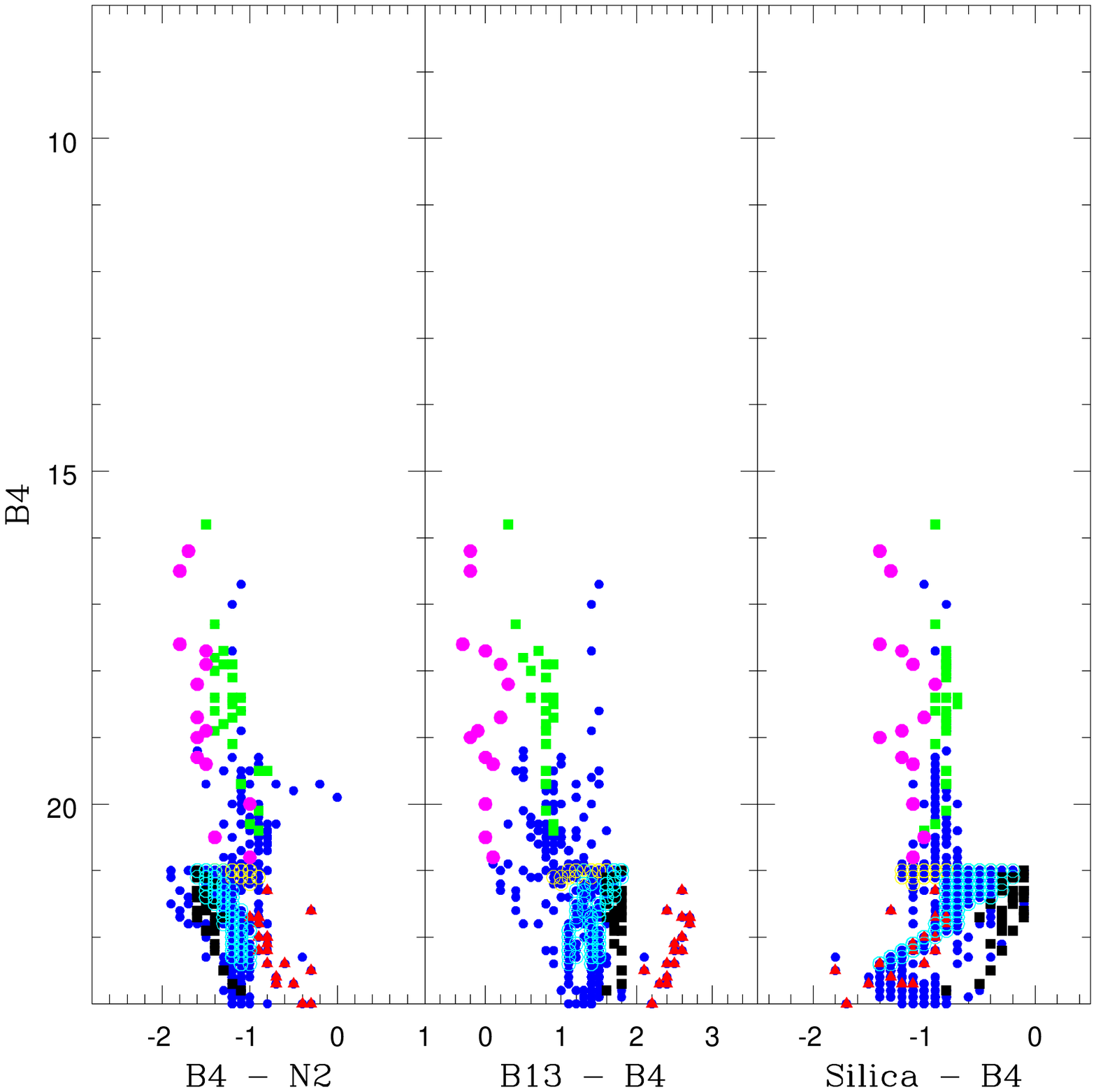}} \qquad
\caption{(i)NUV CMD of M67 (top left), NGC 188 (top right) and NGC 6791 (bottom) (a) B4$-$N2 (b) B13$-$B4 (c) Silica$-$B4 vs B4. \label{fig10}}
\end{figure}

The B4 CMDs are shown in figure \ref{fig10}. It can be seen that the CMD using the B13$-$B4 colour separates most of the sequences. On the other hand, the B4$-$N2 has a larger colour range. The colour derived using the silica filter has issues as mentioned above.

We do not present the CMDs using the B15 filter due to the poor efficiency of the filter, even though we have simulated them. We have presented the simulated CMDs in the FUV filters and NUV filters separately. It will not be a good idea to combine FUV with NUV, as only the blue stragglers and  WDs are detected in the FUV. Hence, in order to study or distinguish between blue stragglers or  WDs, it may be useful to combine any one of the FUV filters with the NUV filters. 
It will be a good idea to combine the NUV filter magnitudes with V magnitude to create CMDs, as it is possible to detect a good fraction of the upper MS in the NUV filters. We present the CMD using NUV silica filter and V magnitude in the next section.    
%


\subsection{NUV Silica vs V CMD}
\begin{figure} [here]
\centerline{\includegraphics[width=7.5cm]{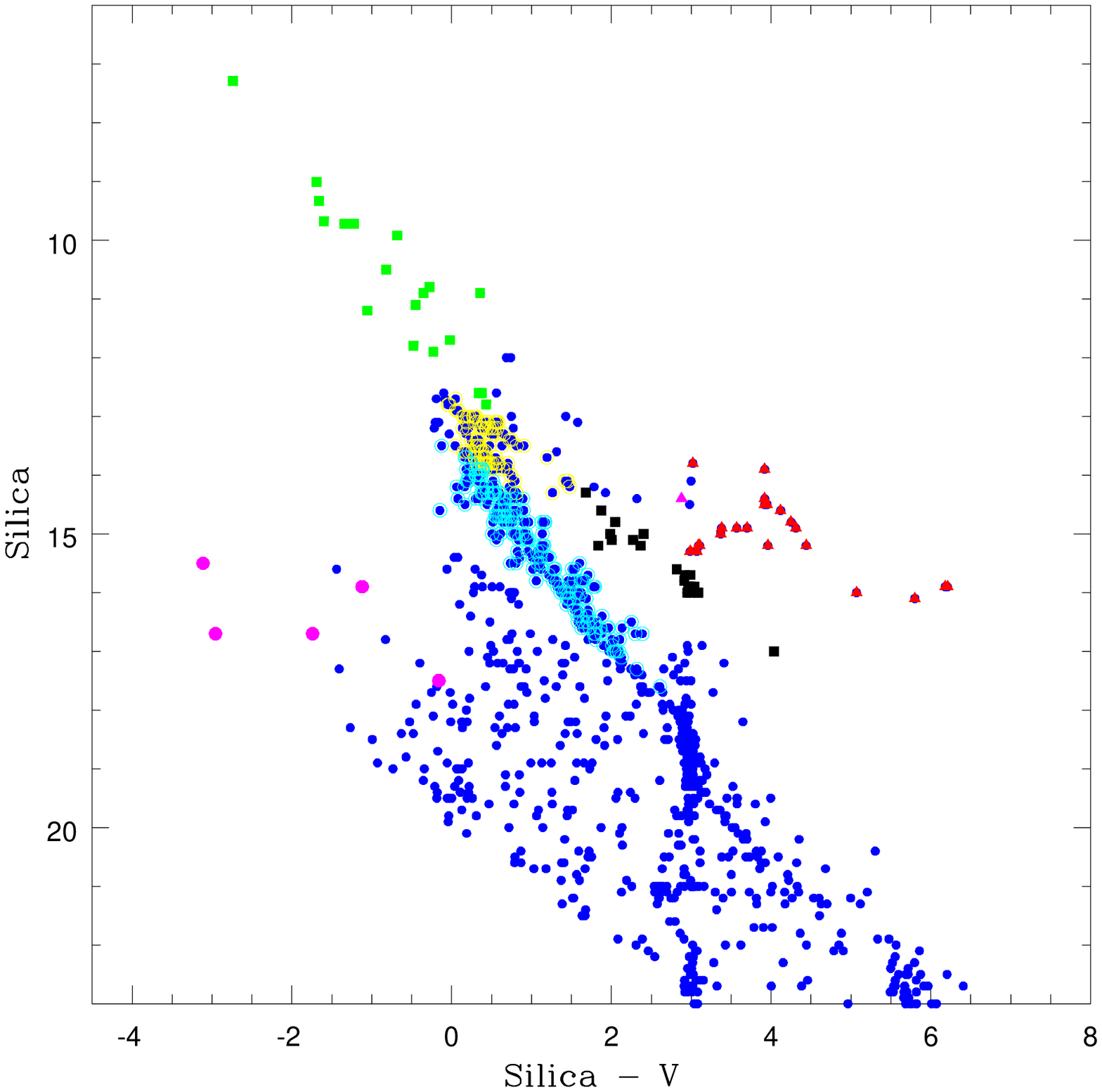} \qquad
     \includegraphics[width=7.5cm]{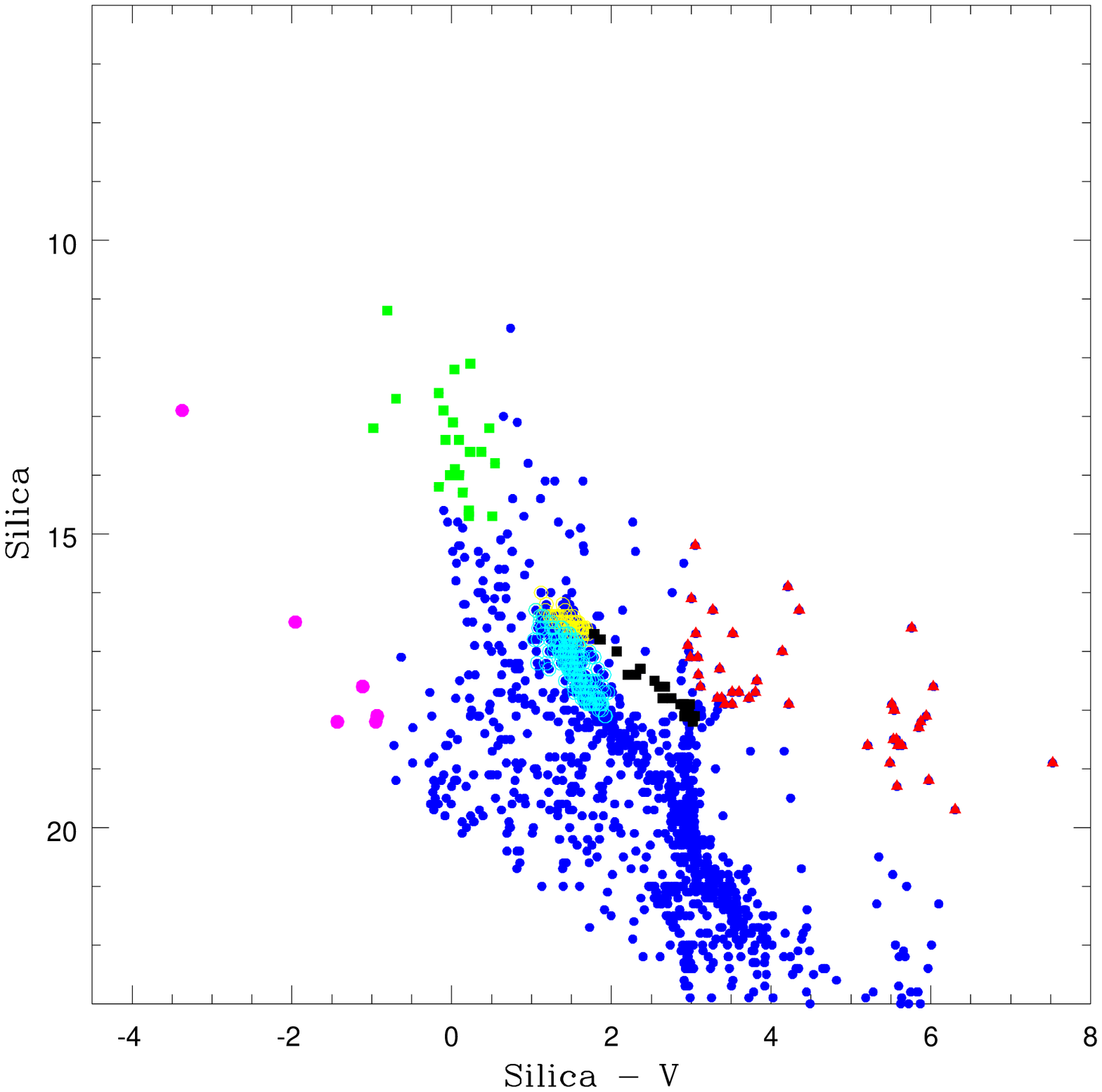}}
\centerline{\includegraphics[width=7.5cm]{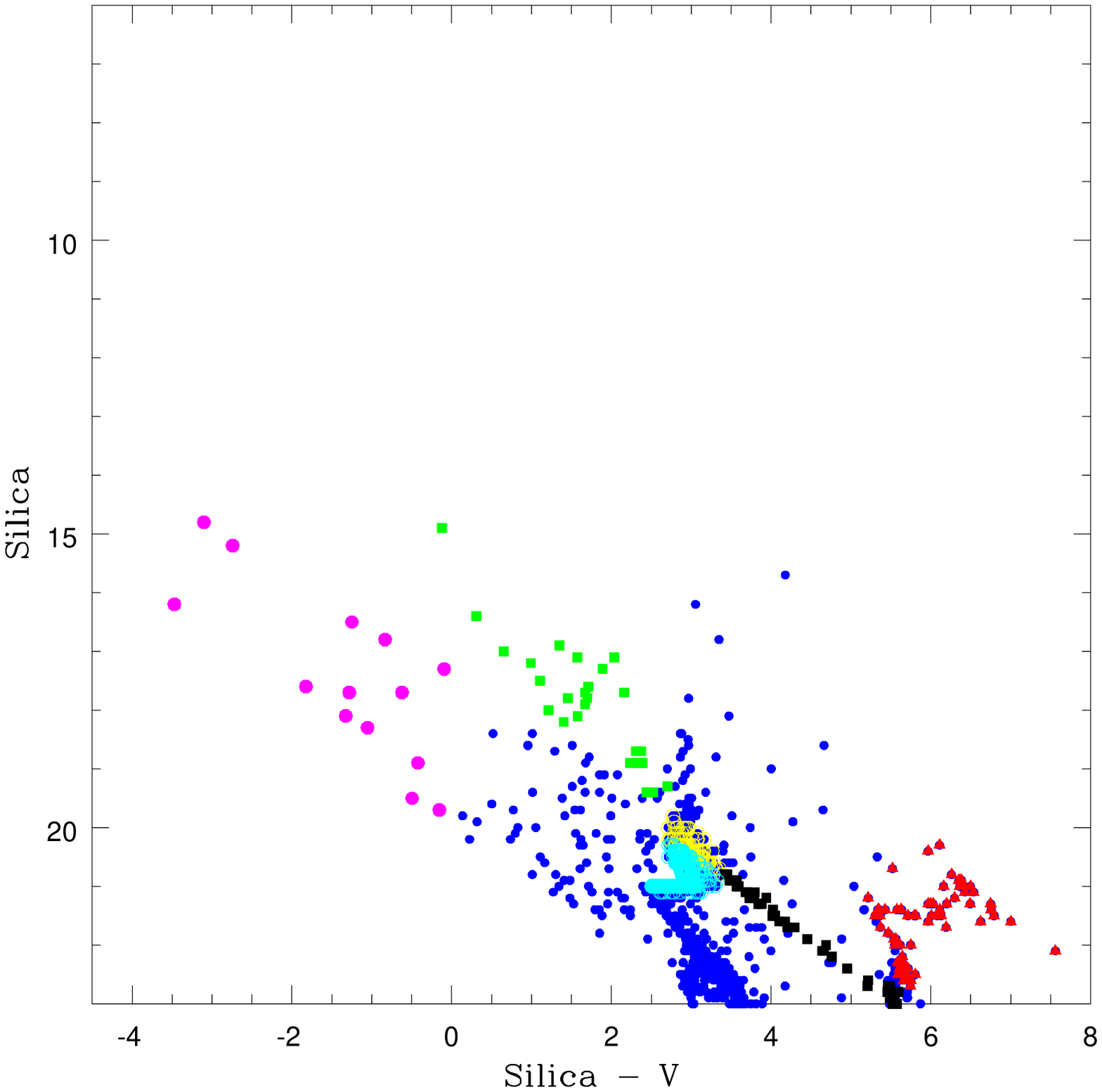}} \qquad
\caption{(i)NUV-V CMD of M67 (top left), NGC 188 (top right) and NGC 6791 (bottom) of Silica$-$V vs Silica \label{fig11}}

\end{figure}

For the NUV-V CMD, we chose the Silica filter in NUV as it has the largest wavelength coverage. We have shown the Silica vs (Silica$-$V) CMD in figure \ref{fig11} for the three clusters studied here. We obtain a better and broader colour separation when compared to the FUV, NUV and optical CMDs. We observe all the stellar sequence that is observed in the optical, but having a wider colour separation.

The  WDs and the blue stragglers have shifted to bluer colour index, also they appear brighter in this filter combination and hence studying them in this filter combination is most suitable. The upper MS, turn-off, sub giants and the red giants can be separated due to the large colour range. These figures suggest that the NUV-V CMDs will be quite useful in separating stars of various evolutionary sequence, which can then be used in the analysis of the FUV and NUV CMDs. It is thus important to have V observations of the clusters and then combine with the UVIT observations.

\section{ Colour - Colour Diagrams of M67}
\begin{figure} [here]

\centerline{\includegraphics[width=7.5cm]{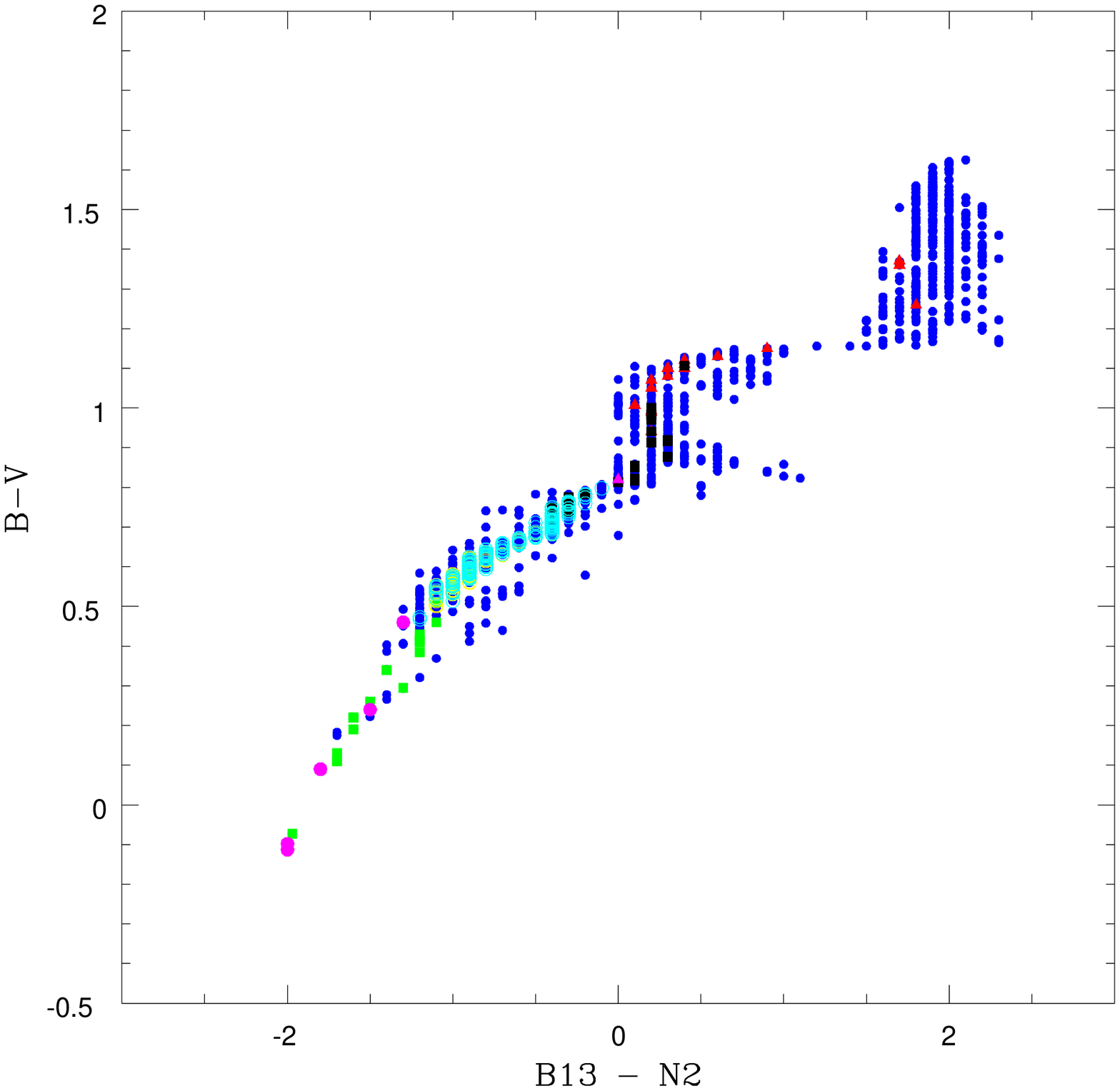} \qquad
     \includegraphics[width=7.5cm]{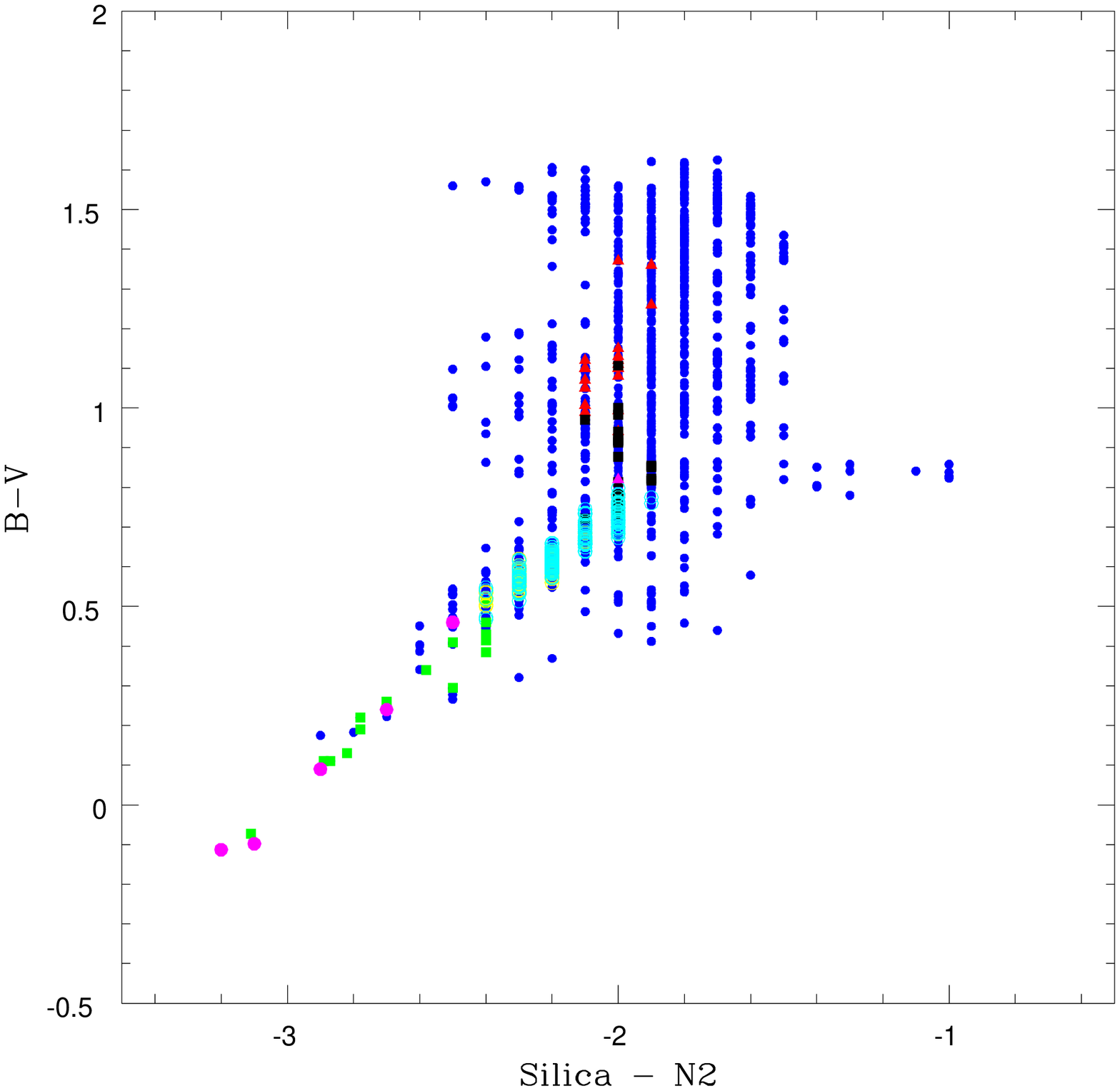}}
\centerline{\includegraphics[width=7.5cm]{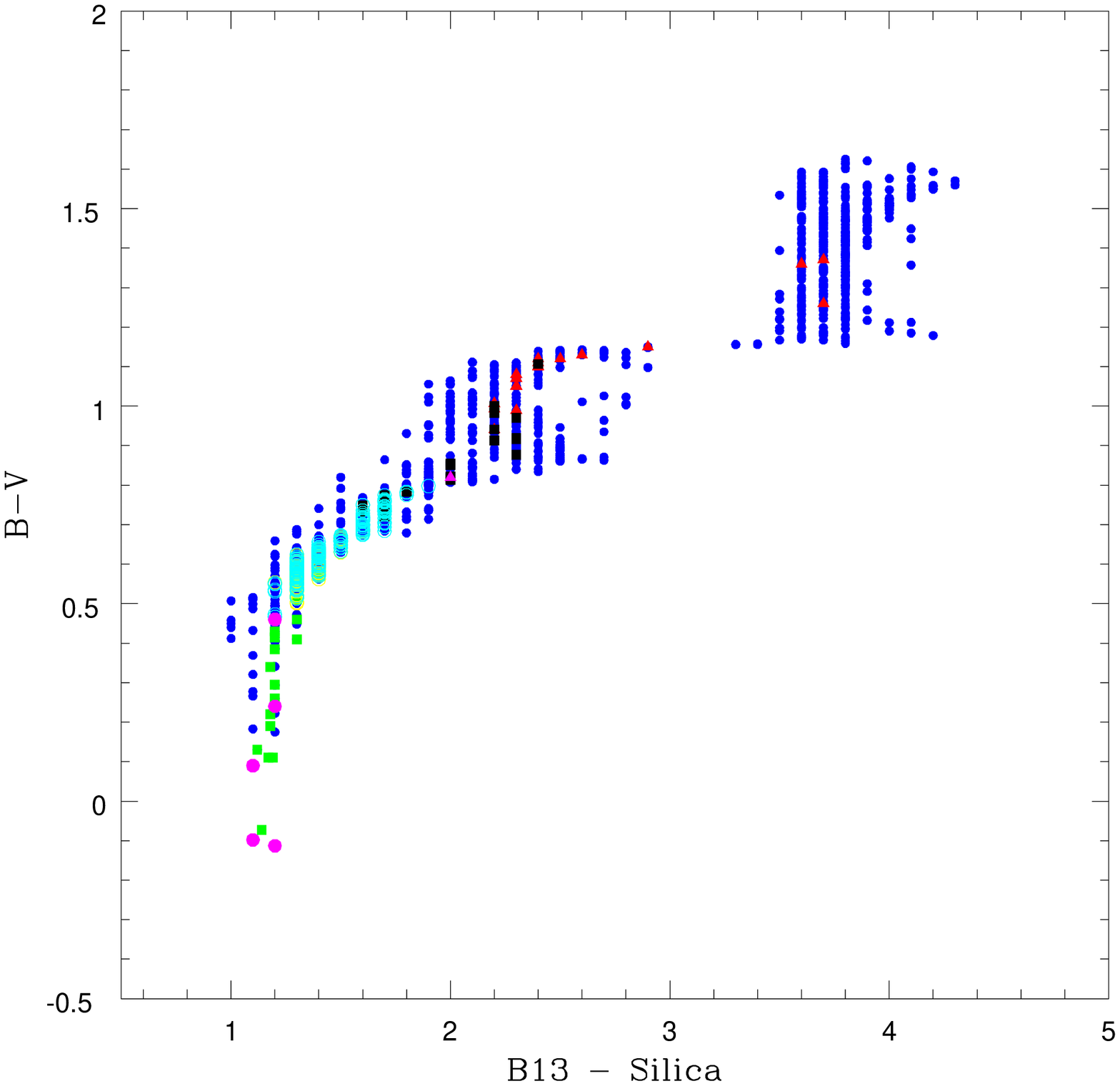}} \qquad

\caption{Colour$-$Colour diagram of M67 with NUV filters and B$-$V colours(i) B13$-$N2 (top left) (ii) Silica$-$N2 (top right) and (iii) B13$-$Silica(bottom) vs. B$-$V  \label{fig12}}
\end{figure}

\begin{figure}
\centerline{\includegraphics[width=7.5cm]{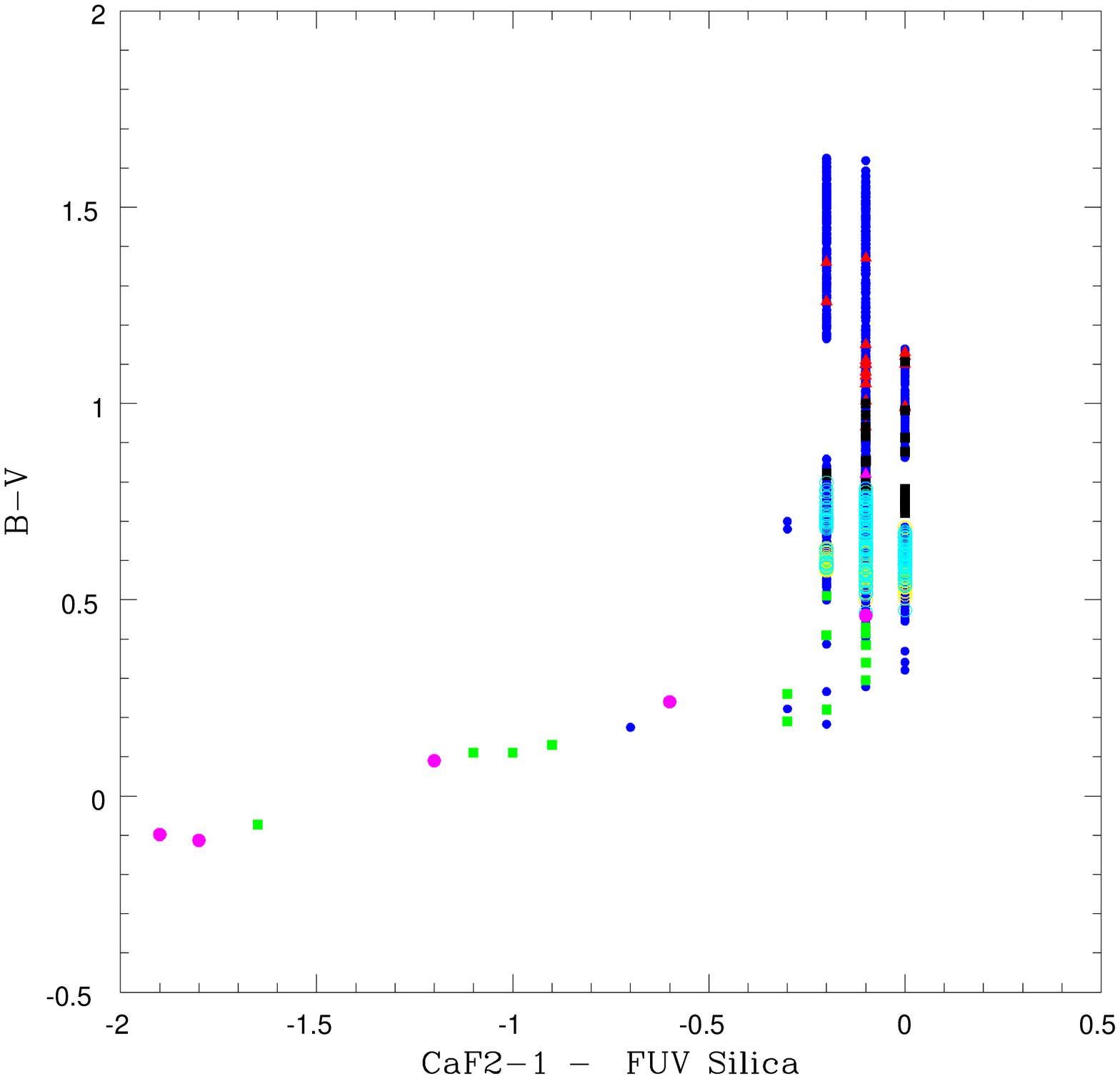} \quad
     \includegraphics[width=7.5cm]{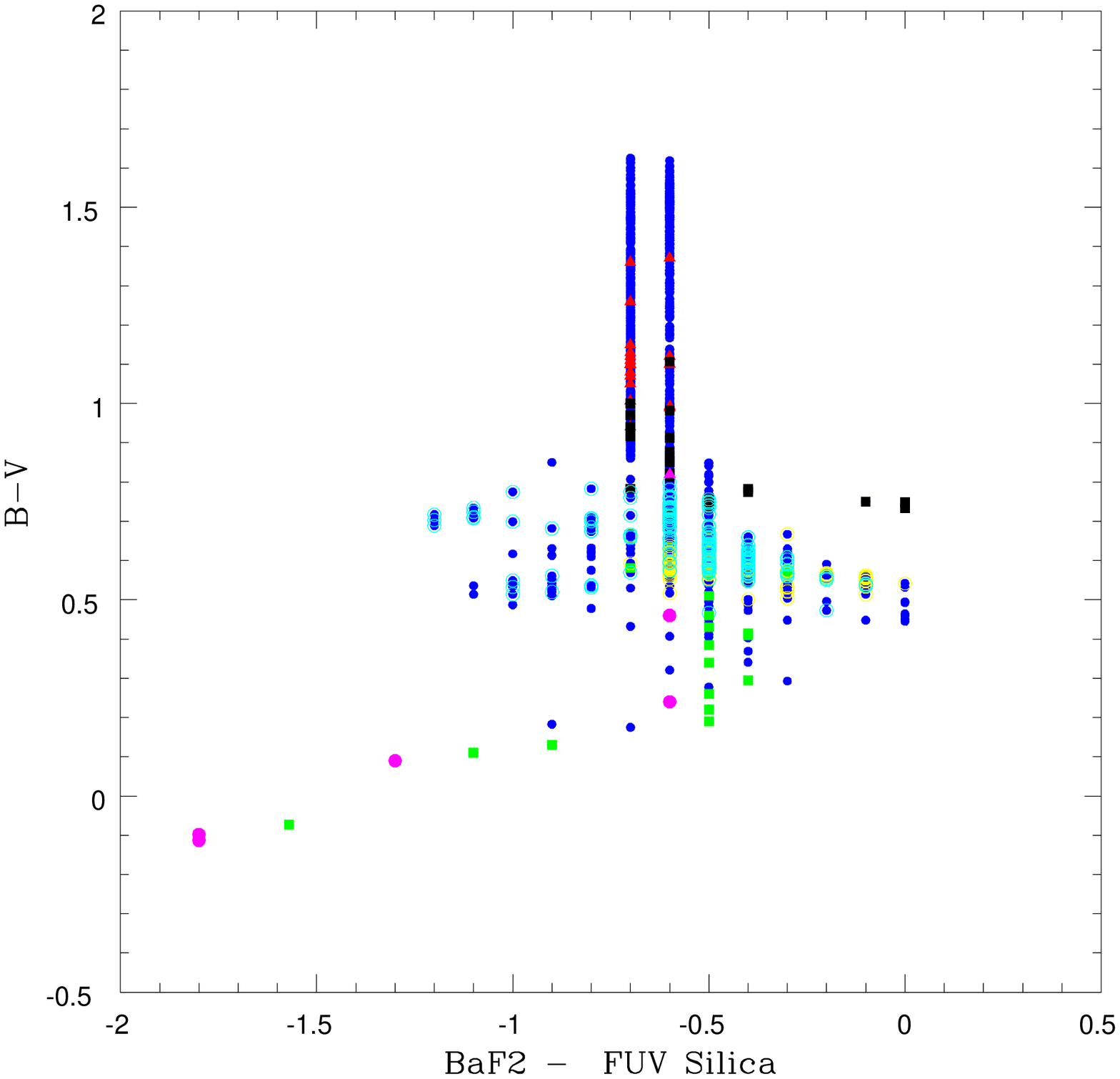}}
   \caption{ Colour - Colour diagram  of M67 with FUV filters and B$-$V colour. Left: CaF2$-$FUV Silica vs. B-V.  Right: BaF2- FUV Silica vs B-V \label{fig13}}
\end{figure}

\begin{figure}
\centerline{\includegraphics[width=7.5cm]{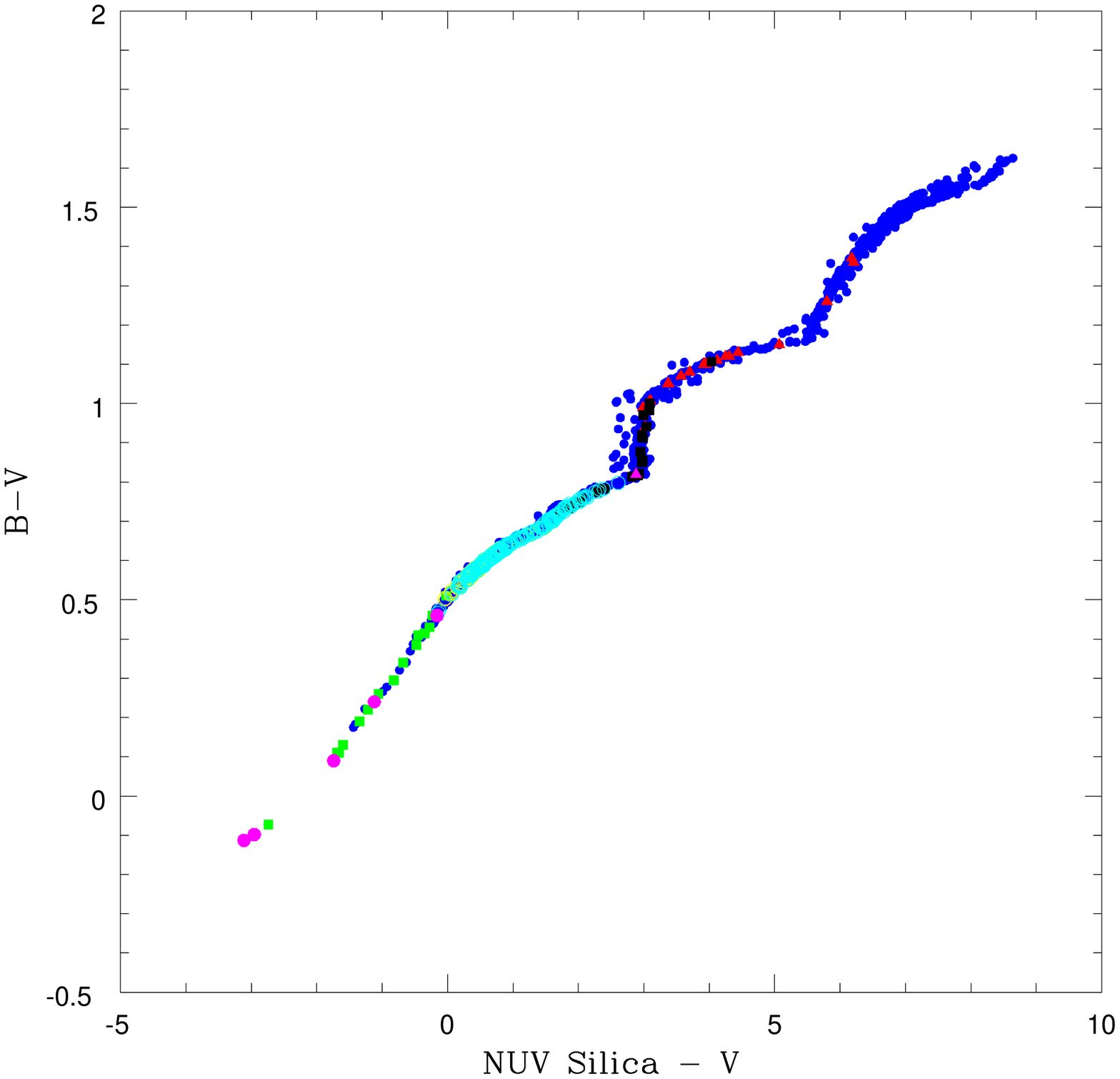} \quad
     \includegraphics[width=7.5cm]{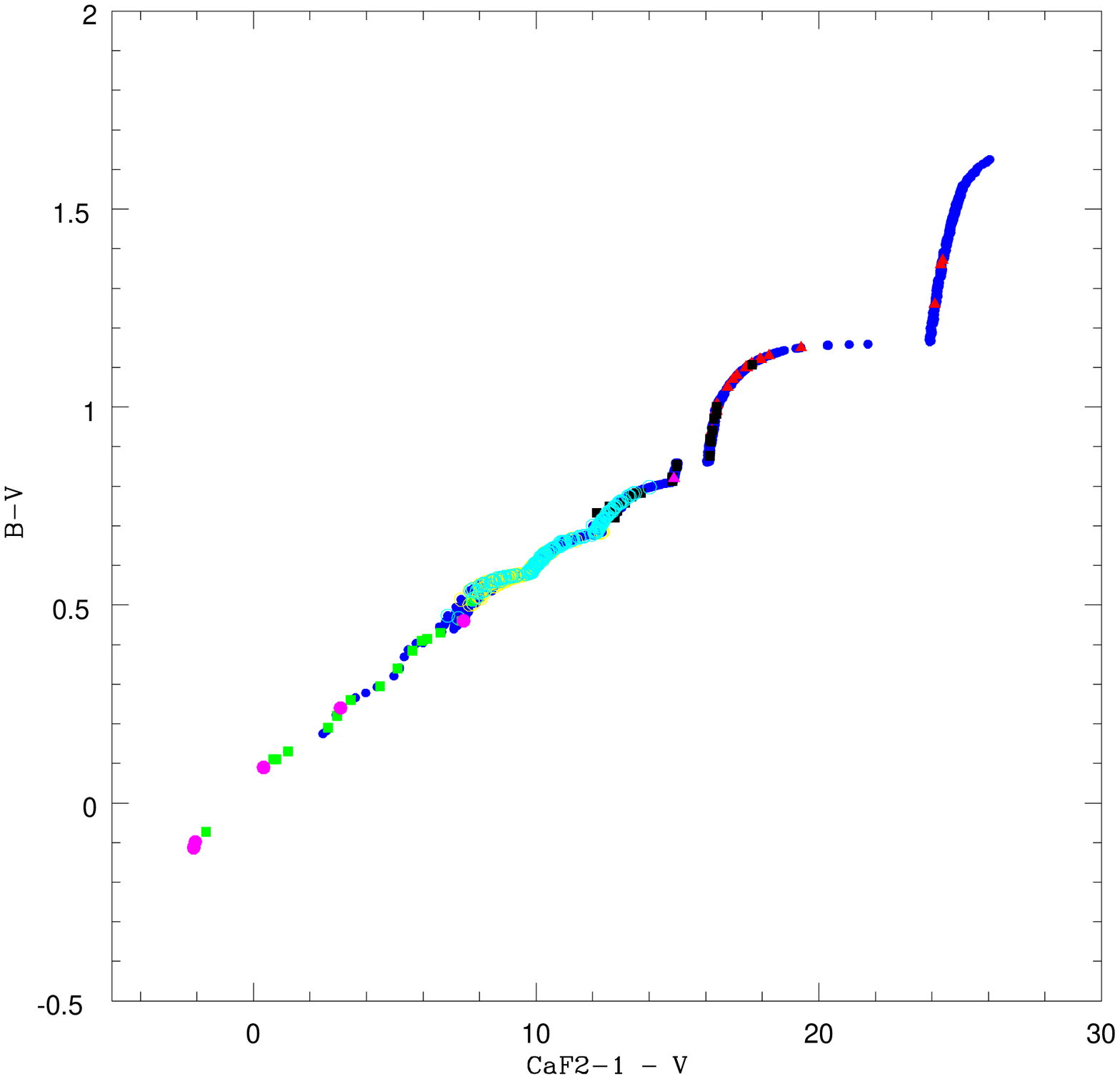}}
   \caption{ Colour - Colour diagram  of M67 using FUV and NUV filters $-$ V colour with B$-$V colour Left: NUV Silica$-$V vs. B$-$V.  Right: FUV CaF2$-$V vs. B$-$V \label{fig14}}
\end{figure}

As the UVIT has a large number of filters, it will be a good idea to check whether the colour-colour diagrams can be used as diagnostic diagrams to identify any specific evolutional phase, or separate stars of various evolutionary phases. 
In figures \ref{fig12}, \ref{fig13} and \ref{fig14}, we have shown various UV colours as a function of (B$-$V) colour for the cluster M67. We see that many
 of the diagrams separate WDs and blue stragglers from the MS 
 stars and red giants. On the other hand, the blue stragglers and WDs
 occupy similar location, also MS and red giants occupy similar location in most of the colour-colour diagram. In combination with the optical CMD, 
the colour-colour diagrams may be used effectively to separate hot
 stars in the cluster. We also plotted similar diagrams for the other two clusters studied here, which also have similar characteristics. As M67 has larger number of stars, we have shown only colour-colour diagrams for M67.

\section{Spatial appearance of M67, NGC 188 and NGC 6791 in FUV and NUV filters}
\label{sect:Spatial}
\begin{figure} [here]
\centerline{\includegraphics[width=7.5cm]{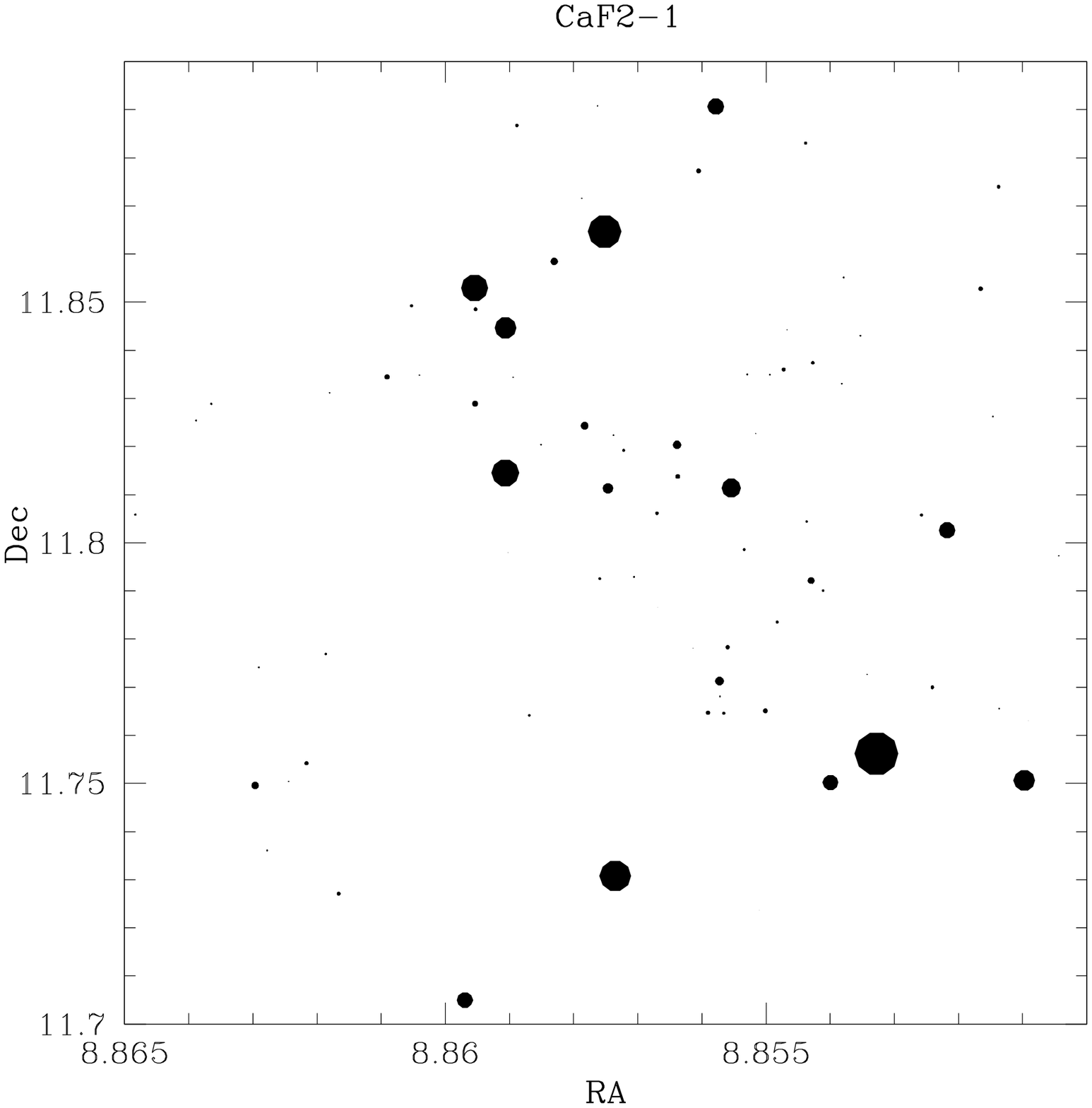} \qquad
     \includegraphics[width=7.5cm]{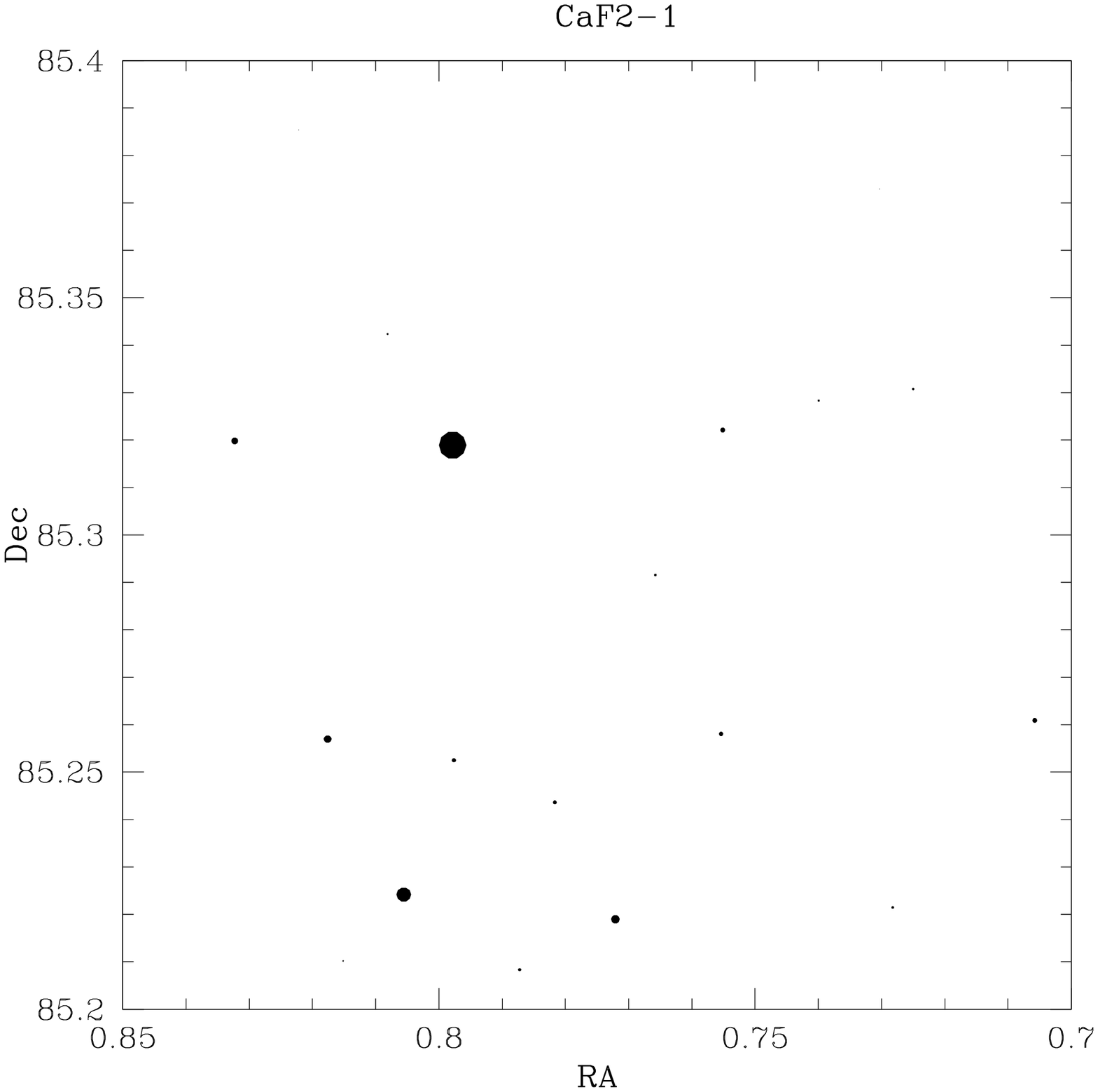}}
\centerline{\includegraphics[width=7.5cm]{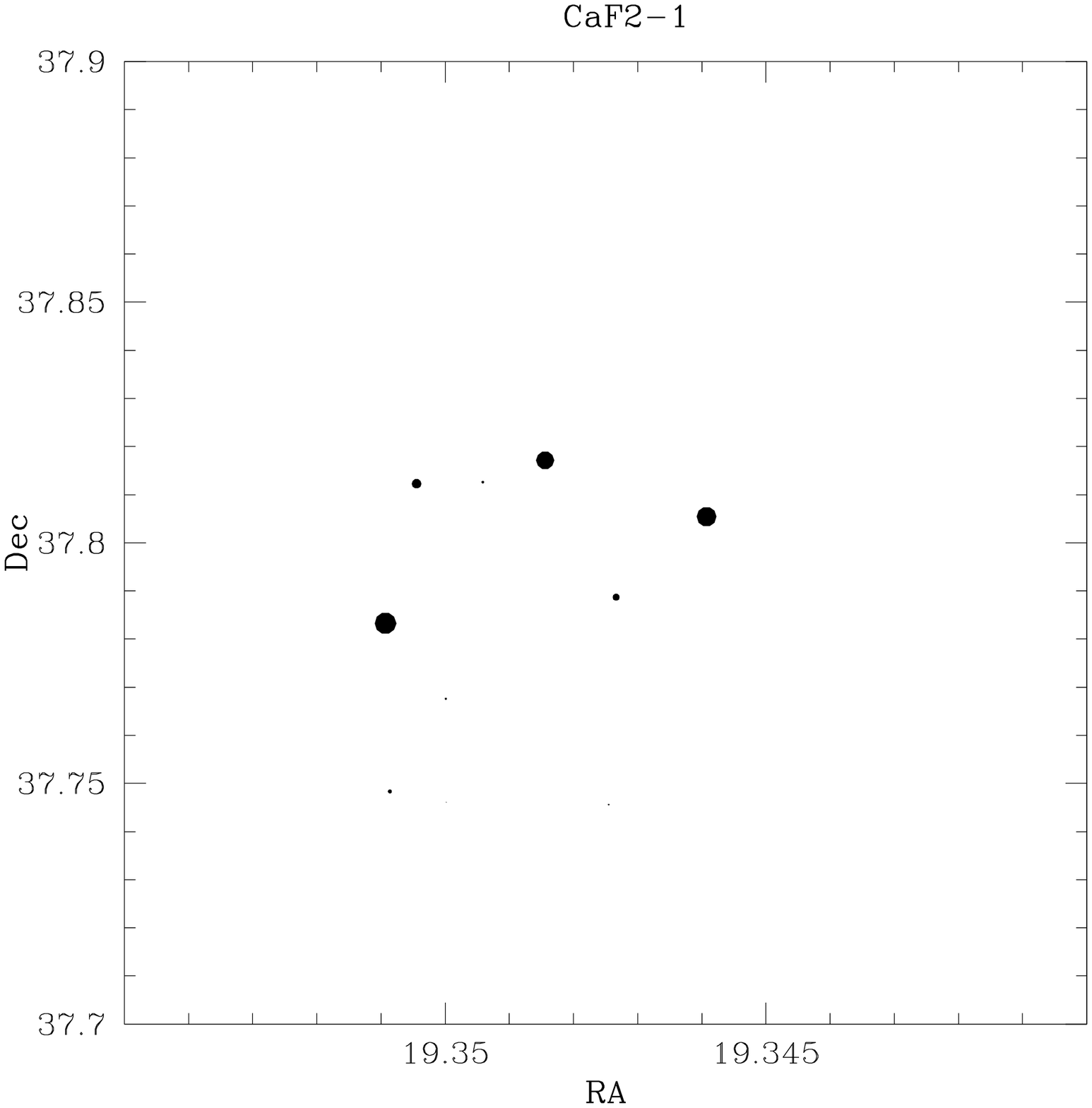}} \qquad
\caption{Spatial appearance of (i) M67 (top left) (ii) NGC 188 (top right) and (iii) NGC 6791(bottom) in the FUV CaF2-1 filter \label{fig15}}
\end{figure}

\begin{figure} [here]
\centerline{\includegraphics[width=7.5cm]{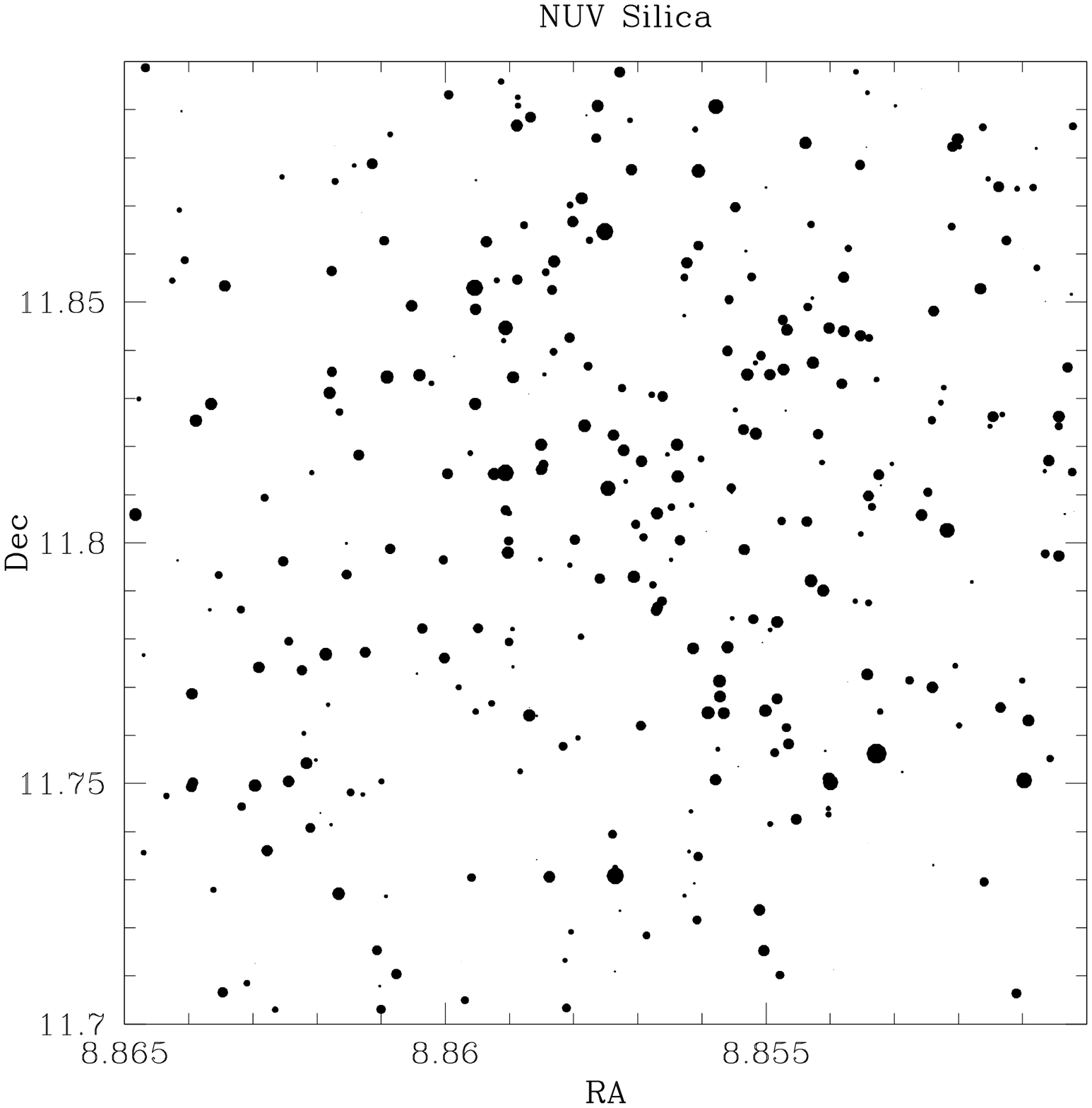} \qquad
     \includegraphics[width=7.5cm]{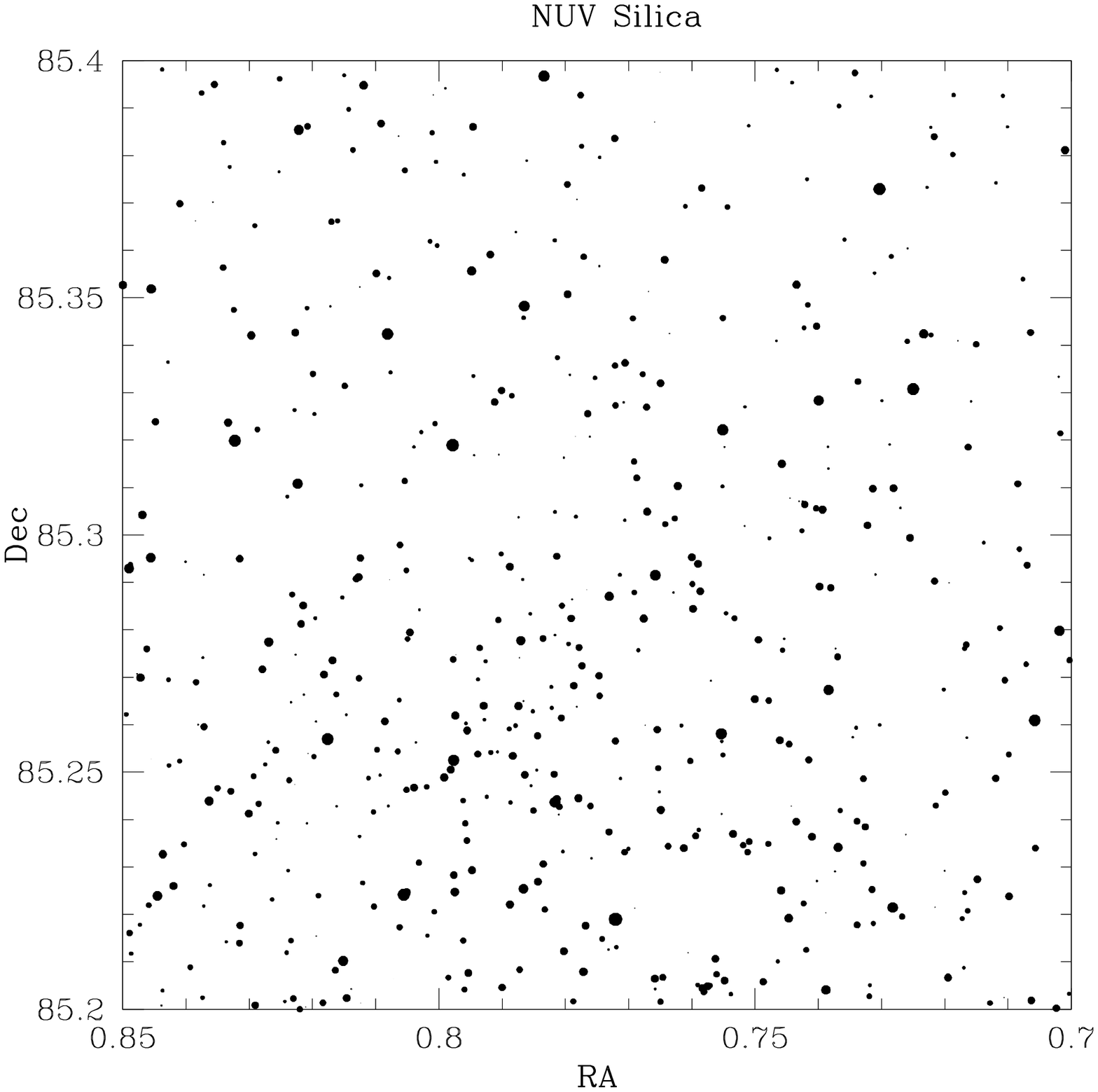}}
\centerline{\includegraphics[width=7.5cm]{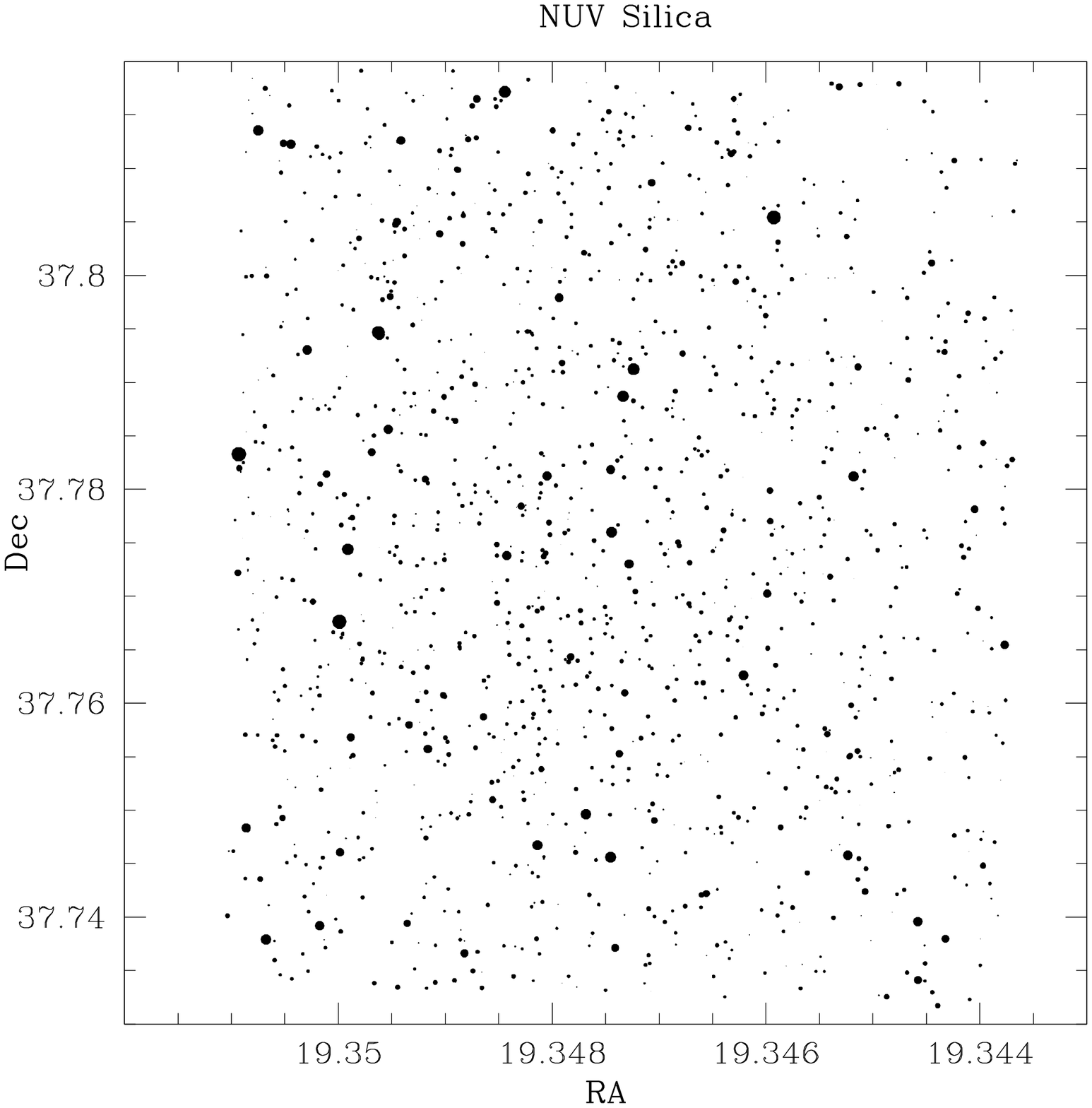}} \qquad
\caption{Spatial appearance of (i) M67 (top left) (ii) NGC 188 (top right) and (iii) NGC 6791 (bottom) in the NUV Silica filter \label{fig16}}

\end{figure}

\begin{figure} [here]
\centerline{\includegraphics[width=7.5cm]{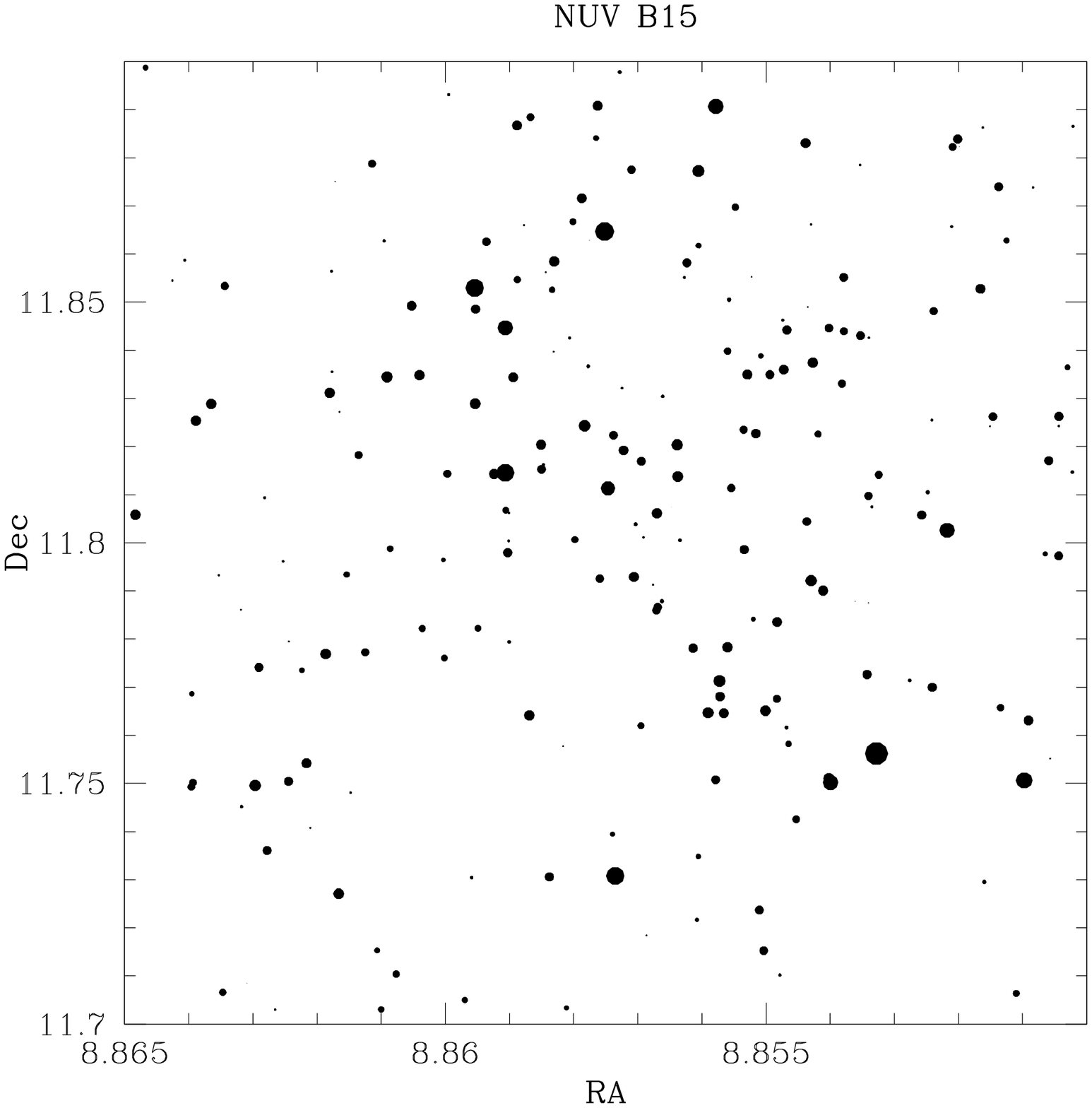} \qquad
     \includegraphics[width=7.5cm]{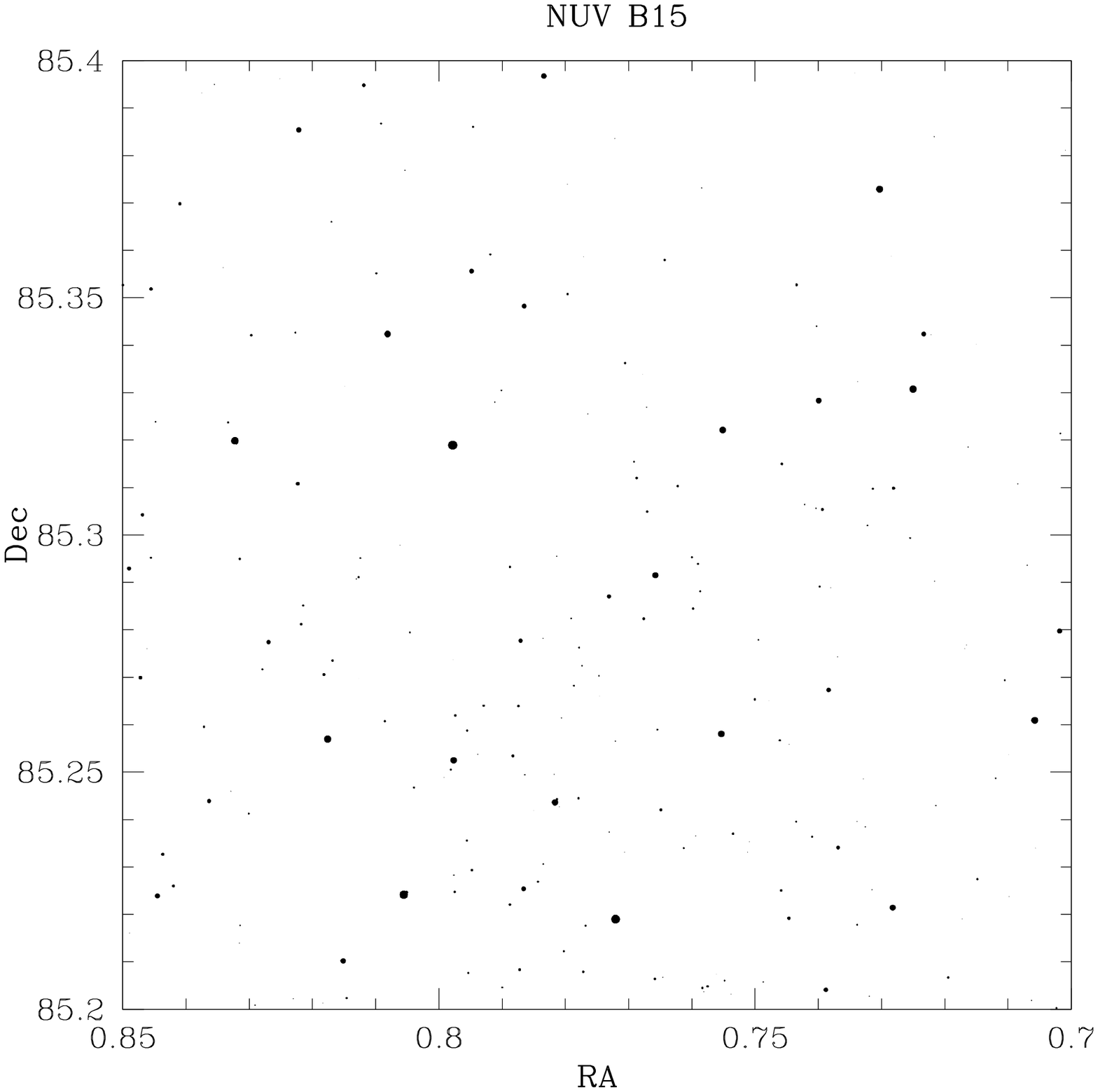}}
\centerline{\includegraphics[width=7.5cm]{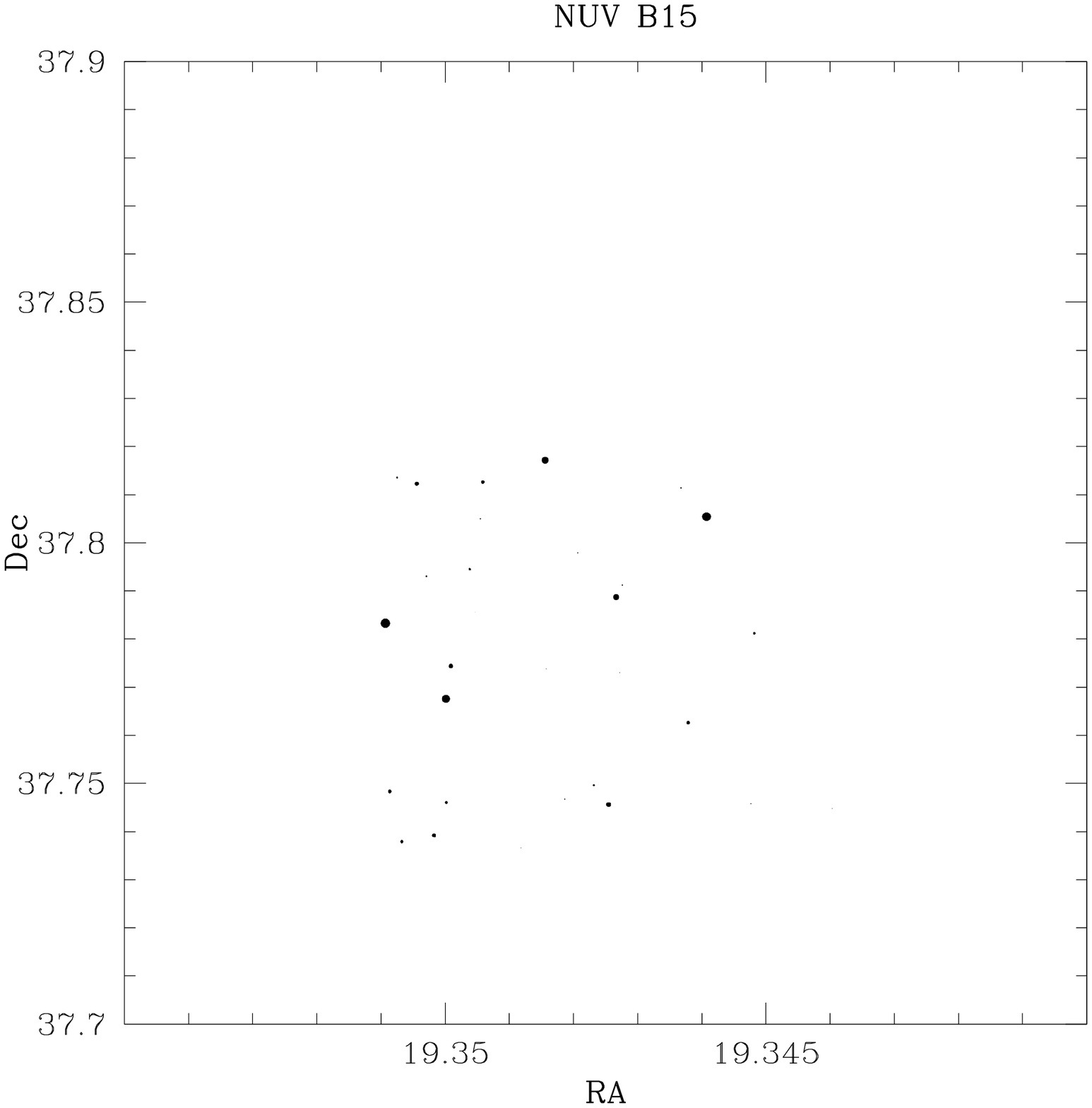}} \qquad
\caption{Spatial appearance of (i) M67 (top left) (ii) NGC 188 (top right) and (iii) NGC 6791(bottom) in the NUV B15 filter \label{fig17}}

\end{figure}

It is important to have a finding chart of the object to be studied using UVIT. A finding chart will thus provide the immediate confidence that the field observed is correct. The finding charts for the 3 clusters studied here are prepared using the magnitudes estimated in CaF2-1 filter for the FUV, Silica and B15 for the NUV region. The plots are made such that the size of the stars are scaled according to the magnitude such that brighter stars appear bigger. The spatial appearance of all three clusters are shown in figures  \ref{fig15}, \ref{fig16} and \ref{fig17}.

We observe that there is  a relatively lesser crowding of stars in the FUV filters. Each point in the plot indicates a star, and only stars which are brighter appear in the diagram. Therefore, in the FUV, as suggested by the CMDs, all those we detect will be  WDs and blue stragglers. Thus, figure \ref{fig15} ,\ref{fig16},\ref{fig17} shows primarily the location of these stars and will be useful while doing the actual observations. 

We have presented the appearance in two NUV filters, Silica and B15. The spatial appearance of the cluster in these filters appears similar, but Silica will be more efficient due to its larger wavelength coverage. The appearance in NUV is very similar to the optical, except that there is a variation in the relative brightness. Thus, the NUV field connects the optical and FUV field, such that the  WDs and blue stragglers get progressively brighter. We expect that these finding charts will be helpful during actual observations. Siegel et al. (2014) presented the observed field of NGC 188 in their figure 1. As the field of view of UVOT is much smaller, they presented the central region of the cluster. We are unable to perform a one-to-one comparison of our simulations with these observations, as the filters systems are different and the UVOT filters have high red leaks.

\section{Comparison with GALEX and SWIFT-UVOT}
\begin{figure}
\includegraphics[width=12cm]{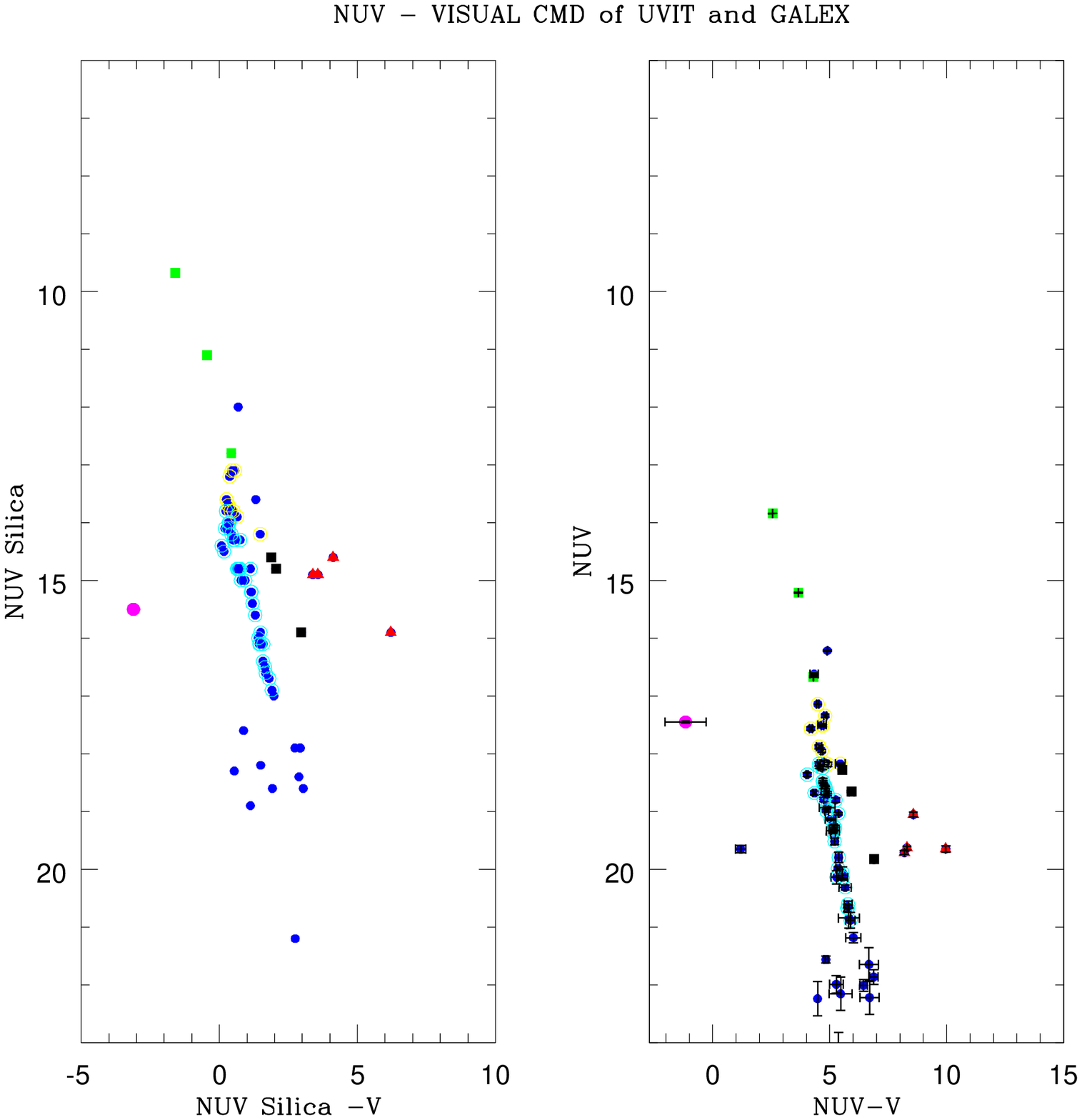} 
\centering
 \caption{Comparison of NUV-V CMD plot of UVIT and GALEX. The errorbars are not shown for few stars, whose error in V are not available.\label{fig18}}
\end{figure}

We compared our simulations with observations of M67 with GALEX, and two clusters with SWIFT-UVOT.
The GALEX has two channels, NUV and FUV. The wavelength coverage of the GALEX NUV matches with the NUV Silica filter in the UVIT.
Similarly, the wavelength coverage of GALEX FUV matches with CaF2 filters in UVIT. As there are only very small number of stars detected in the FUV region, we compare the NUV magnitudes of GALEX with the simulated data for the filter NUV Silica, for the cluster M67. We have shown the M67 simulated CMD and observed CMD from GALEX, side by side, in figure \ref{fig18}. The location of the MS, blue stragglers, sub giants and giants are very similar. This gives us confidence that the  patterns and relative locations of various stars presented in the simulated CMD are likely
to be correct. The main difference between the two plots in figure \ref{fig18}, is the difference in NUV magnitude and hence the colour, such that the simulated magnitudes are brighter. This is due to with the adopted zero-point. We computed the difference between the GALEX NUV and simulated magnitude (NUV Silica) as 4.1$\pm$0.4 mag, for 59 common stars. This difference is same as the difference in the adopted zero-point for UVIT (16 mag) and that of GALEX (20.06 mag). Hence we believe that the shift in the magnitudes can be corrected with the correct estimation of the zero-point, which will be done after the launch during the initial calibration phase. We also computed the difference between the GALEX FUV and the simulated FUV CaF2-1 magnitude, for 8 common stars, which is found to be 2.7$\pm$1.2 mag. Again, the scatter is found to be large, largely due to the filter mismatch. The zero-point for the GALEX FUV filter is 18.82, which is similar within error, when we add the above difference to our zero-point.

\begin{figure} [here]
\centering
\includegraphics[width=12cm]{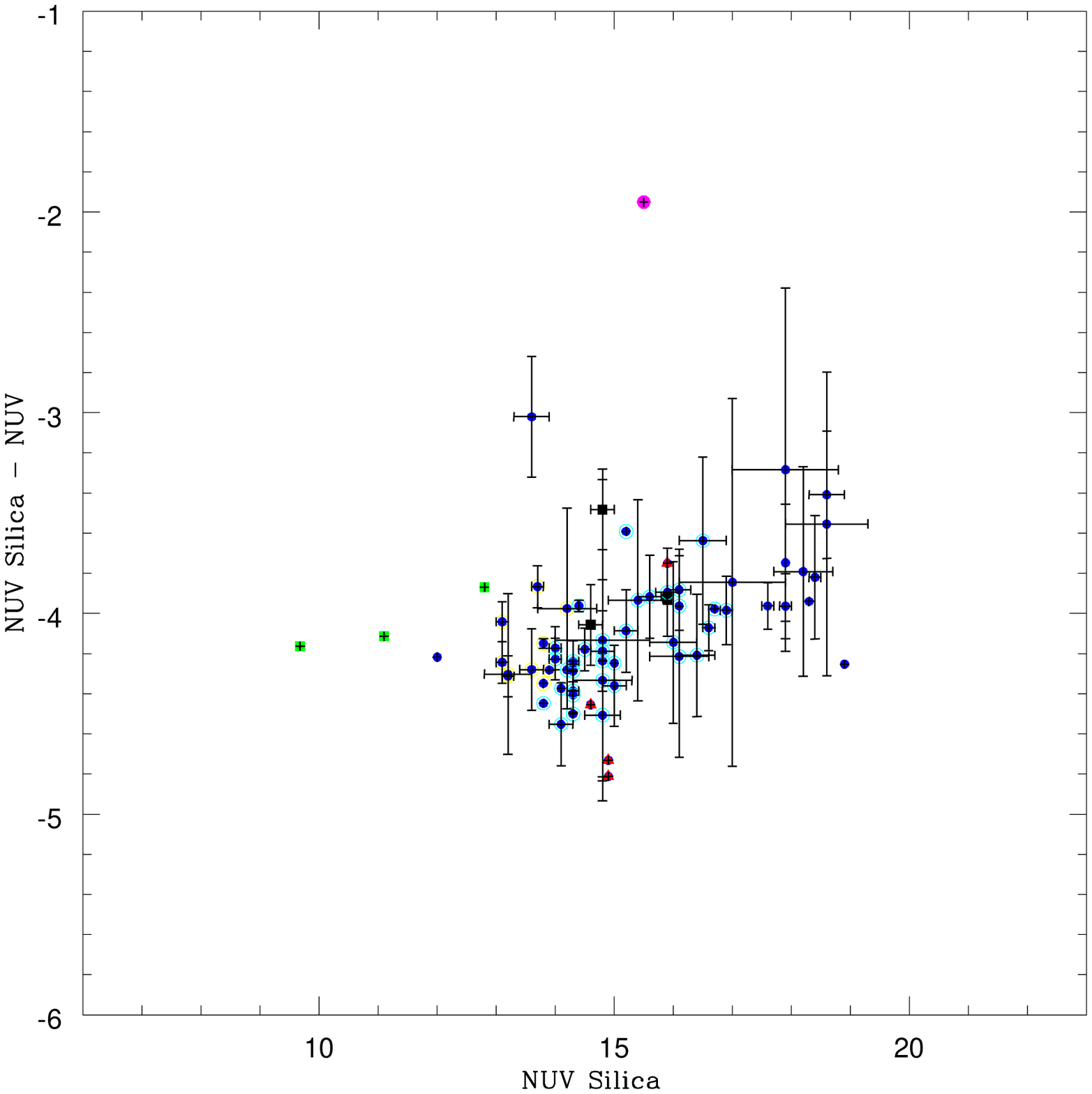} 
 \caption{The simulated NUV Silica magnitude is shown as a function of difference between the simulated NUV Silica magnitude and GALEX NUV magnitude.The errorbars are not shown for few stars, whose error in V are not available.The error bar is not shown for (NUVSilica-NUV) = -1.9502 and NUVSilica =15.5, since it is too large. \label{fig19}}
\end{figure}

\cite{seigel14} presented the CMDs of M67 and NGC 188 using UVOT observations. The features identified in our study and those in their CMDs are similar. For example, in the CMD using the NUV colour, the blue stragglers are bright and have a larger range of magnitude. Also, the MS and the sub giant branch occupy the same space making it difficult to separate the two sequences in the CMDs, similar to  that suggested by our simulations. The other NUV CMD shows the sub giant branch being almost parallel to the MS. The CMDs also show a wider colour separation for the  WDs. Hence, we find that our simulations are comparable to the CMDs presented by \cite{seigel14}, which once again gives confidence to the simulations presented in this study. The large scatter is due to the mismatch between the two filters and its effect on the hotter ( WDs) and cooler (giants) stars, as shown in figure \ref{fig19}.

\section{Discussion and Conclusion}
\label{sect:discussion}
UVIT is one of the state of the art telescopes to be flown in the multi-wavelength ASTROSAT mission. This telescope has an angular resolution much better than the previous missions and it also carries a set of very useful filters. Simulations to predict the expected output from the UVIT payload for various science topics of interest are highly essential. Particularly because it has filters within the FUV and NUV channels for the first time. The results presented in this paper are part of the preparatory exercise to understand and efficiently utilise the telescope and filter capability to maximise the science output. In this study we have presented simulated CMDs of three open clusters and their spatial appearance in the UVIT filters. We have used the ETC as well as a stand alone software to estimate count rates of large number of stars in three clusters. We have assumed a zero-point to convert the count rates to magnitudes, which is compared with the GALEX zero-points. The  CMDs are plotted using all the filter combinations of FUV and NUV. We are able to identify the location of various evolutionary sequences in the cluster  CMD of M67, NGC 188 and NGC 6791. Stars and evolutionary sequences that are identified here are those which are previously known in the optical  CMD. The identification of expected location for stars of known type will help in identifying new candidates in studied and unstudied clusters. The present study recommends most suitable filter combination to study various stellar sequences that are identified.

We compared our simulated CMDs in the NUV with those obtained from GALEX and UVOT. The locations of stars of various evolutionary phases are found to be comparable. The magnitudes of simulated stars match well that of GALEX, in both NUV and FUV, after accounting for the difference in the zero-point. While studying this cluster using the UVIT, there could be a shift in the position of the entire colour and magnitude depending on the photometric zero point value for various filters. But the relative position of the stars in the cluster with respective to each other is expected to remain the same. This indicates that pattern of the CMD that we observe in the simulated FUV CMD, NUV CMD and NUV- Visual would be similar. Hence we would like to stress that the magnitudes should be taken with caution and the CMDs presented here are indicative to the extend that they help us in planning efficient observations. Using the study, we also demonstrate that it is important to prepare diagnostic tools using the available filter combinations to do science with UVIT. The ground calibration estimates as well as ETC related tools are available to execute such studies. We plan to extend such studies to other topics, such as young star clusters, globular clusters, star forming galaxies etc.

We summarise our results below:
\begin{itemize}
 \item[$\bullet$] We have presented simulated CMDs of three old open clusters for various combinations of FUV, NUV and optical (V) filters. It is observed that the FUV CMDs will be able to identify only the blue stragglers and  WDs in these open clusters. They can be separated and grouped with the help of various FUV filters. The upper MS and the turn-off are detected in the NUV CMDs in M67 and NGC 188, but only the turn-off in NGC 6791. The CMDs generated using NUV and V magnitudes are found to be very useful.
\item[$\bullet$] From the analysis of the FUV CMDs we observe that the blue stragglers have increased FUV magnitude range, which would help in grouping them. We point out that the increased range in the FUV magnitude is also likely to help in the study of turn-off of younger open clusters to obtain better resolution in age. We plan to study younger clusters in the future.
\item[$\bullet$] In the NUV CMDs, the upper MS, the turn-off, sub-giants as well as the red giants are located very close to each other, which would make it difficult to differentiate them from one another. Identification of these stars from the optical CMD is thus very essential.
\item [$\bullet$]The spatial appearance  of the three clusters in the broad band FUV and NUV filters are presented, lesser crowding of stars is observed. These can be used as finding charts during the actual observations to  help us identify the field.
\item [$\bullet$] Our simulations compare well with the estimations from GALEX and SWIFT-UVOT. 
\item [$\bullet$] A tool to estimate count rates of a large number of stars in all the UVIT filters is developed as part of this work, which will be made available to the community as a web tool.
 \end{itemize}

\normalem
\begin{acknowledgements}
Sindhu thanks IIA for all the support during her MPhil project.
This research has made use of the VizieR catalogue access tool, CDS, Strasbourg,
France and WEBDA database, operated at the Department of Theoretical Physics and Astrophysics of the Masaryk University

\end{acknowledgements}
  
\bibliographystyle{raa}
\bibliography{bibtex}

\end{document}